\documentclass[journal]{IEEEtran}
\usepackage[usenames, dvipsnames]{color}
\usepackage{amssymb}
\usepackage{amsmath}
\usepackage[makeroom]{cancel}
\usepackage{graphicx}
\usepackage{enumerate}
\usepackage{verbatim}
\usepackage{url}
\urldef{\mailsa}\path|{gurgurka}@boun.edu.tr|
\usepackage{siunitx}
\usepackage{caption}
\usepackage{mathtools}
\usepackage{algorithm}
\usepackage{algorithmic}
\usepackage{mwe}

\usepackage{subfig}
\usepackage{graphicx}

\usepackage{tikz}
\usetikzlibrary{arrows,automata,positioning}
\usetikzlibrary{shapes}

\newcommand{\pbaselocal}[1]{p_{c_{#1}}^{lo}}

\newcommand{\pbasesat}[1]{p_{c_{#1}}^{sat}}

\newcommand{\pbaseBS}[1]{p_{c_{#1}}^{BS}}

\newcommand{\pdtwod}[2]{p_{c_{#1}}^{D(f_1)}(#2)}

\newcommand{\lambdaContent}{\lambda_{c}}
\newcommand{\lambdaci}[1]{\lambda_{HU}^{c_{#1}}}

\newcommand{\probDTwoDTx}[1]{\Pi_{tx}(#1)}
\newcommand{\probDTwoDRec}[1]{\Pi_{rec}(#1)}
\newcommand{\probDTwoDProb}[1]{\Pi_{c_{#1}}}

\newcommand{\rate}[1]{r_{#1}}
\newcommand{\lambdacioverBase}[1]{\lambda_{HU}^{c_{#1}}}

\newcommand{\PerChannelSatTxPower}{P_{sat}^{ch}}
\newcommand{\PerChannelBSTxPower}{P_{BS}^{ch}}
\newcommand{\DeviceTxPower}{P_{dev}^{tx}}

\newcommand{\lambdaPUTer}{\lambda_{PU}^{ter}}

\newcommand{\lambdaHU}{\lambda_{HU}}

\newcommand{\MeanBaseContentSize}{\hat{s(v_b)}}

\newcommand{\SatCacheSize}{Cache_{sat}}
\newcommand{\BSCacheSize}{Cache_{BS}}
\newcommand{\HUDeviceCacheSize}{Cache_{Dev}}

\newcommand{\SatFrequency}{f_{sat}}
\newcommand{\TerFrequency}{f_{ter}}
\newcommand{\SatBandwidth}{W_{sat}}
\newcommand{\TerBandwidth}{W_{ter}}
\newcommand{\NSat}{N_{f_{sat}}}
\newcommand{\NTer}{N_{f_{ter}}}
\newcommand{\RatioSat}{r_{sat}}
\newcommand{\RatioBS}{r_{BS}}
\newcommand{\RatioHUDev}{r_{dev}}

\newcommand{\MeanDistancetoBS}{d_{BS}}
\newcommand{\DistancetoSat}{d_{sat}}
\newcommand{\DistanceDD}{d_{D2D}}

\newcommand{\CapacityHUSat}{C_{HU}^{sat}}

\newcommand{\CapacityHU}{C_{HU}^{x}}

\newcommand{\iPUTer}{i_{PU}^{ter(\overline{f_1})}}
\newcommand{\iPUTerfOne}{i_{PU}^{ter(f_1)}}

\newcommand{\iHUSat}{i_{HU}^{sat}}
\newcommand{\iHUSatUniv}{i_{HU}^{sat(u)}}
\newcommand{\iHUBS}{i_{HU}^{BS}}
\newcommand{\iHUDev}{i_{HU}^{D(f_1)}}
\newcommand{\iHUBSUniv}{i_{HU}^{BS(u)}}

\newcommand{\state}[1]{\textrm{state}_{{\tiny#1}}}
\newcommand{\gammaHUSatU}[2]{\gamma_{HU}^{sat(u)}(#1,#2)}
\newcommand{\gammaHUBSU}[2]{\gamma_{HU}^{BS(u)}(#1,#2)}
\newcommand{\gammaHUSat}[2]{\gamma_{HU}^{sat}(#1,#2)}
\newcommand{\gammaHUBS}[2]{\gamma_{HU}^{BS}(#1,#2)}
\newcommand{\gammaHUD}[2]{\gamma_{HU}^{D(f_1)}(#1,#2)}
\newcommand{\gammaHUSatAgg}[1]{\Gamma_{HU}^{sat}(#1)}
\newcommand{\gammaHUSatAll}{\Gamma_{HU}^{sat}}
\newcommand{\gammaHUBSAgg}[1]{\Gamma_{HU}^{BS}(#1)}
\newcommand{\gammaHUBSAll}{\Gamma_{HU}^{BS}}

\newcommand{\gammaHUSatUAgg}[1]{\Gamma_{HU}^{sat(u)}(#1)}
\newcommand{\gammaHUSatUAll}{\Gamma_{HU}^{sat(u)}}
\newcommand{\gammaHUBSUAgg}[1]{\Gamma_{HU}^{BS(u)}(#1)}
\newcommand{\gammaHUBSUAll}{\Gamma_{HU}^{BS(u)}}

\newcommand{\gammaHUDAgg}[1]{\Gamma_{HU}^{D(f_1)}(#1)}
\newcommand{\gammaHUDAll}{\Gamma_{HU}^{D(f_1)}}

\newcommand{\floor}[1]{\lfloor #1 \rfloor}

\newcommand{\RBS}{R_{BS}}
\newcommand{\RInt}{R_{Int}}

\newcommand{\DMAX}{D_{max}}

\newcommand{\lambdaNHU}{\lambda_{N_{HU}}}

\newcommand{\meanservicedurationHUSatU}{\Delta_{HU}^{sat(u)}}
\newcommand{\meanservicedurationHUBSU}{\Delta_{HU}^{BS(u)}}
\newcommand{\meanservicedurationHUUniv}{\Delta_{HU}^{x(u)}}

\newcommand{\capHUSatU}{C_{HU}^{sat(u)}}
\newcommand{\capHUBSU}{C_{HU}^{BS(u)}}
\newcommand{\capHUUniv}{C_{HU}^{x(u)}}

\newcommand{\stateiHUSat}[1]{i_{HU}^{sat}(#1)}
\newcommand{\stateiHUBS}[1]{i_{HU}^{BS}(#1)}
\newcommand{\stateiHUSatU}[1]{i_{HU}^{sat(u)}(#1)}
\newcommand{\stateiHUBSU}[1]{i_{HU}^{BS(u)}(#1)}
\newcommand{\stateiHUD}[1]{i_{HU}^{D(f_1)}(#1)}
\newcommand{\stateiPUSat}[1]{i_{PU}^{sat}(#1)}
\newcommand{\stateiPUTer}[1]{i_{PU}^{ter(\overline{f_1})}(#1)}
\newcommand{\stateiPUTerfOne}[1]{i_{PU}^{ter(f_1)}(#1)}

\newcommand{\idleSat}[1]{idle_{s}(#1)}

\newcommand{\idleTerExceptfOne}[1]{idle_{t,\overline{f_1}}(#1)}

\newcommand{\muPUTer}{\mu_{PU}^{ter}}

\newcommand{\muHUSat}{\mu_{HU}^{sat}}
\newcommand{\muHUBS}{\mu_{HU}^{BS}}
\newcommand{\muHUDev}{\mu_{HU}^{D}}

\newcommand{\muHUSatU}{\mu_{HU}^{sat(u)}}
\newcommand{\muHUBSU}{\mu_{HU}^{BS(u)}}
\newcommand{\muHUUniv}{\mu_{HU}^{x(u)}}

\newcommand{\epbHU}{EPB_{HU}}

\newcommand{\BSPower}{P_{BS}}
\newcommand{\BSUnivPower}{P_{BS(u)}}
\newcommand{\DTwoDPower}{P_{D2D}}
\newcommand{\localPower}{P_{local}}
\newcommand{\overallPower}{P_{overall}}

\newcommand{\RATESAT}[1]{R_{sat}(#1)}
\newcommand{\RATEBS}[1]{R_{BS}(#1)}
\newcommand{\RATEDD}[1]{R_{D2D}(#1)}

\newcommand{\goodputHU}{G_{HU}}
\newcommand{\goodputHUSat}{Th_{HU}^{sat}}
\newcommand{\goodputHUSatUniv}{Th_{HU}^{sat(u)}}
\newcommand{\goodputHUBS}{Th_{HU}^{BS}}
\newcommand{\goodputHUBSUniv}{Th_{HU}^{BS(u)}}
\newcommand{\goodputHUDTwoD}{Th_{HU}^{D}}
\newcommand{\goodputHULocal}{G_{HU}^{local}}

\newcommand{\effarrHUsat}{\lambda_{eff(HU)}^{sat}}
\newcommand{\effarrHUsatuniv}{\lambda_{eff(HU)}^{sat(u)}}
\newcommand{\effarrHUBS}{\lambda_{eff(HU)}^{BS}}
\newcommand{\effarrHUBSuniv}{\lambda_{eff(HU)}^{BS(u)}}
\newcommand{\effarrHUdev}{\lambda_{eff(HU)}^{D2D}}

\newcommand{\pdropBS}{p_{drop}^{BS}}
\newcommand{\pdropDTWOD}{p_{drop}^{D2D}}
\newcommand{\plocal}{p_{local}}

\newcommand{\receiveBSparameter}{\theta_{BS}}
\newcommand{\receiveLocalparameter}{\theta_{loc}}

\newcommand{\be}{\begin{equation}}
\newcommand{\ee}{\end{equation}}
\newcommand{\bea}{\begin{eqnarray}}
\newcommand{\eea}{\end{eqnarray}}
\newcommand{\ba}{\begin{array}}
\newcommand{\ea}{\end{array}}
\newcommand{\bt}{\begin{tabular}}
\newcommand{\et}{\end{tabular}}
\newcommand{\bfin}{\begin{figure}}
\newcommand{\bfi}{\begin{figure}[!htb]}
\newcommand{\efi}{\end{figure}}

\begin{document}

\title{Analysis of Content-Oriented Heterogeneous Networks with D2D and Cognitive Communications}

\author{\IEEEauthorblockN{S. Sinem Kaf{\i}lo\u{g}lu, G\"{u}rkan G\"{u}r$^\dagger$ and
Fatih Alag\"{o}z}\\
\IEEEauthorblockA{Department of Computer Engineering, Bogazici University\\ Istanbul, Turkey\\
$^\dagger$Zurich University of Applied Sciences (ZHAW)\\ Winterthur, Switzerland\\ Email: \{sinem.kafiloglu, fatih.alagoz\}@boun.edu.tr}, gueu@zhaw.ch
\thanks{This work was supported by the Scientific and Technical Research Council of Turkey (TUBITAK) under grant number 116E245.}
}

\maketitle
\begin{abstract}
Content-oriented operation with D2D communications and spectrum sharing in the form of cognitive communications are expected to be enablers for cost-efficient next-generation wireless systems with bandwidth-hungry services. To this end, caching is an important building block of these systems. In this work, we model such a heterogeneous network and model a popularity-aware caching algorithm where we try to keep more popular contents in caches of system units with a higher probability compared to less popular contents. 
Moreover, for improving the usage of the channel capacity, overlaying for device-to-device (D2D) transmissions (unless these transmission interfere with each other) is enabled. We model our system with an analytical continuous time Markov chain (CTMC). We have a heterogeneous network architecture consisting of a LEO satellite, base station (BS) and a set of devices within the boundary of BS coverage that can get service from the satellite, from the BS or in D2D mode. Our devices operate in primary mode when they get service from the satellite but in secondary mode when they get service from the BS or some device in D2D mode over the \textbf{terrestrial} link. 
We evaluate our system for energy efficiency (EE) and goodput metrics. We looked how popularity-aware caching, integration of universal source concept and overlaying in D2D mode affect the network performance. Besides, we investigate how resource allocation (RA) and performance metrics are tangled. 
\end{abstract}

\section{Introduction}
\label{sec:introduction}
The reduction of energy consumption in communication networks is desired for cost efficiency and minimized environmental impact~\cite{6046158}. Nevertheless, Future Internet is envisaged to serve multimedia heavy traffic with x1000 capacity improved to current networks. According to Cisco VNI report, globally, IP video traffic will be 82\% of all consumer Internet traffic by 2021~\cite{CISCOWhitePaper}. Evidently, the energy consumption of the content traffic needs to be explored for giving an insight into Future Internet. Now, we will be focusing on a specific sub-domain of Future Internet architecture, that is satellite-terrestrial network with D2D extension. In this architecture, our objective is to improve EE for keeping energy consumption low for reducing energy cost for operators and also keeping battery discharging rate of devices low~\cite{6046158}. But there is a trade-off between EE and QoS. QoS needs to be constrained by a lower bound. 

The intersection of satellite and cellular networks with D2D and cognitive extension is yet to be explored. In our previous works~\cite{7492943,7904715}, the energy efficiency (EE) is inspected for the resource allocation (RA) of contents in such a heterogeneous network system. 
In this work,  we model a content-oriented heterogeneous network with D2D and cognitive communications and model a popularity-aware  caching algorithm where we try to keep more popular contents in system unit caches with a higher probability compared to less popular contents. We aim to alleviate the incomplete treatment of satellite-terrestrial networks with overlaid D2D communication with an optimization perspective for the improvement of EE. 

In our model, we have hybrid users (HUs) that are in primary mode at the satellite link. HUs are in  secondary mode at the terrestrial link and can get service from the BS (BS mode) or operate in D2D mode. First, we model system unit caches. Then, we calculate steady state probability of system units for storing contents. 
In the Markov modelling of resource allocation where continuous time Markov chain (CTMC) is constructed, first we define our state space and then look at state transitions. In the state transition Section, we have special Subsections on PU transitions, D2D operation mode and HU transitions. We specifically look at PU transition as our users are in SU mode at the terrestrial link. We define PU arrival and departure transition rates. Next, in the D2D operation mode overlaying is allowed and we calculate content availability probability for D2D operations in overlaying mode not causing interference to each other. Finally, in HU transitions we define HU arrival and departure transition rates. Arrivals are content requests. For each content request, the mode selection for our HUs are done considering (i) caches of system units (ii) channel states (iii) mode weights. (ii) and (iii) are used in aggregate mode weight functions that will be explained in Subsection~\ref{subsec:HUTransitions}. Next, we define our performance metrics and look at performance evaluation. The contributions of our work are as follows: 
\begin{itemize}
    \item We make use of content request distribution model and model a popularity-aware caching algorithm that keeps popular contents in system caches with a higher probability. This algorithm is investigated in terms of EE and overall network goodput and compared to Least Recently Used (LRU), First In First Out (FIFO) and random caching algorithms with simulations. 
    \item Some contents can not be available in caches of system units in our integrated network. For the sake of completeness in the resource allocation, such contents can be transferred from the universal source. The integration of universal source concept into our analytical model is done. 
    \item In RA, overlaying in D2D that allows the usage of the same frequency by different services without causing interference to each other is enabled for boosting network capacity and improving EE. The improvement in EE is investigated in our performance evaluation. 
\end{itemize}

\section{Related Work}
There is an extensive literature on caching in heterogeneous wireless networks especially from EE and/or QoS perspective. In~\cite{7752740}, EE related to content in cache-enabled D2D network is formulized and the optimal caching strategy for maximizing EE is investigated. 
According to the results, the optimal caching strategy is related to the transmission power level of users that cache contents. 
Yao et al. propose an algorithm that considers the energy-delay tradeoff by applying sleeping control and power matching method for single BS scenario~\cite{Yao2015}. Zhang et al. propose a satellite-terrestrial sofware defined network and investigates for coverage probability, spectral efficiency and EE in sparse and ultra-dense network types~\cite{7919304}. The proposed hybrid scheme improves EE for both sparse and ultra-dense networks compared to LTE and it also improves coverage. Xu et. al. investigated content transmission focusing on acceptable QoS guarantee in cellular network together with D2D~\cite{7915715}. A caching algorithm is proposed to improve QoS by reducing overflow issue at caches and having sufficient contents cached at devices. Secondly, Xu et. al. come up with an RA algorithm that tries to improve EE constrained by acceptable delays. Li et. al. made a survey on caching in cache-enabled cellular networks~\cite{8327582}. The taxonomy consists of macro-cellular, heterogeneous, device-to-device (D2D), cloud-radio access, and fog-radio access network types. Li et. al. provide literature studies from cache placement, delivery or hybrid aspects looking at performance metrics throughput, backhaul
cost, power consumption, network delay, hitting rate etc. 
In~\cite{8254837}, caching strategies for improving EE are proposed in a cellular and D2D hybrid network taking user request preferences into account. In such networks, there are two scenarios: (i) users collaborate (ii) users do not collaborate. According to their simulations, they observe improvement in performance when user preferences
are used. 

There are also studies on the RA problem from EE and/or QoS aspects. In~\cite{6046158}, it is stated that improvement opportunities for EE in satellite systems are either due to the structure of satellite itself such as solar power usage, lack of terrestrial infrastructure or adapting energy wise efficient hardware and/or energy efficient protocols to the satellite. In~\cite{Yao2014}, the optimal power allocation in cognitive radio based D2D network is studied. The D2D mechanism operates in the spectrum in an opportunistic manner. An optimal power allocation is proposed for the maximization of utility and also attaining requested service quality of devices in D2D mode and not exceeding interference limits to primary users. 
Schmidt et al. investigate EE in cellular networks together with D2D operation mechanism in~\cite{7504216}. According to their results, promoting D2D transmissions decreases power consumption and still preserves aggregate throughput of not completely loaded large networks. Apart from the power saving, it also increases network user rates. Br\"{u}ckner et al. propose a dependency aware reservation approach for mobile satellite communications in~\cite{7565115}. In~\cite{4413145}, they propose medium access control (MAC) over cognitive radio networks with two sensing algorithms where one senses randomly while the other is negotiation-based. The physical layer sensing results are used in a cross-layer manner. They analyze the network for saturated and non-saturated cases analytically. In their work the trade-off between delay and throughput in non-saturated network is demonstrated.
In~\cite{7410054}, a Long-Term Evolution Advanced (LTE-A) network system consisting of cellular users and D2D users is researched for the RA problem. In~\cite{7483447}, they propose RA algorithm for optimizing the capacity of D2D users taking the QoS and transmission power of all users into account. First, they determine cellular D2D users based on SINR requirement of D2D pairs. Then, they optimize the transmission power of all users with Langrange Multiplier technique. According to their results, their algorithm performs better than a greedy algorithm, random allocation and algorithm in~\cite{6029344} in terms of the D2D capacity. In~\cite{7542599}, they have an infrastructure and D2D integrated architecture. They have proposed contract-based approach of pricing and matching algorithms and investigated these algorithms separately for EE. They have shown improvement in EE.
In~\cite{7434298}, a cluster head rotation technique is proposed and this method is shown to be more energy efficient compared  to standard clustering technique in D2D systems. Besides,  it is more balanced in terms of energy consumption of all cluster members. Wang et. al propose a distributed algorithm for content download based on  expected available duration of contents and show that their algorithm outperforms EDF and SRPTF algorithms in terms of the number of downloads via D2D offloading~\cite{7556318}.  

Overall, the literature lacks a complete treatment of integrated satellite and cellular networks with D2D and spectrum sharing for the optimization of EE while QoS is maintained above a threshold. Typically one or a tuple of these aspects are studied. In this work, we elaborate on all of these aspects and provide a holistic view on content-oriented HetNets with D2D and cognitive communications. 

\section{System Model}
\label{sec:analysis}
\begin{figure}[ht]
\centering
\includegraphics[width=0.8\columnwidth]
{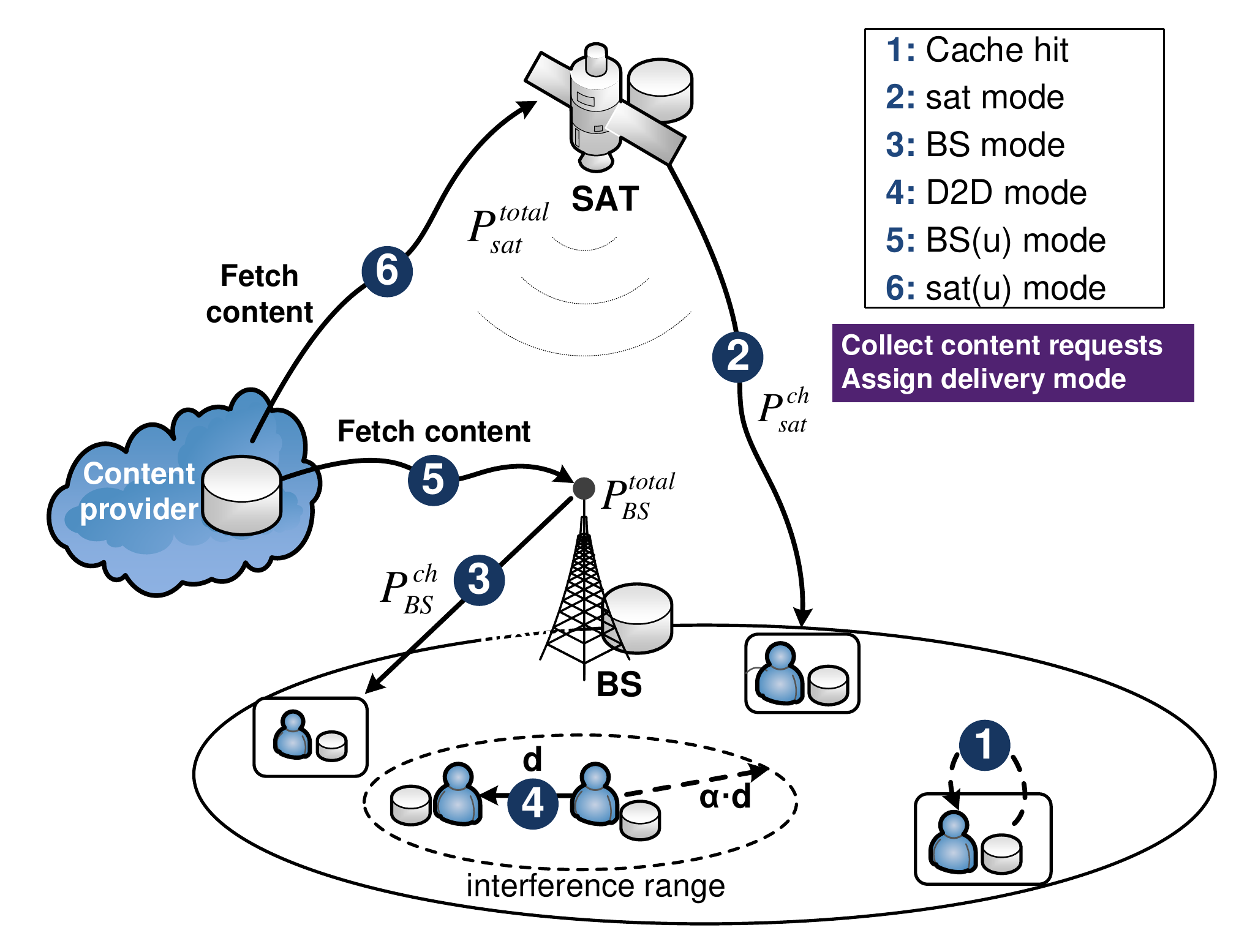}
\caption{System model.}
\label{fig:sysModel}
\end{figure}
To analyze such a system, we consider a heterogeneous network. In Fig.~\ref{fig:sysModel}, the network model is shown.
There exist two distinct frequency ranges in the network: (i) the satellite segment (ii) the terrestrial segment. We assume that there exists one satellite and we focus on the coverage are of one BS only. The terrestrial segment has primary users (PUs) that can fetch contents through the terrestrial frequencies via BS.

We have hybrid users (HUs) that can fetch content from the satellite or BS or some HU device unless it finds the content in its local cache. When it retrieves a content from the satellite, it is in PU mode. When it fetches a content from the BS or HU devices, it accesses terrestrial frequencies opportunistically. Note that content retrievals from the satellite or the BS can be direct or indirect. The direct retrievals occur from the satellite or BS cache to requester HU device (r-HU). The indirect retrievals occur first from the universal source to the satellite or the BS cache and then from there to r-HU. The decision mechanism for the selection of content retrieval unit (the satellite, the BS, some HU device) will be discussed in Section~\ref{subsec:HUTransitions}.

\section{Modelling of Popularity-Aware Caching}\label{sec:contentModel}

In the literature popularity based caching policies draw attention recently. In~\cite{6761239,7841535,7841508} content 
popularities are utilized for the cache management. Popularity-aware caching policy is becoming a  
baseline caching policy for content centric networks. Thereof, we will make use of this policy, model it and integrate into our RA strategy. 
In our setting, contents are chunks to be consumed by users. The content set in the network is C=\{$c_{1}$, $c_{2}$, ... $c_{N}$\} with N being the total number of contents. For each content $c_i$, $i\in\{1,2,...,N\}$ a probability for being requested $p_{c_{i}}(s)$ is assigned according to zipf distribution with parameter s. 
These probabilities are \{$p_{c_{1}}(s)$, $p_{c_{2}}(s)$, ... $p_{c_{N}}(s)$\} where $\sum_{c \in C}^{} {p_c(s)}$ = 1 and $\forall$c$\in$C  $p_c(s)\geq0$ while $p_{c_{i}}(s)$ =  $\frac{\frac{1}{i^s}}{\sum_{j=1}^{N}(\frac{1}{j^s})}$ for some i $\in$ \{1, 2, ..., N\} where s is the zipf distribution parameter.
The content request rate of HUs $\lambdaHU$ is a Poisson process. 
Each content $c_i$ has a request rate $\lambdaci{i}$ proportional to the popularity of that content defined as $p_{c_{i}}(s) \lambdaHU$. The request rates for contents are \{$\lambdaci{1}$ := $p_{c_{1}}(s) \lambdaHU$,$\lambdaci{2}$ := $p_{c_{2}}(s) \lambdaHU$,...,$\lambdaci{N}$ := $p_{c_{N}}(s) \lambdaHU$\}. 
We utilize content request distribution model and model a popularity-aware caching (PAC) strategy that tries to keep more popular contents in system unit caches with a greater probability. The pseudecode of PAC algorithm for any local HU device cache is given in Algorithm~\ref{alg:PAC}.
Compared to Least Recently Used (LRU) and First In First Out (FIFO) approaches that make use of locality in user preferences, popularity-aware caching algorithm exploits global content popularities. In the RA phase, we make use of content availability probabilities at system units. Thereof, we derive these probabilities at local HU device, satellite and BS cache. The content availability for D2D operation at some other HU device within some vicinity to requester is investigated in Section~\ref{subsection:D2D}.

\subsection{Local HU Device Cache}\label{sec:HUDeviceCache}
\begin{figure}[!t]
	\centering
	\scalebox{0.9}{
	\begin{tikzpicture}[-latex ,node distance =1.35 cm and 2.3cm, on grid,
	semithick ,	state/.style ={ ellipse , draw, align=center}]
	
	\node[draw,double][state] (1) {\scriptsize $c_i$ $c_{j \neq i}$}; 

    \node[draw,double][state] (2) [right = of 1]{\scriptsize $c_{i}$};

	\node[state] (7) [above = of 1]{\scriptsize $c_{j \neq i}$};
	
	\path (7) edge[bend left=5] node[right] {$\rate{6}$} (1);
	
	\path (1) edge[bend left=5] node[left] {$\rate{5}$} (7);

	\node[state] (4) [below = of 1]{\scriptsize $c_{t  \neq (i, j)}$ $c_j$};

	\path (1) edge[bend left=5] node[right] {$\rate{7}$} (4);
	
	\path (4) edge[bend left=5] node[left] {$\rate{8}$} (1);

	\node[state] (5) [left = of 1]{\scriptsize $c_i$ $c_{t \neq (i,j)}$};
	
	\path (1) edge[bend left=5] node[below] {$\rate{9}$} (5);
	
	\path (5) edge[bend left=5] node[above] {$\rate{10}$} (1);
	\path (2) edge[bend left=5] node[below] {$\rate{4}$} (1);
	
	\path (1) edge[bend left=5] node[above] {$\rate{3}$} (2);

	\node[draw,double][state] (10) [right = of 2]{\scriptsize \textrm{empty}};
	
	\path (10) edge[bend left=5] node[below] {$\rate{1}$} (2);
	
	\path (2) edge[bend left=5] node[above] {$\rate2$} (10);

	\end{tikzpicture}
	}
\caption{HU device local cache updates.} \label{fig:Local Cache}
\end{figure}
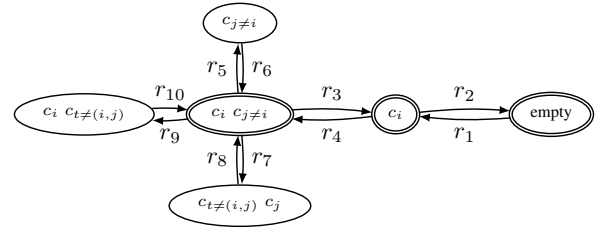

Now we will construct a cache model for some arbitrary HU local cache. Let us start our investigation with HU local cache.
We construct the Markov chain for tracking local cache states and find probabilities for a content being available in the local cache. The local cache size is $\HUDeviceCacheSize$. A set of contents can be stored in the local cache only if the sum of content sizes is less than or equal to $\HUDeviceCacheSize$. The content size distribution is exponential with mean $\MeanBaseContentSize$ = 25 Mbits. 
For the analysis, we assume that at the local cache at most $N_{HU}^{loc}$ contents can be stored. Deriving the balance equations and calculating the steady state probabilities of the double circled states in Fig.~\ref{fig:Local Cache}, all the state space is covered since a state can be in one of these 3 types : (i) idle (ii) contains one content (iii) contains 2 distinct contents. 

Before explaining the state transitions, we need to explain the content size distributions. The parameter of content size distribution  $\lambdaContent$ is equal to $\frac{1}{\MeanBaseContentSize}$ where 
$\MeanBaseContentSize$ = 25 Mbits.

The probability density function for content size is as follows: 
\begin{IEEEeqnarray}{rCl}
f_{c}(x;\lambdaContent) &:=&
\begin{cases}
\lambdaContent e^{-\lambdaContent x} & x \geq 0 \\
0  & \textrm{otherwise}
\end{cases}
\end{IEEEeqnarray}

The cumulative distribution functions for content size is as follows: 
\begin{IEEEeqnarray}{rCl}
F_{c}(x;\lambdaContent) &:=&
\begin{cases}
1- e^{-\lambdaContent x} & x \geq 0 \\
0  & \textrm{otherwise}
\end{cases}
\end{IEEEeqnarray}

The state transition rates are explained as follows: 
If the local cache is empty and a content $c_i$ is requested  for some arbitrary i $\in$ \{1,2,..., N\}, it will transit from empty state to state ($c_i$) if the size of the requested content $S_{c_i}$ is less than or equal to the cache capacity of the HU device local cache ($\HUDeviceCacheSize$) with transition rate $\rate{1} := \lambdaci{i} P(S_{c_i} \leq \HUDeviceCacheSize)$. 
$S_{c_i}$ is the size of the content $c_i$. The content size should be less than or equal to the cache capacity of the HU device local cache.
Note that this probability is equal to the cumulative distribution function $F_{c}(\HUDeviceCacheSize;\lambdaContent)$.  
Transition $\rate{2}$ occurs when a content expires due to  $TTL$. 
Note that $\rate{3}$ and $\rate{5}$ are expiration transitions with same expiration rate with $\rate{2}$ as well. When the local cache has one content and a new content is requested and the total size of these two contents is less than or equal to local cache size, the newly requested content is cached. $\rate{4}$, $\rate{6}$ are such transitions with rates $\lambdaci{j} P(S_{c_i}+S_{c_j}\leq\HUDeviceCacheSize)$ and $\lambdaci{i} P(S_{c_i}+S_{c_j}\leq\HUDeviceCacheSize)$, respectively.

Until now, we considered transitions into or from empty cache state and one content holding cache states ($c_i$) in Fig.~\ref{fig:Local Cache}. We have one more type: Namely 
2 content holder cache states ($c_i$ $c_j$) for i$\neq$j. By writing balance equations for all states looking at the transition rates into and out of these states we derive probabilities of being at some state. Before writing the balance equations, let us focus on the remaining state transitions. 
Transitions $\rate{7}$ and $\rate{9}$ occur when a new content is requested and one of the older cached content should be evicted to be able to cache the newly requested one. In PAC policy, the content to be evicted is selected based on content popularity. More popular contents will be less likely to be preempted compared to less popular ones. For instance, 
the probability for content $c_i$ being requested according to zipf distribution is $p_{c_i}(s)$ and for content $c_j$ is $p_{c_j}(s)$. If $c_i$ is less popular than $c_j$, then $p_{c_j}(s)>p_{c_i}(s)$ and $\scriptstyle{\frac{p_{c_j}(s)}{p_{c_i}(s)+p_{c_j}(s)}}>\scriptstyle{\frac{p_{c_i}(s)}{p_{c_i}(s)+p_{c_j}(s)}}$. Then with larger probability $\scriptstyle{\frac{p_{c_j}(s)}{p_{c_i}(s)+p_{c_j}(s)}}$, less popular content $c_i$ is evicted and with lower probability $\scriptstyle{\frac{p_{c_i}(s)}{p_{c_i}(s)+p_{c_j}(s)}}$ more popular content $c_j$ is evicted. The corresponding transitions are $\rate{7}$ and $\rate{9}$, respectively. Note that newly requested plus the non preempted contents should together fit into the HU device local cache. The rates $\rate{8}$ and $\rate{10}$ have similar logic with $\rate{7}$ and $\rate{9}$. The transition rates are given as follows: 
\begin{IEEEeqnarray}{rCl}
\rate{7} &:=& \scriptstyle{\lambdaci{t} (\frac{p_{c_j}(s)}{p_{c_i}(s)+p_{c_j}(s)})
P(S_{c_t}+S_{c_j} \leq \HUDeviceCacheSize)}\\
\rate{8} &:=& \scriptstyle{\lambdaci{i}
(\frac{p_{c_j}(s)}{p_{c_t}(s)+p_{c_j}(s)})
P(S_{c_i}+S_{c_j} \leq \HUDeviceCacheSize)}\\
\rate{9} &:=& \scriptstyle{\lambdaci{t}
(\frac{p_{c_i}(s)}{p_{c_i}(s)+p_{c_j}(s)})
P(S_{c_i}+S_{c_t} \leq \HUDeviceCacheSize)}\\
\rate{10} &:=& \scriptstyle{\lambdaci{j}
(\frac{p_{c_i}(s)}{p_{c_i}(s)+p_{c_t}(s)})
P(S_{c_i}+S_{c_j} \leq \HUDeviceCacheSize)}
\end{IEEEeqnarray}

After deriving the probability $p_{(empty)}$ of being at empty cache state, the probability $p_{(c_i)}$ of one content holding cache states $(c_i)$ for any i $\in\{1,2,...,N\}$ and the probability $p_{(c_i~c_j)}$ of two distinct content holding cache states $(c_i~c_j)$ for any i $\in\{1,2,...,N\}$ and any j $\in\{1,2,...,N\}$ such that $j\neq i$, the content probabilities being in the local cache are as follows: 
\begin{IEEEeqnarray}{rCl}
\pbaselocal{i} &:=& p_{(c_i)} + \sum_{j=1,j\neq i}^{N} p_{(c_i~c_j)}\label{eq:p_localbase}
\end{IEEEeqnarray}

\begin{algorithm}
\caption{Popularity-Aware Caching (PAC) Strategy}\label{alg:PAC}
\begin{algorithmic}
\STATE \textbf{Procedure PAC($Cache_{Dev},N,s,c_t$)} 
\IF{Local cache is empty and$~S_{c_t}\leq Cache_{Dev}$}
    \STATE Cache newly requested content $c_t$
\ELSIF{Local cache contains one content $c_i$ and $S_{c_i}+S_{c_t}\leq Cache_{Dev}$}
    \STATE Cache newly requested content $c_t$
\ELSE
    \STATE Local cache contains contents $c_i$ and $c_j$. Let $\alpha\in[0,1]$ be random variable from uniform distribution. Calculate the request probabilities $p_{c_i}(s)$ and $p_{c_j}(s)$.
    \IF{$(\alpha \leq \frac{p_{c_j(s)}}{p_{c_i(s)}+p_{c_j(s)}})$} 
        \STATE Evict content $c_i$ and cache newly requested content $c_t$ when $S_{c_t}+S_{c_j}\leq Cache_{Dev}$ 
    \ELSE
        \STATE Evict content $c_j$ and cache newly requested content $c_t$ when $S_{c_i}+S_{c_t}\leq Cache_{Dev}$
    \ENDIF
\ENDIF
\end{algorithmic}
\end{algorithm}

\subsection{Satellite Cache}

In popularity-aware caching policy, more popular contents will be less likely to be preempted compared to less popular ones in the satellite cache. We construct the Markov chain for tracking the satellite cache states and find probabilities for a content being available in the satellite cache. We assume that the satellite cache stores at most $\floor{\frac{\SatCacheSize}{\MeanBaseContentSize}} = \floor{\frac{125 Mbits}{25 Mbits}} = 5$ distinct contents. As the satellite is a main source for content retrieval it will be occupied with larger number of distinct contents with greater probability. Thereof, we omit the cache states that have less than five contents.

\begin{figure}[!t]
	\centering
	\scalebox{0.9}
	{
	\begin{tikzpicture}[-latex ,node distance =1.7 cm and 3.1cm, on grid,
	semithick ,	state/.style ={ ellipse , draw, align=center}]
	
	\node[draw, rounded rectangle,double] (1) {\scriptsize $c_i^b$ $c_{j}^b$ $c_k^b$ $c_l^b$ $c_m^b$};

    \node[draw, rounded rectangle] (2) [above = of 1]{\scriptsize $c_i^b$ $c_{j}^b$ $\textbf{c}_\textbf{t}^\textbf{b}$ $c_l^b$ $c_m^b$};

	\node[draw, rounded rectangle] (4) [right = of 1]{\scriptsize $c_i^b$ $c_{j}^b$ $c_k^b$ 	\scriptsize $c_l^b$ $\textbf{c}_\textbf{t}^\textbf{b}$};

	\path (1) edge[bend left=5] node[above] {$\rate{5}$} (4);
	
	\path (4) edge[bend left=5] node[below] {$\rate{10}$} (1);

	\node[draw, rounded rectangle] (5) [right = of 2]{\scriptsize $c_i^b$ $c_{j}^b$ $c_k^b$ $\textbf{c}_\textbf{t}^\textbf{b}$ $c_m^b$};
	
	\path (1) edge[bend left=5] node[left] {$\rate{4}$} (5);
	
	\path (5) edge[bend left=5] node[right] {$\rate{9}$} (1);
	
	\node[draw, rounded rectangle] (11) [left = of 1]{\scriptsize $\textbf{c}_\textbf{t}^\textbf{b}$ $c_{j}^b$ $c_k^b$ $c_l^b$ $c_m^b$};
	
	\path (11) edge[bend left=5] node[above] {$\rate{6}$} (1);
	
	\path (1) edge[bend left=5] node[below] {$\rate{1}$} (11);

	\node[draw, rounded rectangle] (7) [above = of 11]{\scriptsize $c_i^b$ $\textbf{c}_\textbf{t}^\textbf{b}$ $c_k^b$ $c_l^b$ $c_m^b$};
	
	\path (7) edge[bend left=5] node[right] {$\rate{7}$} (1);
	
	\path (1) edge[bend left=5] node[left] {$\rate{2}$} (7);

    \path (2) edge[bend left=5] node[right] {$\rate{8}$} (1);
	
    \path (1) edge[bend left=5] node[left] {$\rate{3}$} (2);

	\end{tikzpicture}
	}
\caption{The satellite cache updates.} \label{fig:Satellite Cache}
\end{figure}
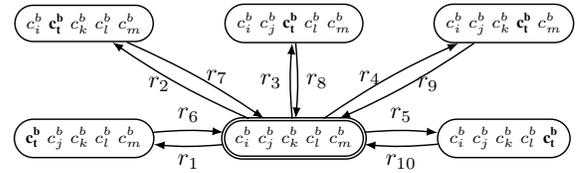

Let us start our inspection of the satellite cache analysis. 
The Markov chain of the satellite cache is given in Fig.~\ref{fig:Satellite Cache}. Let us start with the state transition explanations: 
Assume the current satellite state is the double circled state in Fig.~\ref{fig:Satellite Cache} and no duplicate contents exist in the cache. i.e. each content is different from the others (i$\neq$j$\neq$k$\neq$l$\neq$m).
With probability $p_{c_t}(s)$ (calculated according to the content popularity as given above, if $c_t$ is a popular content it will attain a higher value, if $c_t$ is a non-popular content it will have a lower value), the content $c_t$ that is not currently available in the satellite cache will be cached in the satellite. For the content $c_t$ to be stored in the satellite cache,
one of the contents $c_i$, $c_j$, $c_k$, $c_l$, $c_m$ should be replaced taking their popularities into account. For instance, assume $c_i$ is the least popular content among contents in the satellite cache. Then, the probability for $c_i$ being requested according to zipf distribution $p_{c_i}(s)$ is smaller than the probabilities of other contents within the satellite cache ($c_j$, $c_k$, $c_l$, $c_m$) being requested.
The transition $r_1$ where $c_i$ is replaced will have the highest rate among all transitions $r_\alpha$, $\alpha\in$\{1,2,3,4,5\}. So, the least popular content $c_i$ will be preempted for the sake of new comer content $c_t$ with the highest probability. Note that the summation of transitions $\sum_{i=1}^{5} r_i$ equals to $p_{c_t}(s)$. 
The transition rates $\rate{\alpha}$'s for $\alpha \in \{1,2,...5\}$ are given as follows: 
\begin{IEEEeqnarray}{rCl}
\rate{1} &:=& \scriptstyle{p_{c_t}(s) (\frac{x_i}{\sum_{\theta \in \{i,j,k,l,m\}} x_\theta})} \\
\rate{2} &:=& \scriptstyle{p_{c_t}(s) (\frac{x_j}{\sum_{\theta \in \{i,j,k,l,m\}} x_\theta})} \\
\rate{3} &:=& \scriptstyle{p_{c_t}(s) (\frac{x_k}{\sum_{\theta \in \{i,j,k,l,m\}} x_\theta})} \\
\rate{4} &:=& \scriptstyle{p_{c_t}(s) (\frac{x_l}{\sum_{\theta \in \{i,j,k,l,m\}} x_\theta})} \\
\rate{5} &:=& \scriptstyle{p_{c_t}(s) (\frac{x_m}{\sum_{\theta \in \{i,j,k,l,m\}} x_\theta})}\\
x_i &:=& \scriptstyle{p_{c_j}(s)\times p_{c_k}(s)\times p_{c_l}(s)\times p_{c_m}(s)}\\
x_j &:=& \scriptstyle{p_{c_i}(s)\times p_{c_k}(s)\times p_{c_l}(s)\times p_{c_m}(s)}\\
x_k &:=& \scriptstyle{p_{c_i}(s)\times p_{c_j}(s)\times p_{c_l}(s)\times p_{c_m}(s)}\\
x_l &:=& \scriptstyle{p_{c_i}(s)\times p_{c_j}(s)\times p_{c_k}(s)\times p_{c_m}(s)}\\
x_m &:=& \scriptstyle{p_{c_i}(s)\times p_{c_j}(s)\times p_{c_k}(s)\times p_{c_l}(s)}
\end{IEEEeqnarray}

The incoming transitions $\rate{\alpha}$ for $\alpha\in\{6,7,...10\}$ into double circled state are constructed according to the same caching policy.

By deriving balance equations and calculating the steady state probabilities of the double circled states in Fig.~\ref{fig:Satellite Cache}, we cover all the state space. 
The steady state probability of satellite cache states that contain content x are summed to find the probability of content x being available in the satellite cache as follows:
\begin{IEEEeqnarray}{rCl}
& p_{c_x}^{sat} := ~~~~ 
\sum_{\mathclap{\substack{i,j,k,l,m \in \{1,2,... N\}\\
\footnotesize{i \neq j \neq k \neq l \neq m} \\
x \in \{i,j,k,l,m\}}}}
& ~~~~p_{(c_i~c_j~c_k~c_l~c_m)}\label{eq:p_satbase}
\end{IEEEeqnarray}

\subsection{BS Cache}
For the availabilities of contents at the BS cache, a similar Markov chain of BS cache is constructed. The BS cache size is smaller than the satellite cache and therefore the number of distinct contents that can be stored in the BS cache is less.
The BS cache stores exactly $\floor{\frac{\BSCacheSize}{\MeanBaseContentSize}} = \floor{\frac{100 Mbits}{25 Mbits}} = 4$ distinct contents. We skip the explanation for the BS cache state construction and transitions since the logic is the same with the satellite case. The steady state probability of BS cache states that contain content x are summed to find the probability of content x being available in the BS cache as follows:
\begin{IEEEeqnarray}{rCl}
p_{c_x}^{BS} := ~~~~
\sum_{\mathclap{\substack{i,j,k,l \in \{1,2,... N\}\\
\footnotesize{i \neq j \neq k \neq l} \\
x \in \{i,j,k,l\}}}}
& ~~~~p_{(c_i~c_j~c_k~c_l)}
\label{eq:p_BSbase}
\end{IEEEeqnarray}

\section{Markov Modelling of Resource Allocation}

\begin{table}
\centering
    \renewcommand*{\arraystretch}{1.3}
	\caption{Analysis parameters.}
	\scalebox{0.85}{
	\begin{tabular}{ | p{1.12cm} |p{6.3 cm}|l| }
		\hline
		\textbf{Parameter} & \textbf{Explanation} & \textbf{Value} \\ \hline
		$N$ & Total number of contents & 20\\ \hline
		$N_{HU}^{loc}$ & The maximum number of contents that can be stored in any HU device cache & 2 \\ \hline
		$\MeanBaseContentSize$    & Mean content size requested by a hybrid user & 25 Mbs \\ \hline
		s & Zipf parameter & 1.2 \\ \hline
		$\NSat$ & Total number of satellite frequencies & 2\\ \hline
		$\NTer$ & Total number of terrestrial frequencies & 3 \\ \hline
		$\lambdaPUTer$ & Arrival rate of primary users at terrestrial link & 0.03 $\frac{user}{sec}$ \\ \hline
		$\lambdaHU$ & Arrival rate of hybrid users for content request at the system & 2.4 $\frac{user}{sec}$ \\ \hline
		$\PerChannelSatTxPower$ & Per channel transmission power of the satellite & 48 W \\ \hline
		$\PerChannelBSTxPower$ & Per channel transmission power of the BS & 6 W  \\ \hline
			$\DeviceTxPower$ & Transmission power of a hybrid user device & 80 mW \\ \hline
			$\DistancetoSat$(LEO) & Distance from satellite to earth & 300 km \\ \hline
			$\MeanDistancetoBS$ & Mean distance of a PU and/or HU to the BS & 150 m \\ \hline
			$\DistanceDD$ & Mean distance between receiver and sender HU devices & 30 m   \\ \hline
			$\TerBandwidth$ & Bandwidth of terrestrial link & 2 MHz \\ \hline
			$\SatBandwidth$ & Bandwidth of satellite link & 36 MHz\\ \hline			
			$\SatFrequency$ & Frequency of satellite link & 20 GHz  \\ \hline
			$\TerFrequency$ & Frequency of terrestrial link & 700 MHz \\ \hline
			$\SatCacheSize$ &Satellite cache size &  125 Mbs \\ \hline
			$\BSCacheSize$  & Base station cache size & 100 Mbs   \\ \hline
			$\HUDeviceCacheSize$ & Hybrid device cache size & 50 Mbs \\ \hline
		$\receiveBSparameter$     & Per channel reception power consumption of the BS parameter (Per channel reception power of the BS is $\frac{\PerChannelBSTxPower}{\receiveBSparameter}$) & 5       \\ \hline
		$\receiveLocalparameter$ & Parameter for HU device power consumption when content is found in local cache (Power consumption of HU device when content is found in local is $\frac{\DeviceTxPower}{\receiveLocalparameter}$) & 2 \\ \hline
		$\capHUSatU$ & The average channel capacity between the satellite and universal source & 1 Mbps  \\ \hline
		$\capHUBSU$ & The average channel capacity between the BS and universal source &  10 Mbps \\ \hline
		$\lambdaNHU$ & The mean density of HUs located in the BS cell & 0.0018 $\frac{user}{m^2}$\\ \hline
		$\DMAX$ & The maximum number of
concurrent D2D operations allowed by the network & 5 \\ \hline
        $\RBS$  & The radius of the BS cell & 300 m\\ \hline
		$\RInt$ & The radius of the transmission range of an HU device that does interfere with other HU receivers operating at terrestrial frequency $f_1$ in D2D mode. & 60 m\\ \hline 
	\end{tabular}
	}
	\label{table:Analysis Parameters}
\end{table}

In this Section, we delve into the resource allocation (RA) problem in this heterogeneous network. We do rigorous analytical analysis on the channel usage of HUs for content retrievals. 
Our assumptions are as follows: 
We do not have control over baseline users.
However, we have knowledge about their network traffic characteristics. 
This can be actualized by central capabilities
of the BS as a component of cognitive operation \cite{Nekovee}. We assume that BS performs the centralized RA function.
It can detect PU arrivals and departures, and allocate resources to HUs knowing PU activities. 
An entire content is assumed to be at the same place. i.e. It is not allowed that some portion is in the satellite and the rest in an HU device. The request of a content is an arrival following Poisson distribution and completion of a content retrieval is an exponentially distributed departure. The resource allocation of HUs for content retrieval is done in a non-time slotted manner and modeled as CMTC.
The Markov property, that is a state transition depends only on the current state but not the previous ones, is satisfied by HUs not continuing content fetch from another system unit after interruption due to PU appearance at the current channel. 
If there exists some idle frequency, HU continues operation from the same unit.
Otherwise, the request is dropped in a bufferless operation. PUs have priority over SU mode HUs, hence for fully utilized systems a PU arrival entails the preemption of a low priority user. In our previous work~\cite{7492943}, all frequencies are used in non-overlay setting. In this work, we improve our model: the network has a non-overlay setting in satellite and BS modes but it operates in overlay setting in D2D mode. 
Since the introduction of every new frequency for D2D operation increases state space complexity exponentially, one terrestrial frequency is used for HUs in D2D mode in our work without loss of generality. 

As noted in Section~\ref{sec:introduction}, our main foci are the caching modelling, the universal content server integration, the introduction of overlay in D2D and the investigation of RA scheme in our heterogeneous network. We ignore transmission and sensing errors to track the essence of system behaviour. The system and analysis parameters are listed in Table~\ref{table:Analysis Parameters}.

\begin{table}[t]
\centering
	\caption{State definition.}
	\scalebox{1}{
	\begin{tabular}{ | l | p{6.7cm} | }
			\hline
			\textbf{Notation} & \textbf{Definition}  \\ \hline
			$\iHUSat$ & The number of satellite frequencies used by HUs for retrieving contents directly from the satellite cache  \\ \hline
			$\iHUSatUniv$ & The number of satellite frequencies used by HUs for retrieving contents across the satellite from the universal source \\ \hline
			$\iPUTer$ & The number of terrestrial frequencies used by PUs except for terrestrial frequency $f_1$  \\ \hline
			$\iHUBS$ & The number of terrestrial frequencies used by HUs for retrieving contents directly from the BS cache\\ \hline
			$\iHUBSUniv$ & The number of terrestrial frequencies used by HUs for retrieving contents across the BS from the universal source\\ \hline
			$\iPUTerfOne$ & The indicator for terrestrial frequency $f_1$ whether it is used by PU or not \\ \hline
			$\iHUDev$ & The number of concurrent D2D transmissions used for content retrieval via terrestrial frequency $f_1$ \\ \hline
	\end{tabular}
	}
	\label{table:state definition}
\end{table}

\begin{table*}[h]
    \centering
	\caption{Transitions originated at a generic state $state_0$($s_0$) due to PU activities.}
	\scalebox{0.9}{
	\begin{tabular}{ | l | l | l | l |}
	\hline
	\textbf{Link} &
	\textbf{Activity} &
	\textbf{Destination State} &
	 \textbf{Transition Rate} \\ \hline
	 Ter &
	PU arr. &
	$s_1$ := $\state{(\iPUTer+1)}$ &
	$\frac{(\NTer-1)\lambdaPUTer}{\NTer}1_{(\idleTerExceptfOne{s_0}>0)}$\\[1ex] \hline
	Ter &
	PU arr. &
	$s_2$ := $\state{(\iPUTer+1,\iHUBS-1)}$ &
	$\frac{(\NTer-1)\lambdaPUTer}{\NTer}\times\frac{\stateiHUBS{s_0}}{(\NTer-1)-\stateiPUTer{s_0}}[1_{((\idleTerExceptfOne{s_0}==0)\land((\NTer-1)-\stateiPUTer{s_0}>0))}]$
	\\[1ex] \hline
    Ter &
	PU arr. &
	$s_3$ := $\state{(\iPUTer+1,\iHUBSUniv-1)}$ &
	$\frac{(\NTer-1)\lambdaPUTer}{\NTer}\times \frac{\stateiHUBSU{s_0}}{(\NTer-1)-\stateiPUTer{s_0}}[1_{((\idleTerExceptfOne{s_0}==0)\land((\NTer-1)-\stateiPUTer{s_0}>0))}]$
	\\[1ex] \hline
	Ter &
	PU arr. &
	$s_4$ := $\state{(\iPUTerfOne+1)}$&
	$\frac{\lambdaPUTer}{\NTer}1_{((\stateiPUTerfOne{s_0}==0)\land(\stateiHUD{s_0}==0))}$\\[1ex] \hline
    Ter &
    PU arr. &
	$s_5$ := $\state{(\iPUTerfOne+1,\iHUDev=0)}$ &
	$\frac{\lambdaPUTer}{\NTer}1_{(\stateiHUD{s_0}>0)}
	$\\[1ex] \hline
	Ter &
	PU dept. &
	$s_6$ := $\state{(\iPUTer-1)}$ &
	$\stateiPUTer{s_0}\muPUTer$
\\[1ex] \hline
	Ter &
	PU dept. &
	$s_7$ := $\state{(\iPUTerfOne-1)}$ &
	$\stateiPUTerfOne{s_0}\muPUTer$
\\[1ex] \hline
	\end{tabular}
	}
	\label{table:Outgoing PU transitions from s0$}
\end{table*}

We calculate the channel capacities for content fetch in bits per second (bps) by Shannon's capacity formula under Additive White Gaussian Noise (AWGN) according to free space path model.
The service rate of PUs from terrestrial link is calculated by the division of the channel capacity of terrestrial link over average content size $\MeanBaseContentSize$, given as $\mu_{PU}^{ter}:=\frac{C_{PU}^{ter}}{\MeanBaseContentSize}$. The service rate of HUs that retrieve the requested content over different system units such as the satellite, the BS or in D2D operation mode is as follows: 
\begin{IEEEeqnarray}{rCl}\label{eq:muHUs}
\mu_{HU}^x &:=& \frac{C_{HU}^x}{\MeanBaseContentSize} \textrm{ where } x \in \{sat, BS,D\}
\end{IEEEeqnarray}

We introduce a universal source concept into our analytical model for a more realistic model since some of the contents can not be available in caches of system units and they need to be fetched from external repositories, namely universal source.
We calculate the mean aggregate service durations $\meanservicedurationHUSatU$ and $\meanservicedurationHUBSU$  for the requester HU devices (r-HUs) that retrieve the requested content over the \textbf{universal source} across the satellite or the BS 
in~(\ref{eq:meanservictetimeHUUniv}). The average channel capacity between the satellite and universal source and between the BS and universal source are given in Table~\ref{table:Analysis Parameters}.
Content travels from the universal source to the satellite with mean  content transmission duration $\frac{\MeanBaseContentSize}{\capHUSatU}$. Afterwards, with mean service duration $\frac{\MeanBaseContentSize}{\CapacityHUSat}$ it can be transmitted from the satellite to the r-HU. By adding these durations as shown in~(\ref{eq:meanservictetimeHUUniv}), we get the aggregate service duration of a content retrieval from the universal source across the satellite to the r-HU. The aggregate service duration of some content retrieval from the universal source across the BS to the r-HU is calculated similarly given in~(\ref{eq:meanservictetimeHUUniv}). 
\begin{IEEEeqnarray}{rCl}
\meanservicedurationHUUniv &:= &
\frac{\MeanBaseContentSize}{\capHUUniv} + 
\frac{\MeanBaseContentSize}{\CapacityHU}
\label{eq:meanservictetimeHUUniv}
\end{IEEEeqnarray}

where $x\in\{sat,BS\}$. Now, the service rate of HUs that fetch content from the universal source across the satellite or the BS to the r-HU are given in Equation~\ref{eq:meanservicerateHUUniv}.
\begin{IEEEeqnarray}{rCl}
\muHUUniv &:=&\frac{1}{\meanservicedurationHUUniv} \label{eq:meanservicerateHUUniv} \text{ where } x \in \{sat, BS\}
\end{IEEEeqnarray}

\section{State Definition}

Now, let us look at the definitions of a channel state. 
The only terrestrial frequency used for D2D mode is the terrestrial frequency $f_1$ and the other terrestrial frequencies operate in BS mode. 
A state denoted as $state_0$ consists of seven parts as shown in Fig.~\ref{fig:NewState}. The definitions of these parts are given in Table~\ref{table:state definition}.
\tikzset{font=\footnotesize}
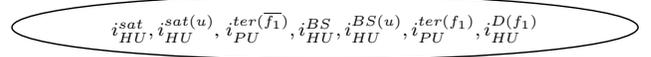
\begin{figure}[h]
	\centering
	\scalebox{0.9}{
	\begin {tikzpicture}[-latex ,node distance =3 cm and 3cm, on grid,
	semithick ,
	state/.style ={ ellipse , draw, align=center}]
	\node[state] (1) {$\iHUSat, \iHUSatUniv$, $\iPUTer, \iHUBS, \iHUBSUniv, \iPUTerfOne, \iHUDev$};
    \end{tikzpicture}
    }
\caption{Channel state $state_{0}$.} \label{fig:NewState}
\end{figure}

\section{State Transitions}\label{sec:state_transitions}

A channel state transition occurs due to PU/HU  arrival/departure. If a user arrives to the system 
we increment the corresponding type of user in the channel state. If a content is completely retrieved then user departs the channel. 
During RA, we look at the content availability and 
for the selection among system units where the content is available, we look at the channel states and for each available frequency we assign mode weight and decide accordingly. By tuning mode weights, we investigate how EE and overall system goodput is affected. In our previous work~\cite{7904715}, we observe that EE is improved by increasing D2D operation mode. As new contributions, we integrate universal source concept, model caching mechanism and enable overlaid D2D operation in this work. With a more realistic network model, we perform rigorous performance analysis through our improved analytical model. We first define utility functions: $\idleSat{x}$ gives us the number of idle frequencies at the satellite link segment at some arbitrary state x and $\idleTerExceptfOne{x}$ the number of idle frequencies at terrestrial link except for the terrestrial frequency $f_1$ at an arbitrary state x. $1_{(\theta)}$ function is the indicator function. 

\begin{IEEEeqnarray}{rCl}
\small{\idleSat{x}} &:=& \scriptstyle{\NSat-\stateiPUSat{x}-\stateiHUSat{x}-\stateiHUSatU{x}}\\
\footnotesize{\idleTerExceptfOne{x}} &:=&\scriptstyle{(\NTer-1)-\stateiPUTer{x}-\stateiHUBS{x}-\stateiHUBSU{x}}\\
1_{(\theta)} &:=&
\begin{cases}
\scriptstyle{1, \textrm{  if } \theta \textrm{ is true}}\\
\scriptstyle{0, \textrm{  otherwise}}
\end{cases}
\end{IEEEeqnarray}

\subsection{PU Transitions}

Before delving into the technical details of PU transitions, 
PUs are active only at the \textbf{terrestrial link}.
At the terrestrial frequency $\boldsymbol{f_1}$ HUs operate \textbf{only in D2D mode}. 
For the \textbf{terrestrial frequencies except $\boldsymbol{f_1}$}, HUs can fetch content \textbf{only from the BS or across the BS from the universal source (BS mode)}. The HU modes are different for terrestrial frequency $f_1$ and other terrestrial frequencies. For analytically processing the preemptions of distinct HU modes, we define two constituents for PUs at the channel state as follows:
(i) $\iPUTerfOne$: the indicator for terrestrial frequency $f_1$ if it is used by PU or not (ii) $\iPUTer$: the number of terrestrial frequencies used by PUs except terrestrial frequency $f_1$.

Now, let us look at the PU state transitions. 
We denote $state_0$ in Fig.~\ref{fig:NewState} as $s_0$. First, we elaborate PU arrivals originated at $s_0$ in Table~\ref{table:Outgoing PU transitions from s0$}. The table consists of four columns. The first column is the link, second activity type(PU arrival/PU departure). The third column is the destination state, the final one the corresponding transition rate. The destination states have the form $state_x$. Compared to our generic state $s_0$, the incremented and/or decremented parts of $s_0$ are represented with $x$. In $s_1$ only $\iPUTer$ is incremented by one upon PU arrival at the satellite link. Therefore the destination state has form $state_{(\iPUTer+1)}$. Now, we look at transitions originated at $s_0\xrightarrow[]{} s_i$ for $i\in\{1,2,...,7\}$ in Table~\ref{table:Outgoing PU transitions from s0$}. 

When a PU arrives at the terrestrial link except for terrestrial frequency $f_1$ and no HU is disrupted, $s_0 \xrightarrow[]{} s_1$ transition with rate $\scriptstyle{\frac{(\NTer-1)\lambdaPUTer}{\NTer}}$ occurs in two scenarios: 1) PU selects an idle frequency among terrestrial frequencies $f_2$, ..., $f_{\NTer}$ and starts operation. 2) PU selects an HU occupied frequency (BS mode) and that HU preempts current frequency and continues its operation in the same mode from an idle frequency among terrestrial frequencies $f_2$, ..., $f_{\NTer}$. In both scenarios, at least one idle terrestrial frequency among terrestrial frequencies $f_2$, ..., $f_{\NTer}$ is needed, which is checked by $1_{(\idleTerExceptfOne{s_0}>0)}$.

When a PU arrives at the terrestrial link except for terrestrial frequency $f_1$ and the service of an HU retrieving content directly from the BS cache is disrupted, $s_0 \xrightarrow[]{} s_2$ transition with corresponding rate in Table~\ref{table:Outgoing PU transitions from s0$}
occurs. $\scriptstyle{\frac{\stateiHUBS{s_0}}{(\NTer-1)-\stateiPUTer{s_0}}}$ means from the set of terrestrial frequencies except for terrestrial frequency $f_1$ that have no PU activity, PU selects a frequency where 
an HU retrieves content directly from the BS cache. The function $\scriptstyle{1_{((\NTer-1)-\stateiPUTer{s_0}>0)}}$ is utilized to make sure $\scriptstyle{\frac{\stateiHUBS{s_0}}{(\NTer-1)-\stateiPUTer{s_0}}}$ is well-defined. Note that the HU operation disrupts because 
no suitable idle frequency exists for continuing the operation in the same mode which is confirmed by $1_{(\idleTerExceptfOne{s_0}==0)}$.
The case for $s_0 \xrightarrow[]{} s_3$ is similar to $s_0 \xrightarrow[]{} s_2$. The only difference is that instead of disrupting the service of an HU retrieving content directly from the BS cache, PU preempts an HU that fetches content across the BS from the universal source.

Until now, PU arrivals at the terrestrial link except for terrestrial frequency $f_1$ are interrogated. Now, we will inspect PU arrivals at the terrestrial frequency $f_1$. 

When a PU arrives at the terrestrial frequency $f_1$ and no HU is disrupted, $s_0 \xrightarrow[]{} s_4$ transition occurs. Corresponding rate is $\frac{\lambdaPUTer}{\NTer}$. Such a transition occurs when the terrestrial frequency $f_1$ is idle (no PU operation and no HU in D2D mode operation at terrestrial frequency $f_1$ occurs.), which is checked by $\scriptstyle{1_{((\stateiPUTerfOne{s_0}==0)\land(\stateiHUD{s_0}==0))}}$. 

When a PU arrives at the terrestrial frequency $f_1$ and HU operation(s) in D2D mode are disrupted, $s_0 \xrightarrow[]{} s_5$ transition with rate $\frac{\lambdaPUTer}{\NTer}$ occurs. This can happen only if the terrestrial frequency $f_1$ is occupied by HU(s) in D2D mode which is checked by $1_{(\stateiHUD{s_0}>0)}$.

When a PU completes service over the terrestrial frequencies except for $f_1$, $s_0 \xrightarrow[]{} s_6:= \scriptstyle{\state{(\iPUTer-1)}}$ transition with rate $\stateiPUTer{s_0}\muPUTer$ occurs. 
When a PU completes service at the terrestrial frequency $f_1$, $s_0 \xrightarrow[]{} s_7:= \scriptstyle{\state{(\iPUTerfOne-1)}}$ transition with rate $\stateiPUTerfOne{s_0}\muPUTer$ occurs. $\stateiPUTerfOne{s_0}$ is an indicator to monitor whether the terrestrial frequency $f_1$ is used by a PU or not. So, it can be either zero or one. If it is zero, there is no service given to a PU so the service completion rate is zero. If $\stateiPUTerfOne{s_0}$ is one, there is a service to a PU and the corresponding service completion rate is $\muPUTer$.

Until now, we investigated PU activities originated at $s_0$. For getting the complete set of balance equations, we also need the transitions destined to $s_0$. Their links, transition types and transition rate explanations are similar to what we did up until now. Thereof, we omit the explanation of transitions due to PU activities destined to our generic state $s_0$.

\subsection{D2D Operation Mode}\label{subsection:D2D}
Due to mobility in the wireless network, device locations show stochastic behaviour. In modelling such dynamic behaviour, in various works such as~\cite{6955966} and~\cite{7562057}, devices are distributed by Poisson Point Process (PPP) in the spatial domain. Also in our study, our users HUs are randomly located in the BS cell following a PPP with mean density $\lambdaNHU$. The D2D operations are handled at the terrestrial frequency $f_1$ in an \textbf{overlaid} manner. We need an proper way to model interference issue in this system. For this purpose, we calculate the probability of a content $c_i$ being available in some HU device within some vicinity of the requester, where the transmitter HU will not cause interference to other HUs in D2D mode and the receiver HU not being interfered by other HUs in D2D mode. This probability $\pdtwod{i}{x}$ taking the current state x as a parameter will be used in Subsection~\ref{subsec:HUTransitions}. 
It is defined as follows: 
\begin{IEEEeqnarray}{rCl}
\pdtwod{i}{x} &:=& 
\scriptstyle{1_{(0 \leq \stateiHUD{x} < \DMAX)}\probDTwoDRec{x} \probDTwoDTx{x} \probDTwoDProb{i}}
\label{eq:pdtwod}\\
\probDTwoDRec{x} &:=& \scriptstyle{\frac{max\{0,(\pi \RBS^2) - (\stateiHUD{x} \pi \RInt^2)\}}{\pi \RBS^2}}\label{eq:D2DRec}\\
\probDTwoDTx{x}  &:=& \scriptstyle{\frac{max\{0,(\pi \RBS^2) - (\stateiHUD{x} \pi (2\RInt)^2)\}}{\pi \RBS^2}}\label{eq:D2DTx}\\
\probDTwoDProb{i} &:=& 1-(1-\pbaselocal{i})^{\floor{
\lambdaNHU \pi \RInt^2}}\label{eq:D2DContent}
\end{IEEEeqnarray}

Note that $\pdtwod{i}{x}$ is zero if the number of receiver HUs in D2D mode attains the maximum value $\DMAX$. Otherwise, it is $\probDTwoDRec{x} \probDTwoDTx{x} \probDTwoDProb{i}$ where $\probDTwoDRec{x}$, $\probDTwoDTx{x}$ and $\probDTwoDProb{i}$ are given in (\ref{eq:D2DRec}), (\ref{eq:D2DTx}) and (\ref{eq:D2DContent}) respectively. 
First let us define $\RInt$ and $\DMAX$. In~\cite{8292555} for modelling interference, interference range is defined where concurrent transmissions in this range interfere with each other while no interference occurs out of this range. 
Similarly, in our work, we define $\RInt$ as the radius of the transmission range of an HU device that interferes with other HU receivers operating at the terrestrial frequency $f_1$ in D2D mode. The interference to the D2D transmissions at the terrestrial frequency $f_1$ out of this range is assumed to be negligible. $\DMAX$ is the maximum number of concurrent D2D operations that is allowed by the network. It has a physical upper bound as shown in~(\ref{eq:DMax}). 
We subtract the maximum number of active HU devices that retrieve contents from the satellite or the BS (directly from cache or over the universal source) ($\NSat+(\NTer-1)$) from the average number of HU devices in BS cell ($\lambdaNHU \pi \RBS^2$) to guarantee that even if all non-D2D HU operations are active there is still room for new D2D operations. In a D2D operation, two HUs are actively used and blocked for new operation. Thereof, we divide the maximum number of HUs in D2D mode by two.  
\begin{IEEEeqnarray}{rCl}
\DMAX \leq \llcorner \frac{(\lambdaNHU \pi \RBS^2)-\NSat-(\NTer-1)}{2} \lrcorner \label{eq:DMax}
\end{IEEEeqnarray}
\vspace{-0.4cm}
\begin{figure}[h]
\centering
\includegraphics[width=0.75\columnwidth]{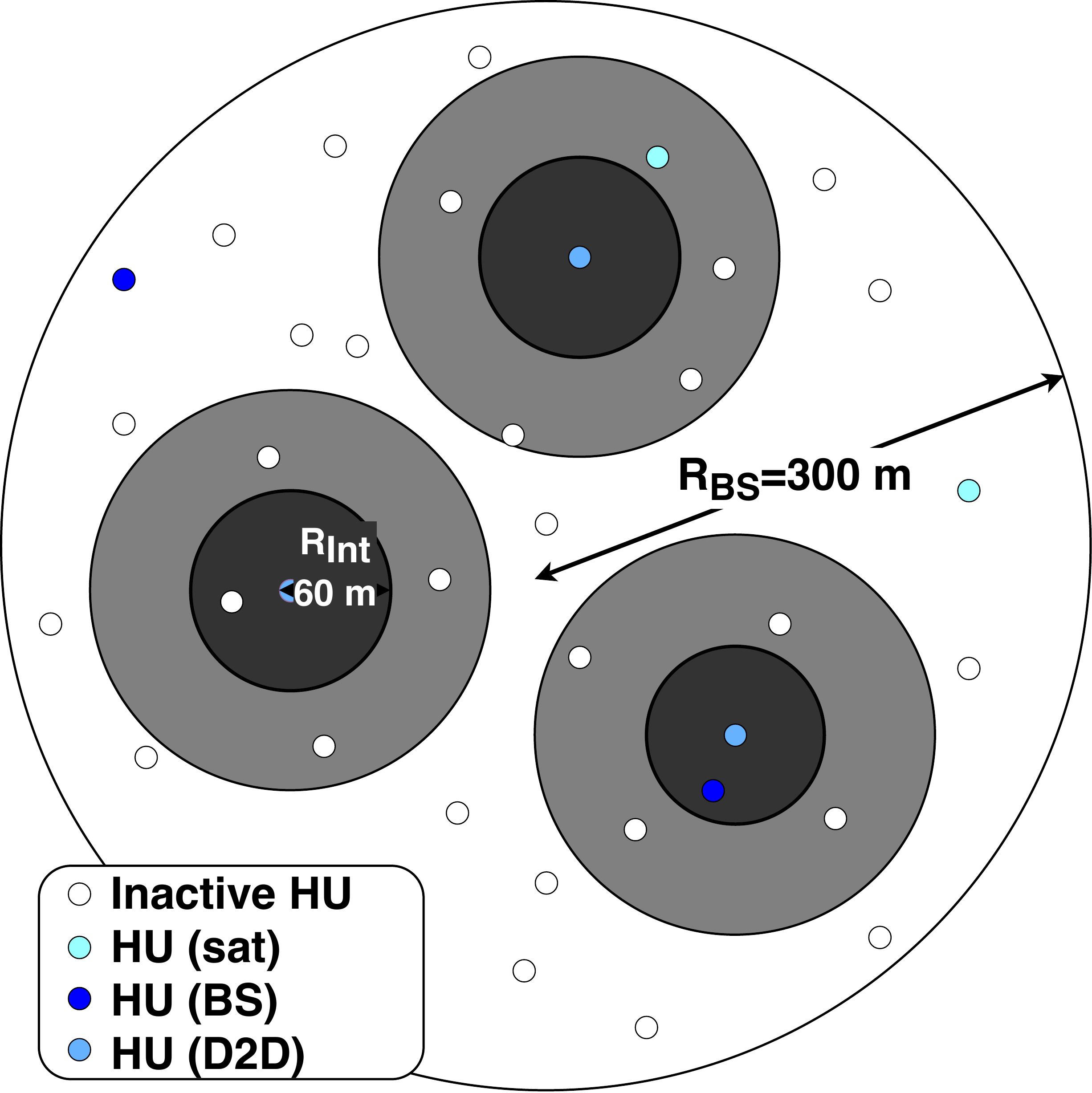}
\caption{D2D spatial stochastic model.}
\label{fig:D2D}
\end{figure}

To calculate the function $\probDTwoDRec{x}$ in~(\ref{eq:D2DRec}), we subtract the interference ranges of active HU devices in D2D mode (dark shaded areas in Fig.~\ref{fig:D2D} $\xrightarrow{} \stateiHUD{x} \pi \RInt^2$) from the cell area ($\pi \RBS^2$) and divide over $\pi \RBS^2$, to get the probability of our receiver HU device not being in the interference range of some active HU device. 
The max function is used to assure that the probability $\probDTwoDRec{x}$ attains a non-negative value.

Note that a transmitter should avoid to be in the reception range of some active receiver HU device in order not to cause interference. To satisfy this, $\stateiHUD{x} \pi (2\RInt)^2$ area should be avoided. The receiver of some D2D operation is within at most $\RInt$ meter away from its transmitter to be able to receive the requested content. If simultaneously another transmitter that is $\RInt$ or closer is also active then this may lead to collision and both transmissions will deteriorate. To avoid such collisions at each receiver HU, 
$\pi (2\RInt)^2$ area with the origin at the HU transmitter is prohibited for other HUs actively transmitting in D2D mode. We aggregate these ranges for all receiver HUs by multiplying with the number of active D2D operations $\stateiHUD{x}$. By subtracting this aggregated zone (dark shaded area+light shaded area in Fig.~\ref{fig:D2D} $\xrightarrow{}\stateiHUD{x} (\pi 2\RInt)^2$) from the cell area ($\pi \RBS^2$) and dividing over $\pi \RBS^2$, we get the probability of our transmitter HU device not deteriorating any active D2D transmission. This probability is denoted by $\probDTwoDTx{x}$ in~(\ref{eq:D2DTx}). Again max function is used to ensure $\probDTwoDTx{x}$ attains a non-negative value.

For receiving a content, that content should be available at some HU device within the reception range. 
Multiplying $\lambdaNHU$ with the reception range of requester HU device r-HU ($\pi \RInt^2$) gives us the average number of HUs in that range. $(1-\pbaselocal{i})^{\floor{
\lambdaNHU \pi \RInt^2}}$ is the probability of no HU device within the reception range of r-HU having content $c_i$ available in the local cache. The final probability $\probDTwoDProb{i}$ is the probability of finding at least one HU device that has the requested content $c_i$ among the HU devices within the reception range ($\pi \RInt^2$) of the r-HU.

In Fig.~\ref{fig:D2D},
the prohibited areas for a new HU transmitter candidate in D2D mode (dark shaded areas+light shaded areas) can have intersecting portions. So, the prohibited area can have values lower than $(\stateiHUD{x} \pi (2\RInt)^2)$ and $\probDTwoDTx{x}$ being larger. Besides, the interference range of active HUs in D2D mode (dark shaded areas in Fig.~\ref{fig:D2D}) or the prohibited areas for a new HU transmitter candidate in D2D mode (dark shaded areas+light shaded areas in Fig.~\ref{fig:D2D}) can intersect with the boundaries of the BS cell so attain lower values and $\probDTwoDRec{x}$, $\probDTwoDTx{x}$ being larger. So, our calculated $\probDTwoDRec{x}$ and $\probDTwoDTx{x}$ serve as lower bounds for the corresponding probabilities. Thus, $\pdtwod{i}{x}$ is a lower bound for the probability of a content $c_i$ being available in some HU device within some vicinity of the content requester, where that HU device will be a transmitter and not cause interference to other HUs in D2D mode while the HU receiver not being interfered by other HUs in D2D mode.

\subsection{HU Transitions}\label{subsec:HUTransitions}

Our RA schemes for HU content requests consider caches of system units (caches of the satellite, BS, HU devices), channel state and mode weights ($r_{sat},r_{BS},r_{dev}$). These weights are configurable system parameters where $r_{sat}+r_{BS}+r_{dev}=1$. In our RA, we utilize aggregate mode weight functions $\RATESAT{x}$, $\RATEBS{x}$, $\RATEDD{x}$ for some state x. Before resource allocation mechanism exploration, let us define these functions.
\begin{align}
\scriptstyle{\RATESAT{x}} &\scriptstyle{:= \RatioSat   \cdot \idleSat{x}}\\
\scriptstyle{\RATEBS{x}} &\scriptstyle{:= \RatioBS    \cdot \idleTerExceptfOne{x}}\\
\scriptstyle{\RATEDD{x}} &\scriptstyle{:= \RatioHUDev \cdot [1_{(0<\stateiHUD{x}<\DMAX)}+}\\
& \scriptstyle{1_{((\stateiHUD{x}==0)\land(\stateiPUTerfOne{x}==0))}]}\nonumber
\end{align}

These functions are utilized for selection between different modes. Consider at channel state x, some HU requests content $c_i$ which is available in the BS and in the cache of some HU device within reception range. Then $\RATEBS{x}$ assigns BS mode weight ($\RatioBS$) per each idle terrestrial frequency except for terrestrial frequency $f_1$ that corresponds to the aggregate weight of the BS. Similarly, $\RATEDD{x}$ assigns $\RatioHUDev$ to terrestrial frequency $f_1$ if that frequency is idle or used by D2D operations but not at the maximum number of allowed concurrent D2D operations. This assignment is the aggregate weight of D2D. $\frac{\RATEBS{x}}{\RATEBS{x}+\RATEDD{x}}$ gives us the probability of content $c_i$ retrieved in BS mode and $\frac{\RATEDD{x}}{\RATEBS{x}+\RATEDD{x}}$ the probability of $c_i$ retrieved in D2D mode.

Let us look at the HU state transitions. 
For HU transition inspection, we denote $state_0$ in Fig.~\ref{fig:NewState} as $h_0$. First, we analyze the transitions originated at $h_0$ upon HU arrivals as listed in Table~\ref{table:HU arrival transitions at h0}. The table consists of six columns. The first column is the identifier. The second column is link type (satellite/terrestrial), the third one availability of requested content. The fourth column is the destination state while the next one is the probability of the corresponding content availability case. The final column is the transition rate.

\begin{table*}[t]
\centering
	\caption{Transitions originated at a generic state $\state{0}$($h_0$) due to HU Arrivals.}
	    \renewcommand{\arraystretch}{1.3}
	\scalebox{0.81}{
	\begin{tabular}{ | p{0.25cm} | p{0.4cm} | p{1.7cm} |p{2.25cm} | p{15.5cm} |}
	\hline 
	\textbf{} &
	\textbf{Link} &
	\hspace{-0.15cm}\textbf{Content avail.} &
	\textbf{Dest. State} &
	\textbf{Transition Rate} \\ \hline 
	\hspace{-0.15cm}1-a &
	Sat &
	$\emptyset$ &
	$h_{1}$:=$\state{(\iHUSatUniv+1)}$ &
	\hspace{-0.15cm}$\gammaHUSatU{i}{\emptyset}$=$\lambdacioverBase{i}[1-\pbaselocal{i}][1-\pbasesat{i}][1-\pbaseBS{i}][1-\pdtwod{i}{h_0}][\frac{\RATESAT{h_0}}{\RATESAT{h_0}+\RATEBS{h_0}}]1_{(\RATESAT{h_0}+\RATEBS{h_0}>0)}$\\[1ex] \hline
	\hspace{-0.15cm}1-b &
	Sat &
	BS &
	$h_{1}$:=$\state{(\iHUSatUniv+1)}$ &
	\hspace{-0.15cm}$\gammaHUSatU{i}{\{BS\}}$=$\lambdacioverBase{i}[1-\pbaselocal{i}][1-\pbasesat{i}]\pbaseBS{i}[1-\pdtwod{i}{h_0}]1_{([(\idleTerExceptfOne{h_0}==0)\lor (\RatioBS == 0)]\land(\idleSat{h_0} > 0)\land(\RatioSat >0))}$ \\[1ex] \hline
	\hspace{-0.15cm}1-c &
	Sat &
	DEV &
	$h_{1}$:=$\state{(\iHUSatUniv+1)}$ &
	\hspace{-0.15cm}$\gammaHUSatU{i}{\{Dev\}}$=$\lambdacioverBase{i}[1-\pbaselocal{i}][1-\pbasesat{i}][1-\pbaseBS{i}]\pdtwod{i}{h_0}[\frac{\RATESAT{h_0}}{\RATESAT{h_0}+\RATEBS{h_0}}]1_{(\RATESAT{h_0}+\RATEBS{h_0}>0)}$
	$1_{((\RatioHUDev==0)\lor(\stateiHUD{h_0} == \DMAX)\lor (\stateiPUTerfOne{h_0} == 1))} $\\[1ex] \hline
	\hspace{-0.15cm}1-d &
	Sat &
	BS$\&$DEV &
	$h_{1}$:=$\state{(\iHUSatUniv+1)}$ &
	\hspace{-0.15cm}$\gammaHUSatU{i}{\{BS,Dev\}}$=$\lambdacioverBase{i}[1-\pbaselocal{i}][1-\pbasesat{i}]\pbaseBS{i}\pdtwod{i}{h_0}1_{(({\RATEBS{h_0}+\RATEDD{h_0}}==0)\land(\idleSat{h_0}>0)\land(\RatioSat>0))}$
	\\[1ex] \hline
	\hspace{-0.15cm}2-a &
	Ter &
	$\emptyset$ &
	$h_{2}$:=$\state{(\iHUBSUniv+1)}$ &
	\hspace{-0.15cm}$\gammaHUBSU{i}{\emptyset}$=$\lambdacioverBase{i}[1-\pbaselocal{i}][1-\pbasesat{i}][1-\pbaseBS{i}][1-\pdtwod{i}{h_0}][\frac{\RATEBS{h_0}}{\RATESAT{h_0}+\RATEBS{h_0}}]1_{(\RATESAT{h_0}+\RATEBS{h_0}>0)}$\\[1ex] \hline
	\hspace{-0.15cm}2-b &
	Ter &
	SAT &
	$h_{2}$:=$\state{(\iHUBSUniv+1)}$ &
	\hspace{-0.15cm}$\gammaHUBSU{i}{\{sat\}}$=$\lambdacioverBase{i}[1-\pbaselocal{i}]\pbasesat{i}[1-\pbaseBS{i}][1-\pdtwod{i}{h_0}]$
	$1_{([(\idleSat{h_0}==0)\lor(\RatioSat == 0)]\land(\idleTerExceptfOne{h_0} > 0) \land (\RatioBS>0))}$\\[1ex] \hline 
	\hspace{-0.15cm}2-c &
	Ter &
	DEV &
	$h_{2}$:=$\state{(\iHUBSUniv+1)}$ &
	\hspace{-0.15cm}$\gammaHUBSU{i}{\{Dev\}}$=$\lambdacioverBase{i}[1-\pbaselocal{i}][1-\pbasesat{i}][1-\pbaseBS{i}]\pdtwod{i}{h_0}[\frac{\RATEBS{h_0}}{\RATESAT{h_0}+\RATEBS{h_0}}]1_{(\RATESAT{h_0}+\RATEBS{h_0}>0)}$
	$1_{((\RatioHUDev==0)\lor(\stateiHUD{h_0} == \DMAX)\lor (\stateiPUTerfOne{h_0} == 1))}$\\[1ex] \hline
	\hspace{-0.15cm}2-d &
	Ter &
	SAT$\&$DEV &
	$h_{2}$:=$\state{(\iHUBSUniv+1)}$ &
	\hspace{-0.15cm}$\gammaHUBSU{i}{\{sat,Dev\}}$=$\lambdacioverBase{i}[1-\pbaselocal{i}]\pbasesat{i}[1-\pbaseBS{i}]\pdtwod{i}{h_0}1_{(({\RATESAT{h_0}+\RATEDD{h_0}} == 0)\land(\idleTerExceptfOne{h_0}>0)\land(\RatioBS>0))}$
	\\[1ex] \hline
	\hspace{-0.15cm}3-a &
	Sat &
	SAT &
	$h_{3}$:=$\state{(\iHUSat+1)}$ &
	\hspace{-0.15cm}$\gammaHUSat{i}{\{sat\}}$=$\lambdacioverBase{i}[1-\pbaselocal{i}]\pbasesat{i}[1-\pbaseBS{i}][1-\pdtwod{i}{h_0}]1_{((\idleSat{h_0}>0)\land(\RatioSat>0))}$
	\\[1ex] \hline
	\hspace{-0.15cm}3-b &
	Sat &
	SAT$\&$BS &
	$h_{3}$:=$\state{(\iHUSat+1)}$ &
	\hspace{-0.15cm}$\gammaHUSat{i}{\{sat,BS\}}$=$\lambdacioverBase{i}[1-\pbaselocal{i}]\pbasesat{i}\pbaseBS{i}[1-\pdtwod{i}{h_0}][\frac{\RATESAT{h_0}}{\RATESAT{h_0}+\RATEBS{h_0}}]1_{({\RATESAT{h_0}+\RATEBS{h_0}}>0)}$\\[1ex] \hline
	\hspace{-0.15cm}3-c &
	Sat &
	SAT$\&$DEV &
	$h_{3}$:=$\state{(\iHUSat+1)}$ &
	\hspace{-0.15cm}$\gammaHUSat{i}{\{sat,Dev\}}$=$\lambdacioverBase{i}(1-\pbaselocal{i})\pbasesat{i}(1-\pbaseBS{i})\pdtwod{i}{h_0}(\frac{\RATESAT{h_0}}{\RATESAT{h_0}+\RATEDD{h_0}})1_{({\RATESAT{h_0}+\RATEDD{h_0}}>0)}$\\[1ex] \hline
	\hspace{-0.15cm}3-d &
	Sat &
	SAT$\&$BS$\&$DEV &
	$h_{3}$:=$\state{(\iHUSat+1)}$ &
	\hspace{-0.15cm}$\gammaHUSat{i}{\{sat,BS,Dev\}}$=$\lambdacioverBase{i}[1-\pbaselocal{i}]\pbasesat{i}\pbaseBS{i}\pdtwod{i}{h_0}[\frac{\RATESAT{h_0}}{\RATESAT{h_0}+\RATEBS{h_0}+\RATEDD{h_0}}]1_{({\RATESAT{h_0}+\RATEBS{h_0}+\RATEDD{h_0}}>0)}$\\[1ex] \hline
	\hspace{-0.15cm}4-a &
	Ter &
	BS &
	$h_{4}$:=$\state{(\iHUBS+1)}$ &
	\hspace{-0.15cm}$\gammaHUBS{i}{\{BS\}}$=$\lambdacioverBase{i}[1-\pbaselocal{i}][1-\pbasesat{i}]\pbaseBS{i}[1-\pdtwod{i}{h_0}]1_{((\idleTerExceptfOne{h_0}>0)\land(\RatioBS>0))}$\\[1ex] \hline
	\hspace{-0.15cm}4-b &
	Ter &
	SAT$\&$BS &
	$h_{4}$:=$\state{(\iHUBS+1)}$ &
	\hspace{-0.15cm}$\gammaHUBS{i}{\{sat,BS\}}$=$\lambdacioverBase{i}[1-\pbaselocal{i}]\pbasesat{i}\pbaseBS{i}[1-\pdtwod{i}{h_0}][\frac{\RATEBS{h_0}}{\RATESAT{h_0}+\RATEBS{h_0}}]1_{({\RATESAT{h_0}+\RATEBS{h_0}}>0)}$\\[1ex] \hline
    \hspace{-0.15cm}4-c &
	Ter &
	BS$\&$DEV &
	$h_{4}$:=$\state{(\iHUBS+1)}$ &
	\hspace{-0.15cm}$\gammaHUBS{i}{\{BS,Dev\}}$=$\lambdacioverBase{i}[1-\pbaselocal{i}][1-\pbasesat{i}]\pbaseBS{i}\pdtwod{i}{h_0}[\frac{\RATEBS{h_0}}{\RATEBS{h_0}+\RATEDD{h_0}}]1_{({\RATEBS{h_0}+\RATEDD{h_0}}>0)}$\\[1ex] \hline
	\hspace{-0.15cm}4-d &
	Ter &
	SAT$\&$BS$\&$DEV &
	$h_{4}$:=$\state{(\iHUBS+1)}$ &
	\hspace{-0.15cm}$\gammaHUBS{i}{\{sat,BS,Dev\}}$=$\lambdacioverBase{i}[1-\pbaselocal{i}]\pbasesat{i}\pbaseBS{i}\pdtwod{i}{h_0}[\frac{\RATEBS{h_0}}{\RATESAT{h_0}+\RATEBS{h_0}+\RATEDD{h_0}}]1_{({\RATESAT{h_0}+\RATEBS{h_0}+\RATEDD{h_0}}>0)}$\\[1ex] \hline
    \hspace{-0.15cm}5-a &
	Ter &
	DEV &
	$h_{5}$:=$\state{(\iHUDev+1)}$ &
	\hspace{-0.15cm}$\gammaHUD{i}{\{Dev\}}$=$\lambdacioverBase{i}[1-\pbaselocal{i}][1-\pbasesat{i}][1-\pbaseBS{i}]\pdtwod{i}{h_0}[1_{(\RatioHUDev>0)}]$\\
	& & & &
	$[(1_{(0<\stateiHUD{h_0}<\DMAX)})
	+(1_{((\stateiHUD{h_0}==0)\land(\stateiPUTerfOne{h_0}==0))})]$\\[1ex] \hline
    \hspace{-0.15cm}5-b &
	Ter &
	SAT$\&$DEV &
	$h_{5}$:=$\state{(\iHUDev+1)}$ &
	\hspace{-0.15cm}$\gammaHUD{i}{\{sat,Dev\}}$=$\lambdacioverBase{i}[1-\pbaselocal{i}]\pbasesat{i}[1-\pbaseBS{i}]\pdtwod{i}{h_0}[\frac{\RATEDD{h_0}}{\RATESAT{h_0}+\RATEDD{h_0}}]1_{({\RATESAT{h_0}+\RATEDD{h_0}}>0)}$\\[1ex] \hline
    \hspace{-0.15cm}5-c &
	Ter &
	BS$\&$DEV &
	$h_{5}$:=$\state{(\iHUDev+1)}$ &
	\hspace{-0.15cm}$\gammaHUD{i}{\{BS,Dev\}}$=$\lambdacioverBase{i}[1-\pbaselocal{i}][1-\pbasesat{i}]\pbaseBS{i}\pdtwod{i}{h_0}[\frac{\RATEDD{h_0}}{\RATEBS{h_0}+\RATEDD{h_0}}]1_{({\RATEBS{h_0}+\RATEDD{h_0}}>0)}$\\[1ex] \hline
    \hspace{-0.15cm}5-d &
	Ter &
	SAT$\&$BS$\&$DEV &
	$h_{5}$:=$\state{(\iHUDev+1)}$ &
	\hspace{-0.15cm}$\gammaHUD{i}{\{sat,BS,Dev\}}$=$\lambdacioverBase{i}[1-\pbaselocal{i}]\pbasesat{i}\pbaseBS{i}\pdtwod{i}{h_0}[\frac{\RATEDD{h_0}}{\RATESAT{h_0}+\RATEBS{h_0}+\RATEDD{h_0}}]1_{({\RATESAT{h_0}+\RATEBS{h_0}+\RATEDD{h_0}}>0)}$\\[1ex] 
	\hline
	\end{tabular}
	}
	\label{table:HU arrival transitions at h0}
\end{table*}

When with rate $\lambda_{HU}^{c_i}$ some HU requests some arbitrary content $c_i$, our RA mechanism first checks whether $c_i$ is available in the local cache. If so, it does not go into channel and no state transition occurs. 
Otherwise, there are five possible transitions due to HU requesting some arbitrary content $c_i$. 
\begin{itemize}
    \item $\boldsymbol{h_0\xrightarrow{}h_1}$:  $c_i$ is first fetched from the universal source to the satellite cache and then from there to requester HU device (r-HU). 
    \item $\boldsymbol{h_0\xrightarrow{}h_2}$:  $c_i$ is first fetched from the universal source to the BS cache and then from there to r-HU. 
    \item $\boldsymbol{h_0\xrightarrow{}h_3}$:  $c_i$ is fetched from the satellite cache to r-HU.
    \item $\boldsymbol{h_0\xrightarrow{}h_4}$:  $c_i$ is fetched from the BS cache to r-HU. 
    \item $\boldsymbol{h_0\xrightarrow{}h_5}$:  $c_i$ is fetched from the cache of some HU device within reception range to r-HU. 
\end{itemize}

Let us start our exploration with these transitions. When the requested content is fetched from the universal source into the satellite cache and then from the satellite cache to the requester HU device cache, $h_0\xrightarrow{}h_1$ occurs. 
$h_0\xrightarrow{}h_1$ transition has four possible content availability combinations: 
\begin{enumerate}
    \item[1-a)] not available in any system unit 
    \item[1-b)] BS cache only 
    \item[1-c)] some HU device cache within the reception range of r-HU only 
    \item[1-d)] BS cache and some HU device cache within the reception range of r-HU
\end{enumerate}

In the first case among four content availability scenarios    (1-a), the probability of finding requested content $c_i$ nowhere is equal to $[1-\pbaselocal{i}][1-\pbasesat{i}][1-\pbaseBS{i}][1-\pdtwod{i}{h_0}]$. Our RA algorithm will fetch $c_i$ from the universal source either to the satellite or to the BS first. For selecting between the satellite and the BS, $\RATESAT{h_0}$ and $\RATEBS{h_0}$ are used. By function $\RATESAT{x}$ := $\RatioSat   \cdot \idleSat{x}$ we assign a system unit weight ($\RatioSat$) for each idle frequency at the satellite link. Similarly, for function $\RATEBS{x}$ :=  $\RatioBS    \cdot \idleTerExceptfOne{x}$ we assign a system unit weight ($\RatioBS$) for each idle frequency at the terrestrial link except for frequency $f_1$. 
Thus, we assign weights to distinct system units to tune selection probabilities and also consider channel state and the number of idle frequencies at the corresponding links. 
We define $\RATESAT{x}$ and $\RATEBS{x}$ functions as aggregate weights of satellite and BS. 
The proportion of aggregate weight $\RATESAT{h_0}$ over $\RATESAT{h_0}+\RATEBS{h_0}$ gives us the probability of $c_i$ retrieved from the universal source to the satellite cache and then from there to r-HU. This case has transition rate $\gammaHUSatU{i}{\emptyset}$. 

In the second case (1-b), the probability of finding requested content $c_i$ in the BS cache but not in the satellite, the local cache and some other HU device within the reception range is equal to $[1-\pbaselocal{i}][1-\pbasesat{i}]\pbaseBS{i}[1-\pdtwod{i}{h_0}]$. Even though $c_i$ is available in the BS cache, if there is no idle terrestrial frequency except for frequency $f_1$ for getting service or BS mode weight $r_{BS}$ is zero ($1_{([(\idleTerExceptfOne{h_0}==0)\lor (\RatioBS == 0)])}$), $c_i$ can not be retrieved from the BS cache. When the satellite link is available and satellite mode weight is greater than zero ($1_{((\idleSat{h_0} > 0)\land(\RatioSat >0))}$), $c_i$ is fetched from the universal source to the satellite cache first and then transmitted to r-HU. The corresponding rate is denoted as $\gammaHUSatU{i}{\{BS\}}$. 

In the third case (1-c), the probability of finding requested content $c_i$ in some other HU device within the reception range but not in the satellite, the BS and the local cache is equal to $[1-\pbaselocal{i}][1-\pbasesat{i}][1-\pbaseBS{i}]\pdtwod{i}{h_0}$. Even if $c_i$ is available in some HU device cache within the reception range, if the terrestrial frequency $f_1$ is occupied by maximum number of concurrent D2D operations that is allowed by the network $\DMAX$ or the terrestrial frequency $f_1$ is occupied by PU or D2D mode weight is zero ($1_{(\RatioHUDev==0)\lor(\stateiHUD{h_0} == \DMAX)\lor (\stateiPUTerfOne{h_0} == 1)}$), then $c_i$ can not be retrieved in D2D mode. Our RA algorithm will try to fetch the content from the universal source either to the satellite or the BS first. For selecting between the satellite and the BS, $\RATESAT{h_0}$ and $\RATEBS{h_0}$ are utilized. The proportion of aggregate weight $\RATESAT{h_0}$ over $\RATESAT{h_0}+\RATEBS{h_0}$ gives us the probability of $c_i$ retrieved from the universal source to the satellite cache and then from there to r-HU. The corresponding rate is denoted as $\gammaHUSatU{i}{\{Dev\}}$. 

In the fourth case (1-d), the probability of finding requested content $c_i$ in the BS and some other HU device cache within the reception range but not in the local cache and the satellite is equal to $[1-\pbaselocal{i}][1-\pbasesat{i}]\pbaseBS{i}\pdtwod{i}{h_0}$. Although $c_i$ is available, if the terrestrial channel is not usable ($1_{{\RATEBS{h_0}+\RATEDD{h_0}}==0}$), it can not be retrieved over the terrestrial link. If the satellite link is available and satellite mode weight is greater than zero ($1_{((\idleSat{h_0}>0)\land(\RatioSat>0)}$), $c_i$ is fetched from the universal source to the satellite cache first and then transmitted to r-HU. The corresponding rate is denoted as $\gammaHUSatU{i}{\{BS\&Dev\}}$. 

When we sum the transition rate of all four cases, we get the \textbf{aggregate transition rate for the retrieval of content $c_i$ from the universal source across the satellite} as in~(\ref{eq:gammaHUSatUnivAgg}). 
\begin{IEEEeqnarray}{rCl}\label{eq:gammaHUSatUnivAgg}
\gammaHUSatUAgg{i} &:=& \scriptstyle{\gammaHUSatU{i}{\{sat\}}+ \gammaHUSatU{i}{\{sat,BS\}}+}\\ ~&~&\scriptstyle{\gammaHUSatU{i}{\{sat,Dev\}}+ \gammaHUSatU{i}{\{sat,BS,Dev\}}}\nonumber 
\end{IEEEeqnarray}

The \textbf{aggregate transition rate for all contents from the universal source across the satellite} ($h_0\xrightarrow{}h_1$) which corresponds to the service of HUs from the universal source across the satellite 
is given in (\ref{eq:gammaHUSatUAll}). 
\begin{IEEEeqnarray}{rCl}\label{eq:gammaHUSatUAll}
\gammaHUSatUAll &:=& \sum_{i=1}^{N} \gammaHUSatUAgg{i}
\end{IEEEeqnarray}

Let us elaborate on the transition $h_0\xrightarrow{}h_2$. In transition $h_0\xrightarrow{}h_2$ the requested content is fetched from the universal source into the BS cache and then from the BS cache to the requester HU device cache. 
As the transition rate construction is similar to $h_0\xrightarrow{}h_1$ case, we skip the details. The \textbf{aggregate transition rate for the retrieval of content $c_i$ from the universal source across the BS} is as follows:
\begin{IEEEeqnarray}{rCl}\label{eq:gammaHUBSUAgg}
\gammaHUBSUAgg{i} &:=& \scriptstyle{\gammaHUBSU{i}{\{BS\}}+ \gammaHUBSU{i}{\{sat,BS\}}+}\\ ~&~& \scriptstyle{\gammaHUBSU{i}{\{BS,Dev\}}+ \gammaHUBSU{i}{\{sat,BS,Dev\}}}\nonumber 
\end{IEEEeqnarray}
The \textbf{aggregate transition rate for all contents from the universal source across the BS} ($h_0\xrightarrow{}h_2$) which corresponds to the service of HUs from the universal source across the BS is given in Eq.~\ref{eq:gammaHUBSUAll}.
\begin{IEEEeqnarray}{rCl}\label{eq:gammaHUBSUAll}
\gammaHUBSUAll &:=& \sum_{i=1}^{N} \gammaHUBSUAgg{i}
\end{IEEEeqnarray}

Until now, we looked at HU requests that fetch content over the universal source. Now, let us continue with transition $h_0\xrightarrow{}h_3$. In transition $h_0\xrightarrow{}h_3$ the service is given directly from the satellite cache over the satellite link. 
For this transition to occur there are four possible content availability combinations: 
\begin{enumerate}
    \item[3-a)] satellite cache only 
    \item[3-b)] satellite and BS cache 
    \item[3-c)] satellite cache and some HU device cache within the reception range of r-HU 
    \item[3-d)] satellite cache, BS cache and some HU device cache within the reception range of r-HU 
\end{enumerate}

In the first case among four content availability combinations (3-a), the probability of finding requested content $c_i$ only in the satellite is given by $[1-\pbaselocal{i}]\pbasesat{i}[1-\pbaseBS{i}][1-\pdtwod{i}{h_0}]$. $c_i$ can be retrieved from the satellite if the satellite link is available and the satellite mode weight $\RatioSat$ is greater than zero ($1_{((\idleSat{h_0}>0)\land(\RatioSat>0))}$), $c_i$ is fetched from the satellite to r-HU. The corresponding rate is denoted as $\gammaHUSat{i}{\{sat\}}$.

In the second case (3-b), the probability of finding requested content $c_i$ in the satellite and the BS but not in the local cache and some other HU device within the reception range is equal to $[1-\pbaselocal{i}]\pbasesat{i}\pbaseBS{i}[1-\pdtwod{i}{h_0}]$. For the selection between satellite and BS, $\RATESAT{h_0}$ and $\RATEBS{h_0}$ are utilized. 
The proportion of aggregate weight $\RATESAT{h_0}$ over the $\RATESAT{h_0}+\RATEBS{h_0}$ is the probability of 
$c_i$ retrieved from the satellite cache. The corresponding rate is denoted as $\gammaHUSat{i}{\{sat,BS\}}$.  

In the third case (3-c), the probability of finding requested content $c_i$ in the satellite and some other HU device within the reception range but not in the local cache and the BS is equal to $[1-\pbaselocal{i}]\pbasesat{i}[1-\pbaseBS{i}]\pdtwod{i}{h_0}$. For selecting between satellite and some HU device, $\RATESAT{h_0}$ and $\RATEDD{h_0}$ are utilized and with probability $\frac{\RATESAT{h_0}}{\RATESAT{h_0}+\RATEDD{h_0}}$ $c_i$ is retrieved from the satellite. The corresponding rate is denoted as $\gammaHUSat{i}{\{sat,Dev\}}$.  

In the fourth case (3-d), the probability of finding requested content $c_i$ in the satellite, BS and some other HU device within the reception range but not in the local cache is equal to $[1-\pbaselocal{i}]\pbasesat{i}\pbaseBS{i}\pdtwod{i}{h_0}$. $c_i$ is retrieved from the satellite with probability $\frac{\RATESAT{h_0}}{\RATESAT{h_0}+\RATEBS{h_0}+\RATEDD{h_0}}$. The corresponding transition rate is $\gammaHUSat{i}{\{sat,BS,Dev\}}$.

When we sum the transition rates of all four cases, we get the \textbf{aggregate transition rate for the retrieval of content $c_i$ directly from the satellite cache} as given in~(\ref{eq:gammaHUSatAgg}).
\begin{IEEEeqnarray}{rCl}\label{eq:gammaHUSatAgg}
\gammaHUSatAgg{i} &:=& \scriptstyle{\gammaHUSat{i}{\{sat\}}+ \gammaHUSat{i}{\{sat,BS\}}+}\\ ~&~&\scriptstyle{\gammaHUSat{i}{\{sat,Dev\}}+ \gammaHUSat{i}{\{sat,BS,Dev\}}}\nonumber 
\end{IEEEeqnarray}
The \textbf{aggregate transition rate for all contents retrieved directly from the satellite cache} ($h_0\xrightarrow{}h_3$) corresponding to the service of HUs directly from the satellite is given in Eq.~\ref{eq:gammaHUSatAll}.
\begin{IEEEeqnarray}{rCl}\label{eq:gammaHUSatAll}
\gammaHUSatAll &:=& \sum_{i=1}^{N} \gammaHUSatAgg{i}
\end{IEEEeqnarray}

Similar to $h_0\xrightarrow{}h_3$, 
the transition $h_0\xrightarrow{}h_4$ where the service is given directly from the BS cache over the terrestrial frequencies except for terrestrial frequency $f_1$ 
has four content availability combinations. But this time instead of the satellite, the BS always has the requested content. 
As transition rate construction is similar to $h_0\xrightarrow{}h_3$ case, we skip the details. 
The \textbf{aggregate transition rate for the retrieval of content $c_i$ directly from the BS cache} is given in~(\ref{eq:gammaHUBSAgg}). 
\begin{IEEEeqnarray}{rCl}\label{eq:gammaHUBSAgg}
\gammaHUBSAgg{i} &:=& \scriptstyle{\gammaHUBS{i}{\{BS\}}+ \gammaHUBS{i}{\{sat,BS\}}+}\\ ~&~&\scriptstyle{\gammaHUBS{i}{\{BS,Dev\}}+ \gammaHUBS{i}{\{sat,BS,Dev\}}}\nonumber 
\end{IEEEeqnarray}
The \textbf{aggregate transition rate for all contents retrieved directly from the BS cache} ($h_0\xrightarrow{}h_4$) which corresponds to the service of HUs directly from the BS is $\gammaHUBSAll := \sum_{i=1}^{N} \gammaHUBSAgg{i}$

Now, let us inspect the transition $h_0\xrightarrow{}h_5$ where the requested content is given from the cache of some HU device within the reception range of r-HU in D2D operation mode at terrestrial frequency $f_1$. This transition has four possible content availability combinations:
\begin{enumerate}
    \item[5-a)] some HU device cache in the reception range of r-HU only 
    \item[5-b)] satellite cache and some HU device cache within the reception range of r-HU 
    \item[5-c)] BS cache and some HU device cache within the reception range of r-HU 
    \item[5-d)] satellite cache, BS cache and some HU device cache within the reception range of r-HU 
\end{enumerate}

In the first case among four content availability combinations (5-a), the probability of finding requested content $c_i$ only in some HU device within the reception range but not in the satellite, BS and the local cache is equal to $[1-\pbaselocal{i}][1-\pbasesat{i}][1-\pbaseBS{i}]\pdtwod{i}{h_0}$. Then $c_i$ can be retrieved from some HU device cache if D2D mode weight ($r_{dev}$) is greater than zero. Besides, the terrestrial frequency $f_1$ should be idle ($1_{((\stateiHUD{h_0}==0)\land(\stateiPUTerfOne{h_0}==0))}$) or if some HUs are actively operating at that frequency the number of active HUs at that frequency should not exceed the upper bound $\DMAX$ ($1_{(0<\stateiHUD{h_0}<\DMAX)}$). The corresponding rate is denoted as $\gammaHUD{i}{\{Dev\}}$. 
5-b case has rate $\gammaHUD{i}{\{sat,Dev\}}$, 5-c $\gammaHUD{i}{\{BS,Dev\}}$ and 5-d $\gammaHUD{i}{\{sat,BS,Dev\}}$.

When we sum the transition rate of four cases, we get the \textbf{aggregate transition rate for the retrieval of content $c_i$ in D2D operation mode from the terrestrial frequency $f_1$} as given in~(\ref{eq:gammaHUDevAgg}). 
\begin{IEEEeqnarray}{rCl}\label{eq:gammaHUDevAgg}
\gammaHUDAgg{i}&:=& \scriptstyle{\gammaHUD{i}{\{Dev\}}+ \gammaHUD{i}{\{sat,Dev\}}}\\ &+&\scriptstyle{\gammaHUD{i}{\{BS,Dev\}}+ \gammaHUD{i}{\{sat,BS,Dev\}}}\nonumber 
\end{IEEEeqnarray}
The \textbf{aggregate transition rate for all contents in D2D operation mode} ($h_0\xrightarrow{}h_5$) which corresponds to the service of HUs in D2D mode is given in Eq.~\ref{eq:gammaHUDevAll}.
\begin{IEEEeqnarray}{rCl}\label{eq:gammaHUDevAll}
\gammaHUDAll &:=& \sum_{i=1}^{N} \gammaHUDAgg{i}
\end{IEEEeqnarray}

When an HU completes service and departs the channel a transition from our generic state $h_0$ to one of the destination states at the corresponding link with the corresponding transition rate 
in Table~\ref{table:HU departure transitions from h0} occurs.

\begin{table}[t]
\centering
    \renewcommand{\arraystretch}{1.3}
	\caption{Transitions originated at $h_0$ due to HU departures.}
	\begin{tabular}{ | l | l | l | p{11 cm} |}
	\hline
	\textbf{Link} &
	\textbf{Destination State} &
	\textbf{Transition Rate} \\ \hline \hline
	Sat &
	$h_{6}$ := $\state{(\iHUSatUniv-1)}$ &
	$\stateiHUSatU{h_0}\muHUSatU$\\[1ex] \hline
	Ter &
	$h_{7}$ := $\state{(\iHUBSUniv-1)}$ &
	$\stateiHUBSU{h_0}\muHUBSU$\\[1ex] \hline
	Sat &
	$h_{8}$ := $\state{(\iHUSat-1)}$ &
	$\stateiHUSat{h_0}\muHUSat$\\[1ex] \hline
	Ter &
	$h_{9}$ := $\state{(\iHUBS-1)}$ &
	$\stateiHUBS{h_0}\muHUBS$\\[1ex] \hline
	Ter &
	$h_{10}$ := $\state{(\iHUDev-1)}$ &
	$\stateiHUD{h_0}\muHUDev$\\[1ex] 
	\hline
	\end{tabular}
	\label{table:HU departure transitions from h0}
\end{table}

Until now, we looked at HU arrivals originating at our generic state $h_0$. For the balance equations we need $h_0$ destined states and their transition rates. The transitions destined to our generic state $h_0$ due to HU arrivals are listed in Table~\ref{table:HU arrival transitions into h0} while the transitions destined to our generic state $h_0$ due to HU departures are given in Table~\ref{table:HU departure transitions into h0}. The states and the composition of the transition rates are similar to $h_0$ originated states and their transitions. So, we skip the explanations of transitions due to HU arrivals/departures destined to our generic state $h_0$. With the complete set of balance equations, we find the steady state probabilities of being at each channel state. 

\begin{table*}[t]
\centering
	\caption{Transitions destined to a generic state $\state{0}$($h_0$) due to HU Arrivals.}
	\begin{tabular}{ | l | l | l | p{11 cm} |}
	\hline 
	\textbf{Link} &
	\textbf{Content avail.} &
	\textbf{Source State} &
	\textbf{Transition Rate} \\ \hline \hline
	Sat &
	Not available &
	$h_{6}$ := $\state{(\iHUSatUniv-1)}$ &
	$[\lambdacioverBase{i}(1-\pbaselocal{i})(1-\pbasesat{i})(1-\pbaseBS{i})(1-\pdtwod{i}{h_6})(\frac{\RATESAT{h_6}}{\RATESAT{h_6}+\RATEBS{h_6}})$\\
	&~&~&
	$1_{((\RATESAT{h_6}+\RATEBS{h_6}>0)\land(\stateiHUSatU{h_0}>0))}]+[\lambdacioverBase{i}(1-\pbaselocal{i})(1-\pbasesat{i})\pbaseBS{i}$\\
	&~&~&
	$(1-\pdtwod{i}{h_6})1_{\{[(\idleTerExceptfOne{h_6}==0)\lor(\RatioBS == 0)]\land(\stateiHUSatU{h_0}>0)\land(\idleSat{h_6}>0)\land(\RatioSat>0)\}}]$\\
	&~&~&
	$+[\lambdacioverBase{i}(1-\pbaselocal{i})(1-\pbasesat{i})(1-\pbaseBS{i})\pdtwod{i}{h_{6}}\times1_{(\stateiHUSatU{h_0}>0)}\times$\\
	&~&~&
	$1_{((\RatioHUDev==0)\lor(\stateiHUD{h_6}==\DMAX)\lor(\stateiPUTerfOne{h_6}==1))}$\\
	&~&~&
	$\frac{\RATESAT{h_6}}{\RATESAT{h_6}+\RATEBS{h_6}}1_{(\RATESAT{h_6}+\RATEBS{h_6}>0)}]+[\lambdacioverBase{i}(1-\pbaselocal{i})(1-\pbasesat{i})\pbaseBS{i}\pdtwod{i}{h_{6}}$
	\\
	&~&~&
	$1_{(({\RATEBS{h_{6}}+\RATEDD{h_{6}}}==0)\land(\stateiHUSatU{h_0}>0)\land(\idleSat{h_6}>0)\land(\RatioSat>0))}]$
	\\[1ex] \hline
	Ter &
	Not available &
	$h_{7}$ := $\state{(\iHUBSUniv-1)}$ &
	$[\lambdacioverBase{i}(1-\pbaselocal{i})(1-\pbasesat{i})(1-\pbaseBS{i})(1-\pdtwod{i}{h_7})(\frac{\RATEBS{h_7}}{\RATESAT{h_7}+\RATEBS{h_7}})$\\
	&~&~&
	$1_{((\RATESAT{h_7}+\RATEBS{h_7}>0)\land(\stateiHUBSU{h_0}>0))}]+[\lambdacioverBase{i}(1-\pbaselocal{i})\pbasesat{i}(1-\pbaseBS{i})$\\
	&~&~&
	$(1-\pdtwod{i}{h_7})1_{\{[(\idleSat{h_7}==0)\lor(\RatioSat==0)]\land(\stateiHUBSU{h_0}>0)\land(\idleTerExceptfOne{h_7}>0)\land(\RatioBS>0)\}}]$\\
	&~&~&
	$+[\lambdacioverBase{i}(1-\pbaselocal{i})(1-\pbasesat{i})(1-\pbaseBS{i})\pdtwod{i}{h_{7}}\times1_{(\stateiHUBSU{h_0}>0)}\times$\\
	&~&~&
	$1_{((\RatioHUDev==0)\lor(\stateiHUD{h_7}==\DMAX)\lor(\stateiPUTerfOne{h_7}==1))}$\\
    &~&~&	
	$\frac{\RATEBS{h_7}}{\RATESAT{h_7}+\RATEBS{h_7}}1_{(\RATESAT{h_7}+\RATEBS{h_7}>0)}]+[\lambdacioverBase{i}(1-\pbaselocal{i})\pbasesat{i}(1-\pbaseBS{i})\pdtwod{i}{h_{7}}$\\
	&~&~&
	$1_{(({\RATESAT{h_{7}}+\RATEDD{h_{7}}}==0)\land(\stateiHUBSU{h_0}>0)\land(\idleTerExceptfOne{h_7}>0)\land(\RatioBS>0))}]$
	\\[1ex] \hline
	Sat &
	Only in Sat &
	$h_{8}$ := $\state{(\iHUSat-1)}$ &
	$\lambdacioverBase{i}(1-\pbaselocal{i})\pbasesat{i}(1-\pbaseBS{i})(1-\pdtwod{i}{h_8})1_{((\idleSat{h_8}>0)\land(\stateiHUSat{h_0}>0)\land(\RatioSat>0))}$\\[1ex] \hline
	Ter &
	Only in BS &
	$h_{9}$ := $\state{(\iHUBS-1)}$ &
	$\lambdacioverBase{i}(1-\pbaselocal{i})(1-\pbasesat{i})\pbaseBS{i}(1-\pdtwod{i}{h_9})1_{((\idleTerExceptfOne{h_9}>0)\land(\stateiHUBS{h_0}>0)\land(\RatioBS>0))}$\\[1ex] \hline
	Ter &
	Only in dev &
	$h_{10}$ := $\state{(\iHUDev-1)}$ &
	$\lambdacioverBase{i}(1-\pbaselocal{i})(1-\pbasesat{i})(1-\pbaseBS{i})\pdtwod{i}{h_{10}}\times1_{((\stateiHUD{h_0}>0)\land(\RatioHUDev>0))}\times$
	
	$[1_{(0<\stateiHUD{h_{10}}<\DMAX)}
	+(1_{((\stateiHUD{h_{10}}==0)\land(\stateiPUTerfOne{h_{10}}==0))})]$\\[1ex] \hline
	Sat &
	In sat and BS &
	$h_{8}$ := $\state{(\iHUSat-1)}$ &
	$\lambdacioverBase{i}(1-\pbaselocal{i})\pbasesat{i}\pbaseBS{i}(1-\pdtwod{i}{h_8})(\frac{\RATESAT{h_8}}{\RATESAT{h_8}+\RATEBS{h_8}})$
	
	$1_{(({\RATESAT{h_8}+\RATEBS{h_8}}>0)\land(\stateiHUSat{h_0}>0))}$\\[1ex] \hline
	Ter &
	In sat and BS &
	$h_{9}$ := $\state{(\iHUBS-1)}$ &
	$\lambdacioverBase{i}(1-\pbaselocal{i})\pbasesat{i}\pbaseBS{i}(1-\pdtwod{i}{h_9})(\frac{\RATEBS{h_9}}{\RATESAT{h_9}+\RATEBS{h_9}})$
	
	$1_{(({\RATESAT{h_9}+\RATEBS{h_9}}>0)\land(\stateiHUBS{h_0}>0))}$\\[1ex] \hline
	Sat &
	In sat and dev &
	$h_{8}$ := $\state{(\iHUSat-1)}$ &
	$\lambdacioverBase{i}(1-\pbaselocal{i})\pbasesat{i}(1-\pbaseBS{i})\pdtwod{i}{h_8}(\frac{\RATESAT{h_8}}{\RATESAT{h_8}+\RATEDD{h_8}})$
	
	$1_{(({\RATESAT{h_8}+\RATEDD{h_8}}>0)\land(\stateiHUSat{h_0}>0))}$\\[1ex] \hline
	Ter &
	In sat and dev &
	$h_{10}$ := $\state{(\iHUDev-1)}$ &
	$\lambdacioverBase{i}(1-\pbaselocal{i})\pbasesat{i}(1-\pbaseBS{i})\pdtwod{i}{h_{10}}(\frac{\RATEDD{h_{10}}}{\RATESAT{h_{10}}+\RATEDD{h_{10}}})$
	
	$1_{(({\RATESAT{h_{10}}+\RATEDD{h_{10}}}>0)\land(\stateiHUD{h_0}>0))}$\\[1ex] \hline
	Ter &
	In BS and dev &
	$h_{9}$ := $\state{(\iHUBS-1)}$ &
	$\lambdacioverBase{i}(1-\pbaselocal{i})(1-\pbasesat{i})\pbaseBS{i}\pdtwod{i}{h_9}(\frac{\RATEBS{h_9}}{\RATEBS{h_9}+\RATEDD{h_9}})$
	
	$1_{(({\RATEBS{h_9}+\RATEDD{h_9}}>0)\land(\stateiHUBS{h_0}>0))}$\\[1ex] \hline
	Ter &
	In BS and dev &
	$h_{10}$ := $\state{(\iHUDev-1)}$ &
	$\lambdacioverBase{i}(1-\pbaselocal{i})(1-\pbasesat{i})\pbaseBS{i}\pdtwod{i}{h_{10}}(\frac{\RATEDD{h_{10}}}{\RATEBS{h_{10}}+\RATEDD{h_{10}}})$
	
	$1_{(({\RATEBS{h_{10}}+\RATEDD{h_{10}}}>0)\land(\stateiHUD{h_0}>0))}$\\[1ex] \hline
	Sat &
	In sat, BS and dev &
	$h_{8}$ := $\state{(\iHUSat-1)}$ &
	$\lambdacioverBase{i}(1-\pbaselocal{i})\pbasesat{i}\pbaseBS{i}\pdtwod{i}{h_8}(\frac{\RATESAT{h_8}}{\RATESAT{h_8}+\RATEBS{h_8}+\RATEDD{h_8}})$\\
	&~&~&
	$1_{(({\RATESAT{h_8}+\RATEBS{h_8}+\RATEDD{h_8}}>0)\land(\stateiHUSat{h_0}>0))}$\\[1ex] \hline
	Ter &
	In sat, BS and dev &
	$h_{9}$ := $\state{(\iHUBS-1)}$ &
	$\lambdacioverBase{i}(1-\pbaselocal{i})\pbasesat{i}\pbaseBS{i}\pdtwod{i}{h_9}(\frac{\RATEBS{h_9}}{\RATESAT{h_9}+\RATEBS{h_9}+\RATEDD{h_9}})$\\
	&~&~&
	$1_{(({\RATESAT{h_9}+\RATEBS{h_9}+\RATEDD{h_9}}>0)\land(\stateiHUBS{h_0}>0))}$\\[1ex] \hline
	Ter &
	In sat, BS and dev &
	$h_{10}$ := $\state{(\iHUDev-1)}$ &
	$\lambdacioverBase{i}(1-\pbaselocal{i})\pbasesat{i}\pbaseBS{i}\pdtwod{i}{h_{10}}(\frac{\RATEDD{h_{10}}}{\RATESAT{h_{10}}+\RATEBS{h_{10}}+\RATEDD{h_{10}}})$\\
	&~&~&
	$1_{(({\RATESAT{h_{10}}+\RATEBS{h_{10}}+\RATEDD{h_{10}}}>0)\land(\stateiHUD{h_0}>0))}$\\[1ex] 
	\hline
	\end{tabular}
	\label{table:HU arrival transitions into h0}
\end{table*}

\begin{table*}[t]
\centering
    \renewcommand{\arraystretch}{1.3}
	\caption{Transitions destined to a generic state $\state{0}$($h_0$) due to HU departures.}
	\begin{tabular}{ | l | l | l | p{11 cm} |}
	\hline
	\textbf{Link} &
	\textbf{From State} &
	\textbf{ Transition rate} \\ \hline \hline
	Sat &
	$h_{1}$ := $\state{(\iHUSatUniv+1)}$ &
	$\stateiHUSatU{h_1}\muHUSatU1_{(\idleSat{h_0}>0)}$\\[1ex] \hline
	Ter &
	$h_{2}$ := $\state{(\iHUBSUniv+1)}$ &
	$\stateiHUBSU{h_2}\muHUBSU1_{(\idleTerExceptfOne{h_0}>0)}$\\[1ex] \hline
	Sat &
	$h_{3}$ := $\state{(\iHUSat+1)}$ &
	$\stateiHUSat{h_3}\muHUSat1_{(\idleSat{h_0}>0)}$\\[1ex] \hline
	Ter &
	$h_{4}$ := $\state{(\iHUBS+1)}$ &
	$\stateiHUBS{h_4}\muHUBS1_{(\idleTerExceptfOne{h_0}>0)}$\\[1ex] \hline
	Ter &
	$h_{5}$ := $\state{(\iHUDev+1)}$ &
	$\stateiHUD{h_5}\muHUDev1_{((0<=\stateiHUD{h_5}<=\DMAX)\land(0<=\stateiHUD{h_0}<=\DMAX))}$\\[1ex]
	\hline
	\end{tabular}
	\label{table:HU departure transitions into h0}
\end{table*}

\section{Performance Metrics}\label{sec:performanceMetrics}
 
Our analytical model provides an apparatus to investigate our heterogeneous network for its performance characteristics. We elaborate on some system performance metrics in this Section. 

\subsection{Goodput}

We need to grasp an insight into the network system characteristics. For this purpose, we investigate the overall system goodput. First, we calculate the throughput (the rate HU content requests are served) of different system units as listed below: 
\begin{align}
\scriptstyle{\goodputHUSat} &\scriptstyle{:= \effarrHUsat \times \MeanBaseContentSize} \label{eq:goodputHUSat}\\
\scriptstyle{\goodputHUSatUniv} &\scriptstyle{:= \effarrHUsatuniv \times \MeanBaseContentSize} \label{eq:goodputHUSatUniv}\\
\scriptstyle{\goodputHUBS} &\scriptstyle{:= \effarrHUBS \times (1-\pdropBS) \times \MeanBaseContentSize} \label{eq:goodputHUBS} \\
\scriptstyle{\goodputHUBSUniv} &\scriptstyle{:=\effarrHUBSuniv \times (1-\pdropBS) \times \MeanBaseContentSize} \label{eq:goodputHUBSUniv} \\
\scriptstyle{\goodputHUDTwoD} &\scriptstyle{:= \effarrHUdev \times (1-\pdropDTWOD) \times \MeanBaseContentSize} \label{eq:goodputHUD2D}  
\end{align}

The throughput of HUs retrieving contents directly from the satellite cache is given in~(\ref{eq:goodputHUSat}). As HUs in the satellite link are in PU mode, the effective arrival rate $\effarrHUsat$ given in (\ref{eq:effarrHUsat}) is equal to the effective service rate of HUs that are served over the satellite link directly from the satellite cache. Multiplying effective service rate with average content size $\MeanBaseContentSize$, the overall bits served over the satellite link by transmitting requested contents directly from the satellite cache per unit time is calculated.
The throughput of HUs retrieving contents first from the universal source to the satellite cache and then from there to requester HU devices over the satellite link is calculated similarly as given in~(\ref{eq:goodputHUSatUniv}).

The throughput of HUs retrieving contents directly from the BS cache is given in~(\ref{eq:goodputHUBS}). The effective service rate of HUs that are served directly from the BS cache is equal to the arrival rate of HUs to the terrestrial link for retrieving content directly from the BS cache $\effarrHUBS$ times  $1-\pdropBS$ where $\effarrHUBS$ given in (\ref{eq:effarrHUBS}) and $\pdropBS$ in (\ref{eq:pdropBS}). We exclude dropped contents that use directly the BS for service as they do not contribute to successful transmissions. 
The throughput of HUs retrieving contents first from the universal source to the BS cache and then from there to requester HU devices (r-HUs) is calculated similarly as given in~(\ref{eq:goodputHUBSUniv}).

Now, let us look at the throughput for HUs that are in D2D operation mode. 
The throughput of HUs that operate in D2D mode is given in~(\ref{eq:goodputHUD2D}). 
It is equal to the effective service rate of HUs that retrieve requested contents from some HU device within a predefined range successfully times the average content size. 
The effective service rate of HUs that retrieve requested contents in D2D operation mode successfully (without being dropped) is equal to the effective arrival rate of HUs to the frequency $f_1$ that is used for D2D operation mode $\effarrHUdev$ times $1-\pdropDTWOD$ where $\effarrHUdev$ given in (\ref{eq:effarrHUdev}) and $\pdropDTWOD$ in (\ref{eq:pdropDTWOD}).

Until now, we defined the throughput of different system units. For the services that are done by the local cache, we look at the $\goodputHULocal$ value. The effective request rate of HUs over the local cache is equal to the arrival rate of HU requests $\lambdaHU$ times the probability of an HU getting service from the local cache  $\plocal$ given in (\ref{eq:plocal}). For the calculation of service rate in bps, the effective request rate of HUs over the local cache is multiplied by the average content size $\MeanBaseContentSize$.
\begin{align}
\scriptstyle{\goodputHULocal} &\scriptstyle{:= \lambda_{HU} \times \plocal \times \MeanBaseContentSize}
\end{align}

The overall system goodput of HUs is equal to the summation of services taken without using system sources (requested content found in the local cache) ($\goodputHULocal$) and the summation of services given over the network system in bits per second. 
\begin{align}
\scriptstyle{\goodputHU} &\scriptstyle{:= \goodputHULocal + \goodputHUSat + \goodputHUSatUniv + \goodputHUBS + \goodputHUBSUniv + \goodputHUDTwoD} \label{eq:overallGoodputHU}
\end{align}

In performance metrics, we utilize effective arrival rate of HUs to different system units over different link types ($\effarrHUsat$ , $\effarrHUsatuniv$ , $\effarrHUBS$ , etc. ), drop rate of HUs over the terrestrial link due to PU activity ($\pdropBS$, $\pdropDTWOD$) and the probability of an HU getting service from the local cache ($\plocal$). Then, the  HUs effective arrival rate definitions are: 
\begin{align}
\label{eq:effarrHUsat} \scriptstyle{\effarrHUsat} &\scriptstyle{:= \sum_{x \in S} \gammaHUSatAll(x) \pi_{x}}  \\ 
\label{eq:effarrHUsatuniv} \scriptstyle{\effarrHUsatuniv} &\scriptstyle{:= \sum_{x \in S} (\sum_{i = 1}^{N} \gammaHUSatU{i}{x}) \pi_{x}}\\
 \label{eq:effarrHUBS} \scriptstyle{\effarrHUBS} &\scriptstyle{:= \sum_{x \in S}  \gammaHUBSAll(x) \pi_{x}}\\
 \label{eq:effarrHUBSuniv} \scriptstyle{\effarrHUBSuniv} &\scriptstyle{:= \sum_{x \in S} (\sum_{i=1}^{N} \gammaHUBSU{i}{x} ) \pi_{x}}\\
 \label{eq:effarrHUdev}\scriptstyle{\effarrHUdev} &\scriptstyle{:= \sum_{x \in S} \gammaHUDAll(x) \pi_{x}}
\end{align}

The effective arrival rate of HUs that fetch contents directly from the satellite cache over the satellite link is given in~(\ref{eq:effarrHUsat}).
Note that we have defined $\gammaHUSatAll$ above in~(\ref{eq:gammaHUSatAll}) originated at state $h_0$. 
This function evaluates to different results for different channel states, therefore for mathematical correctness, we continue to use $\gammaHUSatAll(x)$ where x denotes the current channel state. $\gammaHUSatAll(x)$ gives us the overall rate of HUs that get service from the cache of the satellite over the satellite link for a given channel state x. Summing $\gammaHUSatAll(x)\pi_x$ over all states, we get the expected effective arrival rate of HUs that fetch content from the satellite cache over the satellite link. This is an effective arrival rate since the blocking of HUs is taken into account during the calculation of $\gammaHUSatAll(x)$ functions.

The effective arrival rate of HUs that retrieve contents first from the universal source to the satellite cache and then from there to the r-HUs over the satellite link is given in~(\ref{eq:effarrHUsatuniv}). 
$\gammaHUSatU{i}{x}$ is the rate of HUs that request $c_i$ and it is served from the universal source over the satellite link (in Table~\ref{table:HU arrival transitions at h0}). By summing $\gammaHUSatU{i}{x}$ over all contents ($\sum_{i = 1}^{N} \gammaHUSatU{i}{x}$) and getting the weighted sum over all channel states ($\sum_{x\in S} (\sum_{i = 1}^{N} \gammaHUSatU{i}{x}) \pi_{x}$), 
we get the expected effective arrival rate of HUs that fetch contents first from the universal source to the satellite cache and then from there to r-HUs over the satellite link. This is again an effective arrival rate since the blocking of HUs is taken into account during the calculation of $\sum_{i = 1}^{N} \gammaHUSatU{i}{x}$ functions. 
The definitions of other effective arrival rates of HUs $\effarrHUBS$,  $\effarrHUBSuniv$, $\effarrHUdev$ are given in~(\ref{eq:effarrHUBS}),~(\ref{eq:effarrHUBSuniv}),~(\ref{eq:effarrHUdev}).

Now, let us look at the dropping probabilities of HUs over the terrestrial link due to PU activity. 
The HUs operate in SU mode in the terrestrial link, so dropping of HUs occurs there. Remember that for HUs the terrestrial frequency $f_1$ is used in D2D mode only and the other terrestrial frequencies are used in BS mode either \textbf{directly} from the BS cache or \textbf{indirectly} first from the universal source to the BS cache and then from there to the requester HU device. First let us look at the dropping probability of HUs in BS mode given in~(\ref{eq:pdropBS}). 
$\pdropBS$ is equal to the division of the forcibly terminated rate of HUs due to PU arrivals at the terrestrial frequency set \{$f_2$, $f_3$ , ... $f_{\NTer}$\} ($\frac{\lambdaPUTer(\NTer - 1)}{\NTer} U$) over the aggregate effective arrival rate of HUs in BS mode ($\effarrHUBS+\effarrHUBSuniv$).
If at the current channel state x, there exists at least one non-PU operation at the terrestrial frequency set $\{f_2$, $f_3$,... $f_{\NTer}\}$ (one HU getting service in BS mode 
$\xrightarrow{}1_{(\stateiPUTer{x} \neq (\NTer-1))}$) and there is no idle frequency in this set ($1_{(\idleTerExceptfOne{x} == 0)}$), due to PU appearance HU cannot find an idle frequency to continue its operation in BS mode and that HU is forcibly dropped. The probabilities of being at channel states that cause such HU drops are summed to U in~(\ref{eq:pdropBSutil}). 
Multiplying U with the arrival rate of PUs to the terrestrial frequency set $\{f_2$, $f_3$,... $f_{\NTer}\}$ ($\frac{\lambdaPUTer(\NTer - 1)}{\NTer}$), we get the forcibly terminated rate of HUs in BS mode due to PU arrivals.  
\begin{align}
    \label{eq:pdropBSutil} U &\scriptstyle{:= \sum_{x \in S}} (\pi_x \cdot 1_{((\idleTerExceptfOne{x} == 0) \land (\stateiPUTer{x} \neq (\NTer-1)))})\\
    \label{eq:pdropBS} \scriptstyle{\pdropBS} &\scriptstyle{:= \frac{\frac{\lambdaPUTer(\NTer - 1)}{\NTer} U }{\effarrHUBS+\effarrHUBSuniv}}
 \end{align}

The dropping probability of HUs that operate in D2D mode is defined in~(\ref{eq:pdropDTWOD}). $\pdropDTWOD$ is equal to the division of the forcibly terminated rate of HUs in D2D mode due to PU arrival at the terrestrial frequency $f_1$  $\scriptstyle{\frac{\lambdaPUTer}{\NTer} \cdot
 (\sum_{x \in S}
 (\stateiHUD{x} \cdot \pi_x \cdot 1_{(\stateiPUTerfOne{x} == 0)}))}$ over the effective arrival rate of HUs in D2D mode to the terrestrial frequency $f_1$ ($\effarrHUdev$). 
Due to overlaying in D2D mode, more than one D2D operation can occur at the terrestrial frequency $f_1$ and upon a PU arrival on this frequency all active D2D operations will be dropped. The probability of being at channel states that cause such drops are summed as $\scriptstyle{(\sum_{x \in S} (\stateiHUD{x} \cdot \pi_x \cdot 1_{(\stateiPUTerfOne{x} == 0)}))}$. Note that PU can arrive at terrestrial frequency $f_1$ if no PU exists there at current channel state x ($\stateiPUTerfOne{x} == 0$). Multiplying $\scriptstyle{\sum_{x \in S} (\stateiHUD{x} \cdot \pi_x \cdot 1_{(\stateiPUTerfOne{x} == 0)})}$ with the arrival rate of PUs to the terrestrial frequency $f_1$ ($\frac{\lambdaPUTer}{\NTer}$), we get the forcibly terminated rate of HUs in D2D mode due to PU arrival.
\begin{align}
    \label{eq:pdropDTWOD} \scriptstyle{\pdropDTWOD} & \scriptstyle{:= \frac{\frac{\lambdaPUTer}{\NTer} \cdot (\sum_{x \in S} (\stateiHUD{x} \cdot \pi_x \cdot 1_{(\stateiPUTerfOne{x} == 0)}))}{\effarrHUdev}}
\end{align}

 Let us continue with the definition of the content availability in the local cache $\plocal$. 
 For any content $c_i$, $\pbaselocal{i}$ is the content probability being in the local cache as given in Subsection~\ref{sec:HUDeviceCache}. The request rate for some content $c_i$ is $\lambdacioverBase{i}$. By multiplying these two and summing over all contents ($\sum_{i = 1}^{N}(\lambdacioverBase{i}\pbaselocal{i})$), we get the rate of HUs that are served by local caches. $\sum_{i = 1}^{N} \lambdacioverBase{i}$ is the request rate of HUs. Dividing $\sum_{i = 1}^{N}(\lambdacioverBase{i}\pbaselocal{i})$ over $\sum_{i = 1}^{N} \lambdacioverBase{i}$, we get the probability of an HU getting service from its local cache. 
 \begin{align}
 \label{eq:plocal} \plocal &:= \frac{\sum_{i = 1}^{N}(\lambdacioverBase{i}\pbaselocal{i})}{\sum_{i = 1}^{N} \lambdacioverBase{i}}
\end{align}

\subsection{Energy Efficiency}

For evaluating popularity-aware caching and universal source integrated overlaying in D2D enabled RA strategy, energy consumption of our system is an important criterion. For this purpose, we define energy efficiency as the consumed energy in Joule per successfully transmitted number of bits in~(\ref{eq:EE}) simply. The division of the consumed overall power $\overallPower$ in~(\ref{eq:overallPower}) over the overall system goodput of HUs $\goodputHU$ in~(\ref{eq:overallGoodputHU}) gives the consumed energy in Joule per successfully transmitted number of bits to HUs. 
The decrease in the consumed energy per successfully transmitted bits leads to improvement in energy efficiency.
\begin{align}
    \scriptstyle{\overallPower} & \scriptstyle{:= \BSPower + \BSUnivPower + \DTwoDPower + \localPower} \label{eq:overallPower}\\
    \scriptstyle{\epbHU} & \scriptstyle{:= \frac{\overallPower}{\goodputHU}}\label{eq:EE}
\end{align}

Since the satellite is solar powered, the overall power consumption $\overallPower$ does not include that figure. $\overallPower$ consists of four components: $\BSPower$, $\BSUnivPower$, $\DTwoDPower$ and $\localPower$. Now, we define them respectively.

$\BSPower$ is the transmission power consumption of the BS where some HU services directly from the BS cache are \textbf{completed} and some other HU services directly from the BS cache are 
\textbf{dropped} given in (\ref{eq:BSPower}). 
$\scriptstyle{\effarrHUBS \cdot (1-\pdropBS)}$ is the effective service rate of HUs directly from the BS cache while $\scriptstyle{\frac{\PerChannelBSTxPower}{\muHUBS}}$ is the transmission energy consumed by the BS per each such service in Joule. Multiplying these two, we get the energy consumed by the BS for completed HU services directly from the BS cache per unit time. The energy consumption per unit time is by definition power, so we get the transmission power consumption of the BS for completed HU services directly from the BS cache as $\scriptstyle{\effarrHUBS \cdot (1-\pdropBS) \cdot \frac{\PerChannelBSTxPower}{\muHUBS}}$.

$\scriptstyle{\effarrHUBS \cdot \pdropBS}$ is the rate of HU services directly from the BS cache that are dropped. Assuming no bias, such occurrences durate in average half of a complete HU service directly from the BS cache ($\frac{1}{2 \cdot \muHUBS}$ sec). So, $\frac{\PerChannelBSTxPower}{2 \cdot \muHUBS}$ is the average transmission energy consumed by the BS per each such incomplete HU service. Multiplying $\scriptstyle{\effarrHUBS \cdot \pdropBS}$ and $\scriptstyle{\frac{\PerChannelBSTxPower}{2 \cdot \muHUBS}}$, we get 
the transmission power of the BS for dropped HU services directly from the BS cache. 
\begin{align}
    \label{eq:BSPower} 
    \scriptstyle{\BSPower} \scriptstyle{:= (\effarrHUBS \cdot (1-\pdropBS) \cdot \frac{\PerChannelBSTxPower}{\muHUBS})} \scriptstyle{+ (\effarrHUBS \cdot \pdropBS \cdot \frac{\PerChannelBSTxPower}{2 \cdot \muHUBS})}
\end{align}

$\BSUnivPower$ is the power consumption of the BS where HU requested contents are first retrieved from the universal source to the BS cache and then served from there to the requester HUs given in (\ref{eq:BSUnivPower}). Like $\BSPower$, this power calculation is comprised of two parts: (I) completed services indirectly from the BS  (II) dropped services indirectly from the BS 

Let us start with type-I services. $\scriptstyle{\effarrHUBSuniv \cdot (1-\pdropBS)}$ is the effective rate of HUs that the requested contents are first fetched from the universal source to the BS cache. $\frac{\PerChannelBSTxPower}{\receiveBSparameter} \cdot \frac{\MeanBaseContentSize}{\capHUBSU}$ is the reception energy consumed by the BS for each HU service where the requested HU content is fetched from the universal source to the BS cache first. Multiplying these two, we get the reception power consumption of the BS where the requested contents from HUs are fetched from the universal source to the BS cache first. $\scriptstyle{\{\effarrHUBSuniv \cdot (1-\pdropBS) \cdot (\frac{\PerChannelBSTxPower}{\muHUBS})\}}$ is the transmission power consumption of the BS where the requested contents are transmitted from the BS cache to the HU requester devices.

For type-II services $\scriptstyle{\effarrHUBSuniv \cdot \pdropBS \cdot (\frac{\PerChannelBSTxPower}{\receiveBSparameter} \cdot \frac{\MeanBaseContentSize}{\capHUBSU})}$ is the reception power consumption of the BS where the requested contents by HUs are fetched from the universal source to the BS cache first. $\meanservicedurationHUBSU$ is the aggregate service duration for an HU request where the requested content is first fetched from the universal source to the BS cache and then from there to the r-HU as defined in~(\ref{eq:meanservictetimeHUUniv}). Assuming no bias, the duration of an incomplete service is in average half of a complete service, so $\scriptstyle{\frac{\meanservicedurationHUBSU}{2}}$ is expected duration for an incomplete HU service where the requested content is fetched from the universal source to the BS cache and that content is started to be transmitted from the BS cache to the r-HU but dropped. Excluding  $\scriptstyle{\frac{\MeanBaseContentSize}{\capHUBSU}}$ duration for the content fetching from the universal source to the BS cache from  $\scriptstyle{\frac{\meanservicedurationHUBSU}{2}}$, we get the duration of an incomplete HU service from the point where the content is started to be transmitted from the BS cache to the r-HU until it is dropped. Multiplying this duration with $\PerChannelBSTxPower$, we get the average transmission energy consumed by the BS starting from the transmission of the requested content to the r-HU until it is dropped per each such incomplete HU service. 
Next, we multiply $\frac{\meanservicedurationHUBSU}{2}-\frac{\MeanBaseContentSize}{\capHUBSU}$ with $\effarrHUBSuniv \cdot \pdropBS$ and 
we get the transmission power consumption of the BS starting from the transmissions of the requested contents to the r-HUs from the BS cache until they are dropped.
\begin{align}
    \label{eq:BSUnivPower} 
    \scriptstyle{\BSUnivPower}  & \scriptstyle{:= \{\effarrHUBSuniv \cdot (1-\pdropBS) \cdot  ([\frac{\PerChannelBSTxPower}{\muHUBS}]+[\frac{\PerChannelBSTxPower/\receiveBSparameter}{\MeanBaseContentSize/\capHUBSU}])\}} \\ \nonumber
    & \scriptstyle{+ \{\effarrHUBSuniv \cdot \pdropBS \cdot ([(\PerChannelBSTxPower \cdot (\frac{\meanservicedurationHUBSU}{2}-\frac{\MeanBaseContentSize}{\capHUBSU})]+[\frac{\PerChannelBSTxPower}{\receiveBSparameter} \cdot \frac{\MeanBaseContentSize}{\capHUBSU}])\}}
\end{align}

$\DTwoDPower$ is the transmission power consumption of the HU devices that transmit contents in D2D mode which has a similar construction logic like $\BSPower$. 
\begin{align}
     \label{eq:DTWODPower}
    \scriptstyle{\DTwoDPower}   & \scriptstyle{:= (\effarrHUdev \cdot (1-\pdropDTWOD) \cdot \frac{\DeviceTxPower}{\muHUDev})} \scriptstyle{+ (\effarrHUdev \cdot \pdropDTWOD \cdot \frac{\DeviceTxPower}{2 \cdot \muHUDev})}
\end{align}

Apart from the services across the network, some HU content requests are satisfied by the local caches with power consumption $\localPower$ given in~(\ref{eq:localPower}) where $\lambdaHU\cdot\plocal$ is the effective request rate of HUs that are served by local caches and $\frac{\DeviceTxPower}{\receiveLocalparameter} \cdot \frac{1}{\muHUDev}$ is the average energy consumed per each requester HU device that is served by the local cache.
\begin{align}
    \scriptstyle{\localPower}   & \scriptstyle{:= (\lambdaHU \cdot \plocal) \cdot \frac{\DeviceTxPower}{\receiveLocalparameter} \cdot \frac{1}{\muHUDev}} \label{eq:localPower}
\end{align}

\section{Performance Evaluation}\label{sec:results}
In our study, we look at $\epbHU$ and $\goodputHU$ performance metrics. In each of our performance evaluation, we perform simulations to compare with our analytical results for verifying our system model. 

\subsection{Caching Dynamics and Popularity-Aware Caching (PAC)}

\begin{figure*}
\centering
\subfloat[EE and goodput results (a: analytical, s: simulation).]{\includegraphics[width=0.49\textwidth]{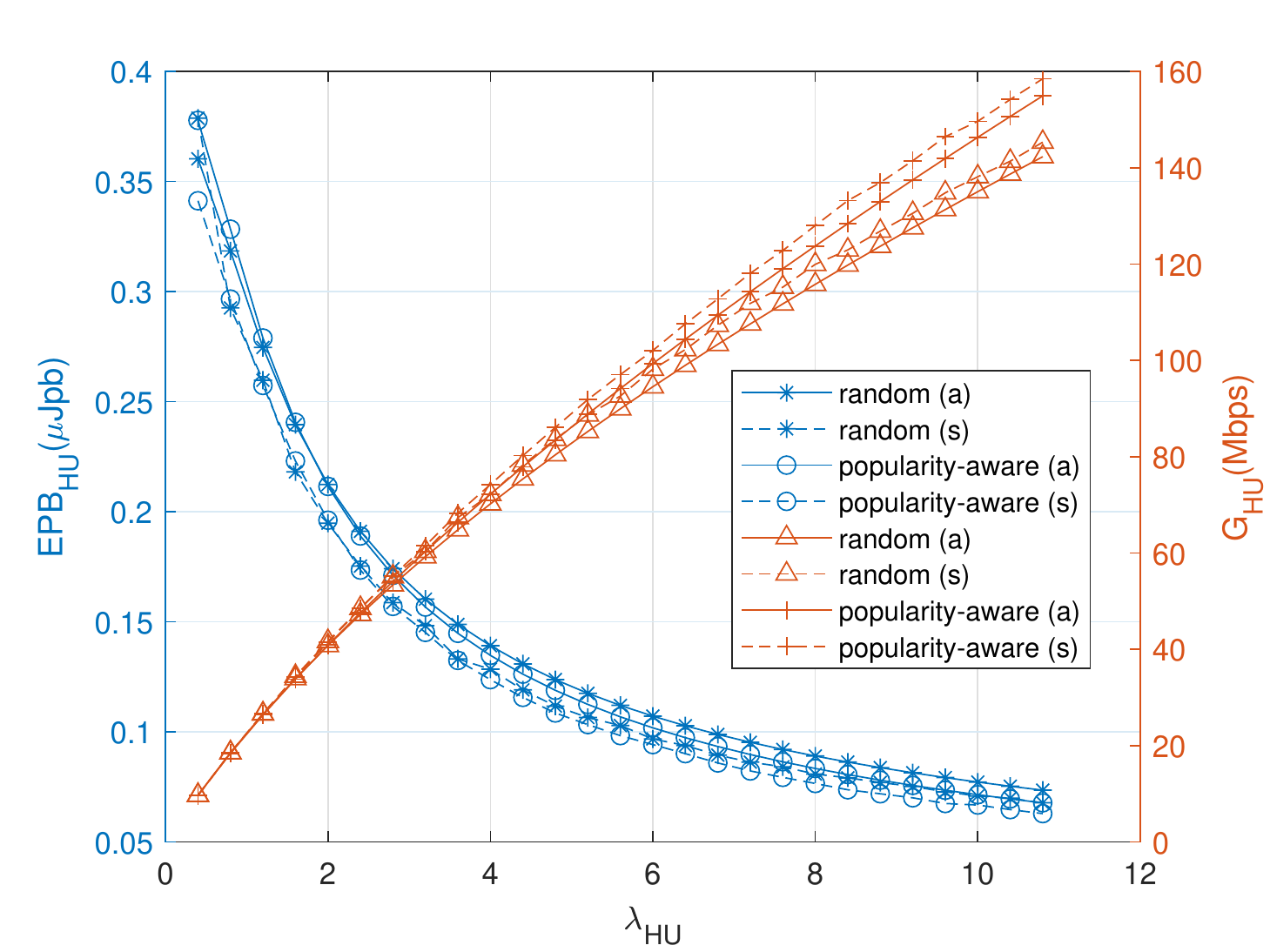}
\label{fig:caching_all_results}} \hfill
\subfloat[Simulation EE and goodput results.]{\includegraphics[width=0.49\textwidth]{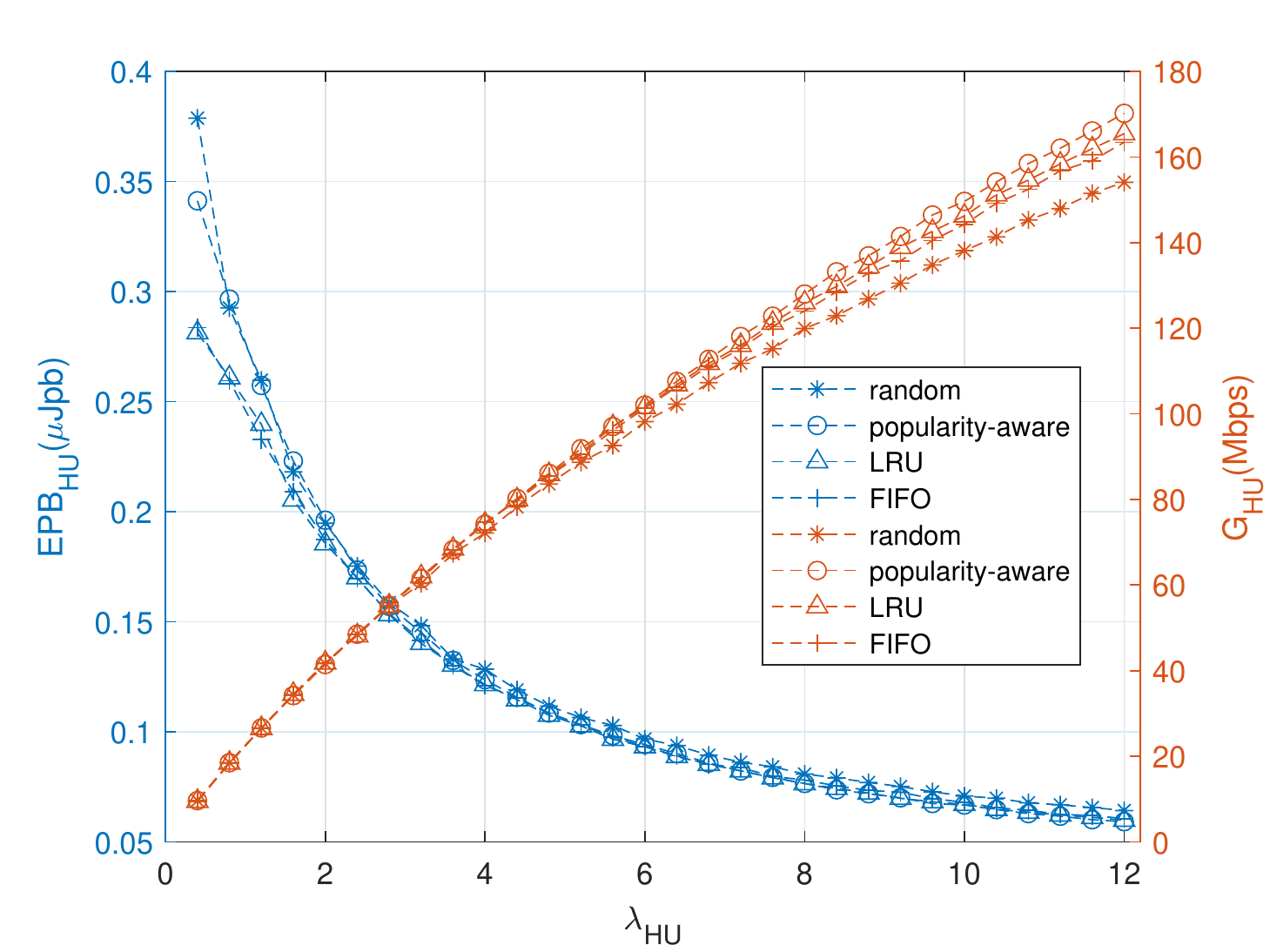}
\label{fig:caching_LRU_FIFO_results}}
\caption{Results for different caching strategies.
\label{fig:cacheAllResults}}
\end{figure*}

Let us look how our proposed popularity-aware caching mechanism PAC affects the performance metrics $\epbHU$ and $\goodputHU$ (Fig.~\ref{fig:caching_all_results}). We compare PAC with random caching as the baseline case. We assume that universal source is on while overlaying mechanism for D2D operation mode is enabled and all mode weights $r_{sat}$, $r_{BS}$, $r_{dev}$ are equal to 1/3. Note that increasing content request rate of HUs $\lambdaHU$, reduces $\epbHU$ and increases $\goodputHU$ for both caching mechanisms. The decrease in the consumed energy per successfully transmitted bits $\epbHU$ leads to improvement in EE. The main reason for the reduction of $\epbHU$ is that as the network gets more crowded with increased rate of HU requests, the probability of satellite channel being idle or the probability of terrestrial frequencies that are usable for operations from the BS being idle decreases (since their service durations are longer compared to D2D mode operations) and thus, D2D mode utilization rate increases. There is not an obvious difference in terms of $\epbHU$ for two different caching methods for any $\lambdaHU$ rate. As $\lambdaHU$ increases, the number of served bits as expected increases for both caching mechanisms, i.e. $\goodputHU$ increases. As shown in Fig.~\ref{fig:caching_all_results}, $\goodputHU$ results for popularity-aware caching is better when compared to random caching especially for larger $\lambdaHU$'s as popularity-aware caching method tries to cache more popular contents with higher probability. The simulation results follow the same trend with the analytical $\epbHU$ and $\goodputHU$ results for both caching mechanisms. Note that for large $\lambdaHU$ values in Fig.~\ref{fig:caching_all_results}, the simulation $\goodputHU$ results are larger than analytical $\goodputHU$ results for both random and PAC mechanisms. 
The reason is that the impact of D2D mode services is greater for large $\lambdaHU$ and $\pdtwod{i}{x}$ is a lower bound for the probability of a content $c_i$ being available in some HU device within the vicinity of requester as explained in Subsection~\ref{subsection:D2D}. In simulations, we are not restricted by lower bounds for mode selection. Thus, we obtain more precise and larger $\goodputHU$ results. Thus, for large $\lambdaHU$ values analytical $\goodputHU$ results are lower than simulation $\goodputHU$ results for both 
caching mechanisms. For simulation results, we run simulations 10 times each for 1200 sec.  

For a more thorough analysis, we also compare simulation results of PAC mechanism with simulation results of caching methodologies \textit{Least Recently Used (LRU)}, \textit{First In First Out (FIFO)} in addition to random caching in Fig.~\ref{fig:caching_LRU_FIFO_results}. Again, we assume that universal source is on, overlaying for D2D mode operation is enabled and all mode weights $r_{sat}$, $r_{BS}$ and $r_{dev}$ are equal to 1/3. Random and popularity-aware caching are time-independent caching strategies. 
They do not depend on the time when contents are cached in system units. 
On the contrary, LRU and FIFO need the cache time of contents in system units for deciding which content to replace for the goal of caching recently requested contents. With increasing content request rate of HUs $\lambdaHU$, $\epbHU$ decreases and $\goodputHU$ increases for all caching strategies. 
For $\lambdaHU$ values less than or equal to 2 $\frac{user}{sec}$, the time-dependent caching strategies (LRU and FIFO) have lower $\epbHU$ results compared to time-independent caching strategies (random and popularity-aware caching). For larger $\lambdaHU$, the gap for $\epbHU$ values of time-dependent and time-independent caching strategies dissolves. 
In Fig.~\ref{fig:caching_LRU_FIFO_results}, the $\goodputHU$ results for PAC, LRU and FIFO caching are better compared to random caching especially for larger $\lambdaHU$'s. The time-dependent caching strategies (LRU and FIFO) try to continue storing recently used or recently cached contents in system unit caches. These caching strategies have a time-dependent insight into user preferences \textbf{locally}. The PAC strategy on the contrary exploits \textbf{global} content popularities globally and tries to keep more popular contents in system unit caches with a greater probability. This approach does not need to process time information of content arrivals or time of content usages, i.e. it is time-independent. All these three caching strategies (LRU, FIFO, PAC) either locally or globally make use of content related information (LRU:time of a request for some content, FIFO:arrival time of content to some system unit cache, PAC:popularity distribution of contents) in order to cache contents efficiently. Thus, they all outperform random caching in terms of $\goodputHU$ for larger $\lambdaHU$. Note that our proposed popularity-aware caching algorithm  attains $\goodputHU$ results slightly better than LRU and FIFO for larger $\lambdaHU$ (For $\lambdaHU=12\frac{user}{sec}$ $\goodputHU$ of FIFO is 163.5 Mbps, $\goodputHU$ of LRU is 165.5 Mbps while $\goodputHU$ of popularity-aware is 170.2 Mbps). The global content awareness outperforms local decisions mechanisms (LRU, FIFO) for caching. Local decisions only consider content request or arrival history (in terms of time). However, our approach fits a content request distribution model and decides according to this popularity model as explained in~Section~\ref{sec:contentModel}. 

\begin{figure*}
\centering
\subfloat[EE results (a: analytical, s: simulation, -: disabled, +: enabled).]{\includegraphics[width=0.49\textwidth]{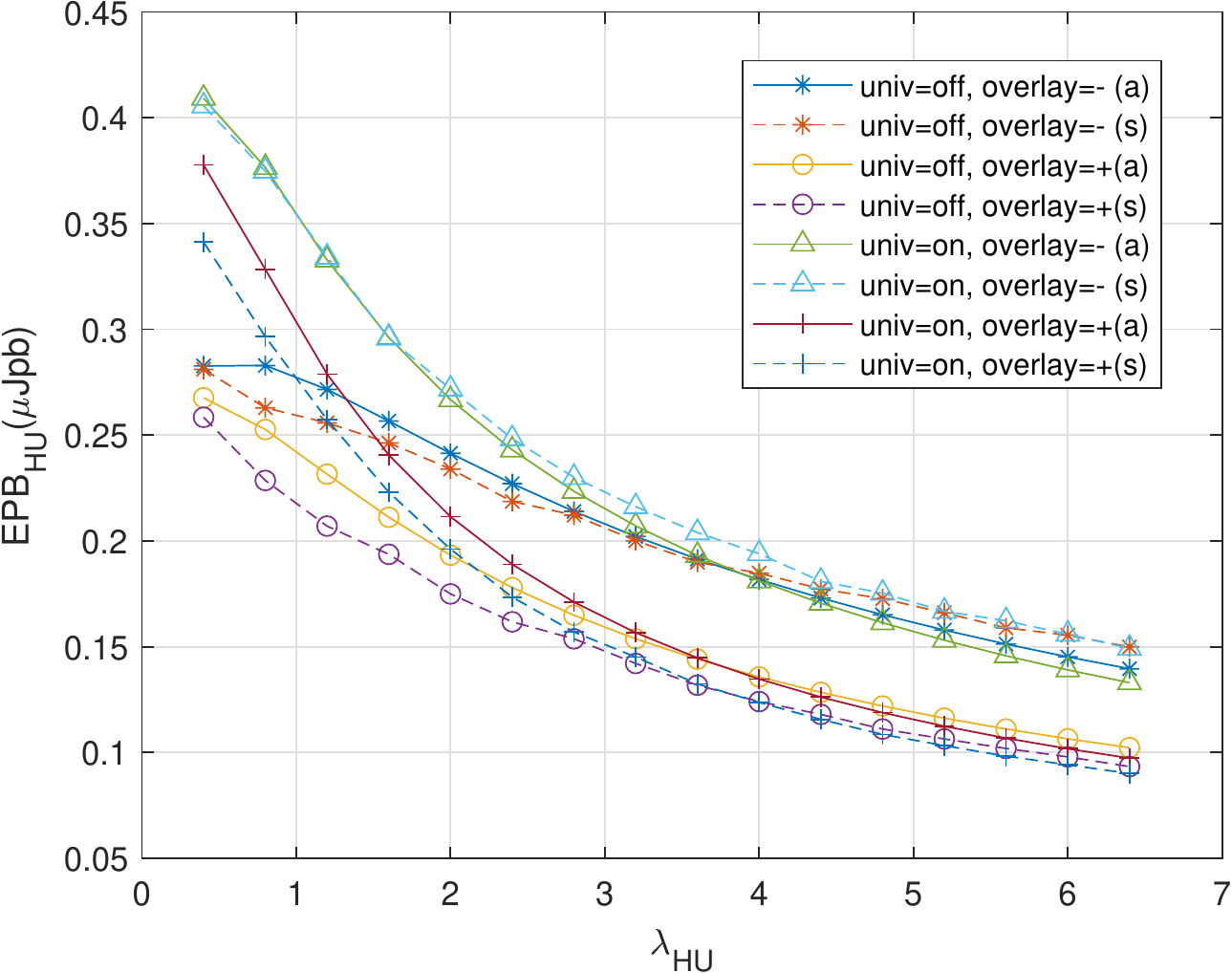}
\label{fig:univ_overlay_EE_results}} 
\hfill
\subfloat[Goodput results (a: analytical, s: simulation, -: disabled, +: enabled).]{\includegraphics[width=0.49\textwidth]{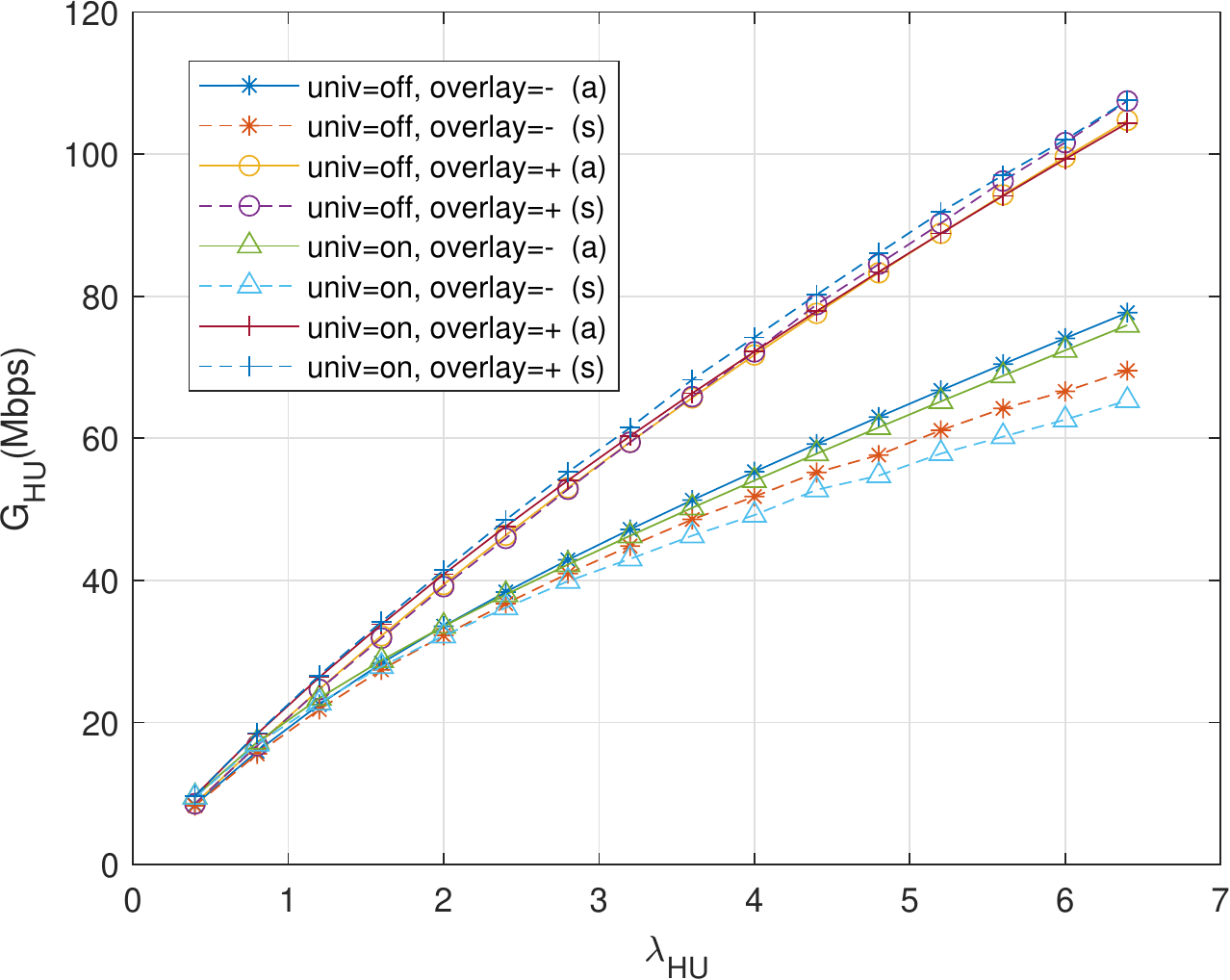}
\label{fig:univ_overlay_thr_results}}
\caption{Universal and overlay scenarios.
\label{fig:univD2DAllResults}}
\end{figure*}

\subsection{Introduction of Universal Source and Overlaying Mechanism for D2D Operation Mode}

We investigate how the introduction of the universal source affects the network performance and how the overlaying mechanism for D2D operation mode affects $\epbHU$ and $\goodputHU$ results. We tune $\DMAX$ for enabling/disabling overlaying mechanism for D2D operation mode. Setting $\DMAX=1$ means only one D2D operation is allowed by the network which corresponds to disabling overlaying. For enabling it, we set $\DMAX>1$, in this section to five. 
Note that popularity-aware caching mechanism is used and all mode weights $r_{sat}$, $r_{BS}$ and $r_{dev}$ are 1/3. 
As you can see, for all settings (universal source on/off, overlaying enabled/disabled) the $\epbHU$ decreases (Fig.~\ref{fig:univ_overlay_EE_results}) and $\goodputHU$ increases (Fig.~\ref{fig:univ_overlay_thr_results}) with increasing $\lambdaHU$ rate. By introducing universal source for both overlaying for D2D mode enabled 
and disabled scenarios, $\epbHU$ value increases for $\lambdaHU$ values lower than 3.2 $\frac{user}{sec}$ as shown in Fig.~\ref{fig:univ_overlay_EE_results}. Unavailable contents are fetched over the universal source with extra reception energy cost at the BS and this leads to reduction in the EE for $\lambdaHU$ values lower than 3.2 $\frac{user}{sec}$.  For larger content request rates, the impact of D2D mode services increases and thereof the energy cost at the BS becomes a less dominant factor in the calculation of $\epbHU$. Accordingly, no more reduction in EE is observed.
By enabling the overlaying for D2D mode in both universal source-on and-off scenarios, the $\epbHU$ value decreases compared to disabled overlaying scenario for any $\lambdaHU$ as shown in Fig.~\ref{fig:univ_overlay_EE_results}. That decrease means for the same amount of successfully served bits less energy is consumed, thus EE is improved. 

With the introduction of universal source for both overlaying for D2D mode enabled and disabled scenarios, $\goodputHU$ results do not change significantly for any $\lambdaHU$ as shown in Fig.~\ref{fig:univ_overlay_thr_results}. The universal source enables unavailable contents to be transmitted so previously unserved contents can then contribute to the goodput and improve goodput. But the services used by universal source are used for larger amount of time, which in turn reduces the probability of these frequencies being idle. So this situation reduces network capacity for the corresponding frequencies and the capacity reduction affects the overall network goodput negatively. Overall, these effects roughly cancel each other and hence the introduction of universal source does not affect the $\goodputHU$ results. However, with the introduction of overlaying for D2D mode operations, the goodput of HUs improves for both universal source on and off scenarios for any $\lambdaHU$ in Fig.~\ref{fig:univ_overlay_thr_results}.

Let us define two settings: (A) universal source on and overlaying for D2D enabled (B) universal source off and overlaying for D2D disabled. For $\lambdaHU\in(0.4,1.2]$ the $\epbHU$ of setting (A) attains larger $\epbHU$ compared to setting (B). With the universal source integration, unavailable contents are retrieved with extra reception energy cost at the BS leading to larger $\epbHU$. Enabling overlaying for D2D is useful for EE and anticipated to reduce $\epbHU$ to compansate for universal source integration. However the network is not in need of concurrent D2D transmissions since the low $\lambdaHU$ value means less content request and the request traffic is not dense enough to use overlaying in D2D. Thereof, for $\lambdaHU \in (0.4,1.2)$, universal source impact is dominant and setting (A) attains larger $\epbHU$ than setting (B). In $\lambdaHU\in(1.2,1.6)$ regime, the $\epbHU$ values of both settings intersect and for $\lambdaHU\in[1.6,6.4)$ regime, setting (A) attains lower $\epbHU$ value than setting (B). With larger $\lambdaHU$ the network becomes needy for concurrent D2D transmissions and hence in setting (A) with overlaying in D2D mode enabled, the D2D services start to dominate the negative EE impact of universal source integration leading to lower $\epbHU$ (improved EE) compared to setting (B). Note that the simulation results follow the same trend with the analytical $\epbHU$ and $\goodputHU$ results for all scenarios. We run simulations 10 times each for 1200 sec.

\begin{figure}
\centering
\includegraphics[width=\columnwidth]{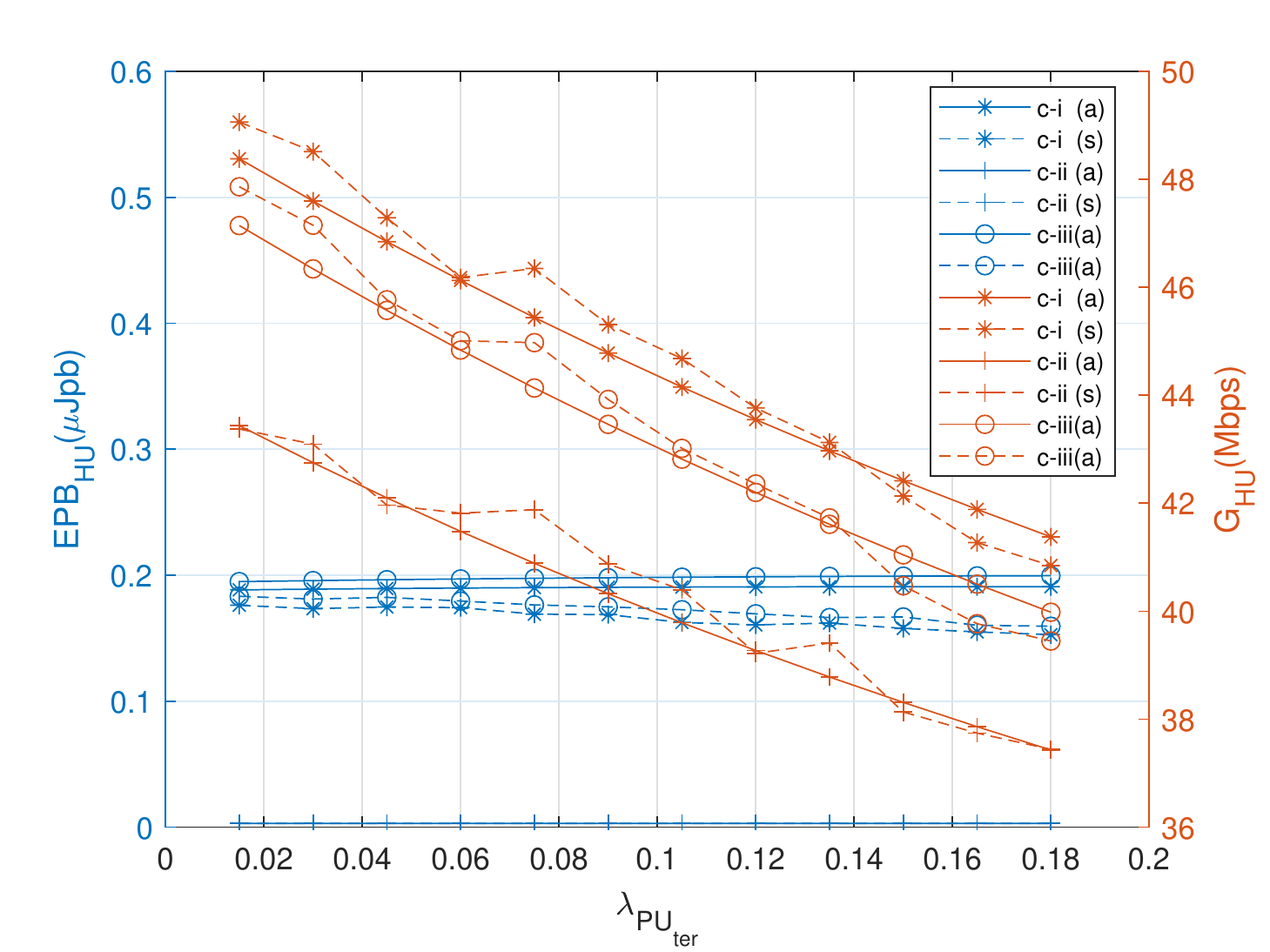}
\caption{EE and goodput results for varying PU arrivals in terrestrial link (a: analytical, s: simulation, c: constellation).}
\label{fig:lambdaPU_ter_EE__thr_results}
\end{figure}

\subsection{Impact of Primary User Activity in Terrestrial Frequencies}

Another important question is how our model behaves for different PU activity. $\epbHU$ and $\goodputHU$ results with respect to increasing $\lambdaPUTer$ are shown in Fig.~\ref{fig:lambdaPU_ter_EE__thr_results}. We look at varying $\lambdaPUTer$ as our HUs are in cognitive mode in the terrestrial link. We assume the universal source is on and overlaying for D2D mode operation is enabled. The arrival rate of HU requests is $\lambdaHU$ =2.4 $\frac{user}{sec}$. The $\lambdaPUTer$ range we investigate is [0.015, 0.18] $\frac{user}{sec}$ as HUs are the driving source of the traffic and thus we assume light PU traffic at the terrestrial link.
We investigate three different mode weight constellations: 
(\textit{i}) all mode weights are equal ($r_{x}$=1/3, $x$ $\in$ $\{sat,BS,dev\}$) 
(\textit{ii}) only D2D mode is on ($r_{dev}$=1) 
(\textit{iii}) the satellite is off while BS and D2D are on with equal weights ($r_{sat}$=0, $r_{BS}$=1/2, $r_{dev}$=1/2).
As shown in Fig.~\ref{fig:lambdaPU_ter_EE__thr_results}, in all constellations we do not observe a significant change in $\epbHU$ with increasing $\lambdaPUTer$. Compared to other two constellations (\textit{constellation-i (c-i)} and \textit{c-iii}), for any $\lambdaPUTer$ value $\epbHU$ is lower in the \textit{c-ii} where only D2D mode is on as shown in Fig.~\ref{fig:lambdaPU_ter_EE__thr_results}. This means EE is better for \lq \lq only D2D mode on" scenario.
However, as depicted in Fig.~\ref{fig:lambdaPU_ter_EE__thr_results}, 
\textit{c-ii} has the lowest $\goodputHU$ among three constellations for any $\lambdaPUTer$.
In all constellations, $\goodputHU$ value decreases with increased $\lambdaPUTer$. 
In the \textit{c-ii} and \textit{c-iii}, with increased $\lambdaPUTer$ the probability of HU requests that are interrupted by PUs and that cannot continue their content retrieval from another idle terrestrial frequency increases. Moreover, the probability of HU requests that cannot be served upon their arrival due to the terrestrial channel being occupied by PUs and/or HUs increases.
Thus, the overall network goodput decreases.

\begin{figure*}
\centering
\subfloat[EE results 
]{\includegraphics[width=0.49\textwidth]{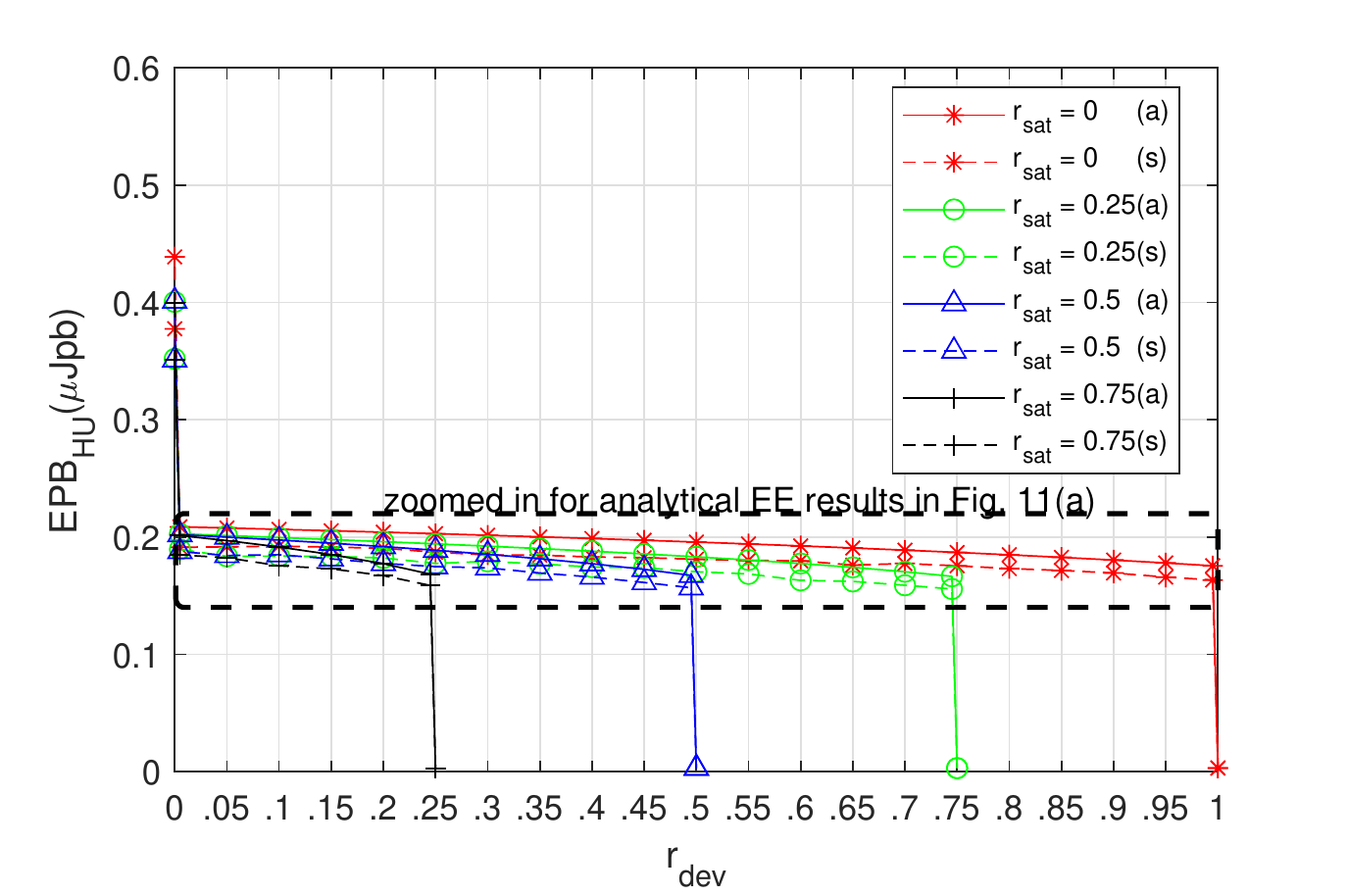}
\label{fig:EE_RSATRBSRDEV_r_sat_fixed_r_dev_x_axis}} 
\hfill
\subfloat[Goodput results
]{\includegraphics[width=0.49\textwidth]{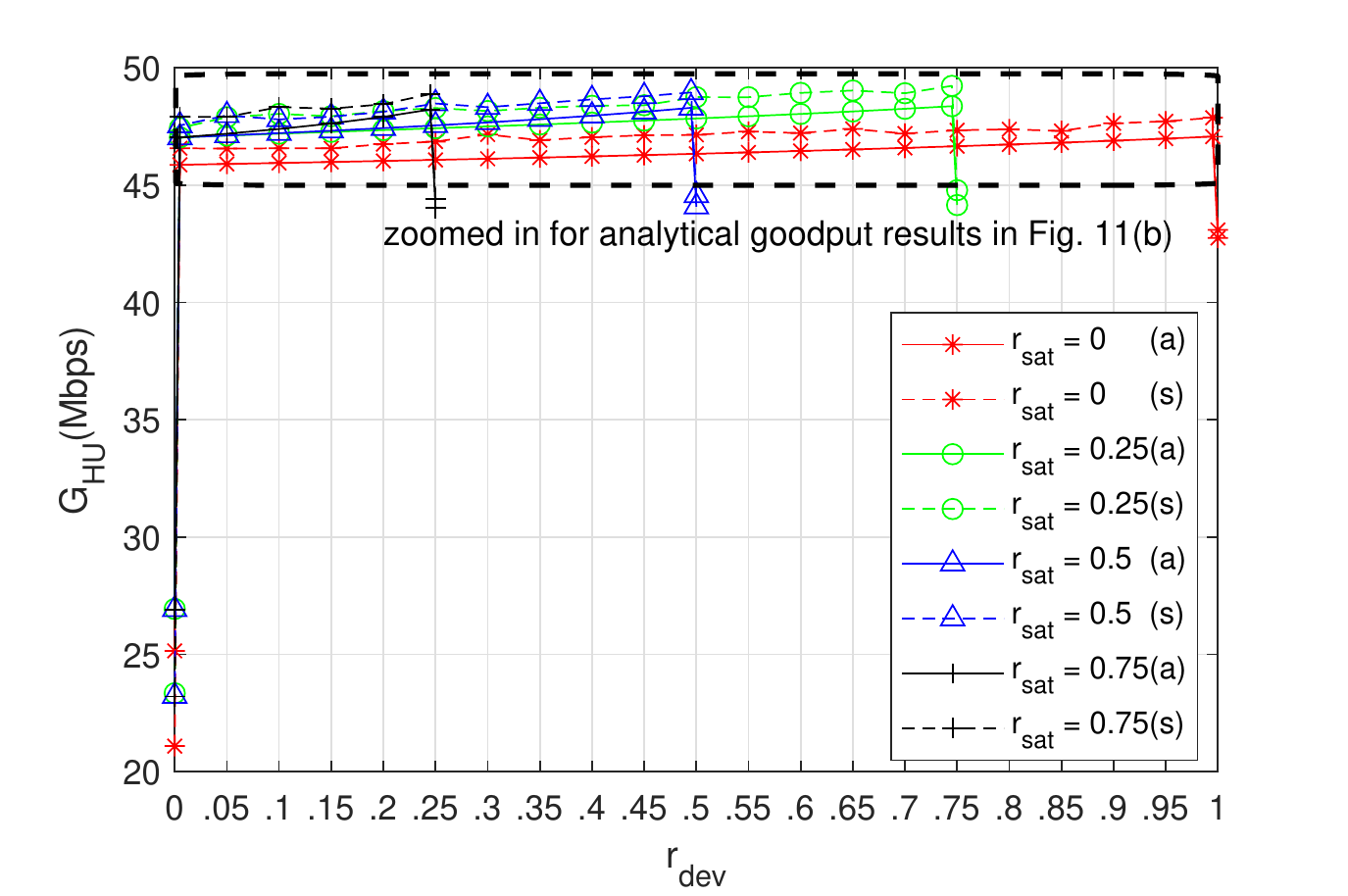}
\label{fig:thr_RSATRBSRDEV_r_sat_fixed_r_dev_x_axis}}
\caption{Results for varying $r_{dev}$ values where $r_{sat}$ is fixed ($r_{dev} = 1 - r_{sat} - r_{BS}$, a: analytical, s: simulation).
}
\end{figure*}

In the \textit{c-i}, the service durations in the satellite link are longer and the satellite link gets saturated rapidly~\cite{7904715}. Therefore, the probability of finding the satellite link idle is low and the increase in the arrival rate of PUs to the terrestrial link $\lambdaPUTer$ decreases the overall network goodput $\goodputHU$. 
After inspecting $\goodputHU$ with increasing $\lambdaPUTer$ for all three constellations, we examine for any fixed $\lambdaPUTer$ how these constellations differ. 
In that case, \textit{c-ii} has the lowest $\goodputHU$ value while \textit{c-i} has the highest $\goodputHU$ value.
The \textit{c-ii} cannot take advantage of relatively large satellite and BS caches compared to HU device caches 
and this reduces the overall network goodput $\goodputHU$.  On the contrary, the \textit{c-i} allows all system units to be used and the system can take advantage of caches of the satellite, BS and HU devices within some proximity of r-HUs. Besides, compared to \textit{c-ii} and \textit{c-iii} both the satellite and terrestrial link can be utilized for HU services in \textit{c-i}. Thus, it attains highest $\goodputHU$ value among all three constellations for any fixed $\lambdaPUTer$. From \textit{c-ii} to \textit{c-iii} BS mode is activated, while from \textit{c-iii} to \textit{c-i} satellite mode is activated. Note that for any fixed $\lambdaPUTer$, as satellite link saturates rapidly~\cite{7904715}, with the activation of satellite mode from \textit{c-iii} to \textit{c-i} less improvement in $\goodputHU$ is observed compared to the activation of BS mode from \textit{c-ii} to \textit{c-iii}. 
The simulation results follow the same trend with the analytical $\epbHU$, $\goodputHU$ results for all scenarios. We run simulations 10 times, each for 1200 sec.

\subsection{Impact of Operation Mode Selection}

For investigating the benefit of heterogeneous architecture, it is crucial to investigate how different operation modes manifest themselves. The model is expected to exhibit beneficial behaviour already presented in various works. We consider a setup where $\NSat$=2 and $\NTer$=3. The first terrestrial frequency $f_1$ is used in D2D operation mode and the other terrestrial frequencies in BS mode as explained in Section~\ref{sec:state_transitions}. Our RA strategy is explained in Subsection~\ref{subsec:HUTransitions}. 
Now, we assume the universal source is on 
and overlaying in D2D is enabled. We examine several mode weight configurations and discuss how they affect $\epbHU$ and $\goodputHU$ performance. The simulation results are consistent with the analytical $\epbHU$ and $\goodputHU$ results for all scenarios. We run simulations 10 times each for 1200 sec. 
For each fixed $r_{sat}$$\in$$\{0, 0.25, 0.5, 0.75\}$,
we inspect the change in $\epbHU$ (Fig.~\ref{fig:EE_RSATRBSRDEV_r_sat_fixed_r_dev_x_axis}) and $\goodputHU$ (Fig.~\ref{fig:thr_RSATRBSRDEV_r_sat_fixed_r_dev_x_axis}) with respect to D2D mode weight $r_{dev}$ where $r_{dev} = 1 - r_{sat} - r_{BS}$.

In $r_{sat}$$\in$$\{0, 0.25, 0.5, 0.75\}$ configurations, when D2D mode is off ($r_{dev}$=0) and the BS mode is on, $\epbHU$ is high (e.g. for $r_{sat}$=0.25, $r_{BS}$=0.75, $r_{dev}$=0, $\epbHU$ attains 0.35 $\mu$Jpb analytically.) as in Fig.~\ref{fig:EE_RSATRBSRDEV_r_sat_fixed_r_dev_x_axis} while $\goodputHU$ is low (e.g. for $r_{sat}$=0.25, $r_{BS}$=0.75, $r_{dev}$=0  $\goodputHU$ attains 26.7 Mbps analytically.) in Fig.~\ref{fig:thr_RSATRBSRDEV_r_sat_fixed_r_dev_x_axis}. In $r_{sat}$$\in$$\{0, 0.25, 0.5, 0.75\}$ configurations, when the BS mode is off ($r_{BS}$=0) and the D2D mode is on in this case, $\epbHU$ attains low values (e.g. for $r_{sat}$=0.25, $r_{BS}$=0, $r_{dev}$=0.75 $\epbHU$ attains 0.003 $\mu$Jpb analytically) which is pleasing in terms of EE. Compared to the previous cases where D2D mode is off \lq \lq the BS mode is off" and \lq \lq D2D mode is on" scenarios are better in terms of $\goodputHU$ values (e.g. for $r_{sat}$=0.25, $r_{BS}$=0, $r_{dev}$=0.75 $\goodputHU$ attains 44.2 Mbps analytically). However, the overall system goodput attains larger values for the scenarios where both BS and D2D modes are on as shown in Fig.~\ref{fig:thr_RSATRBSRDEV_r_sat_fixed_r_dev_x_axis}. Note that the simulation results follow the same trend with the analytical $\epbHU$ and $\goodputHU$ results for all scenarios. For the simulations, we run 10 times each for 1200 sec.

\begin{figure*}
\centering
\subfloat[Analytical EE results 
]{\includegraphics[width=0.49\textwidth]{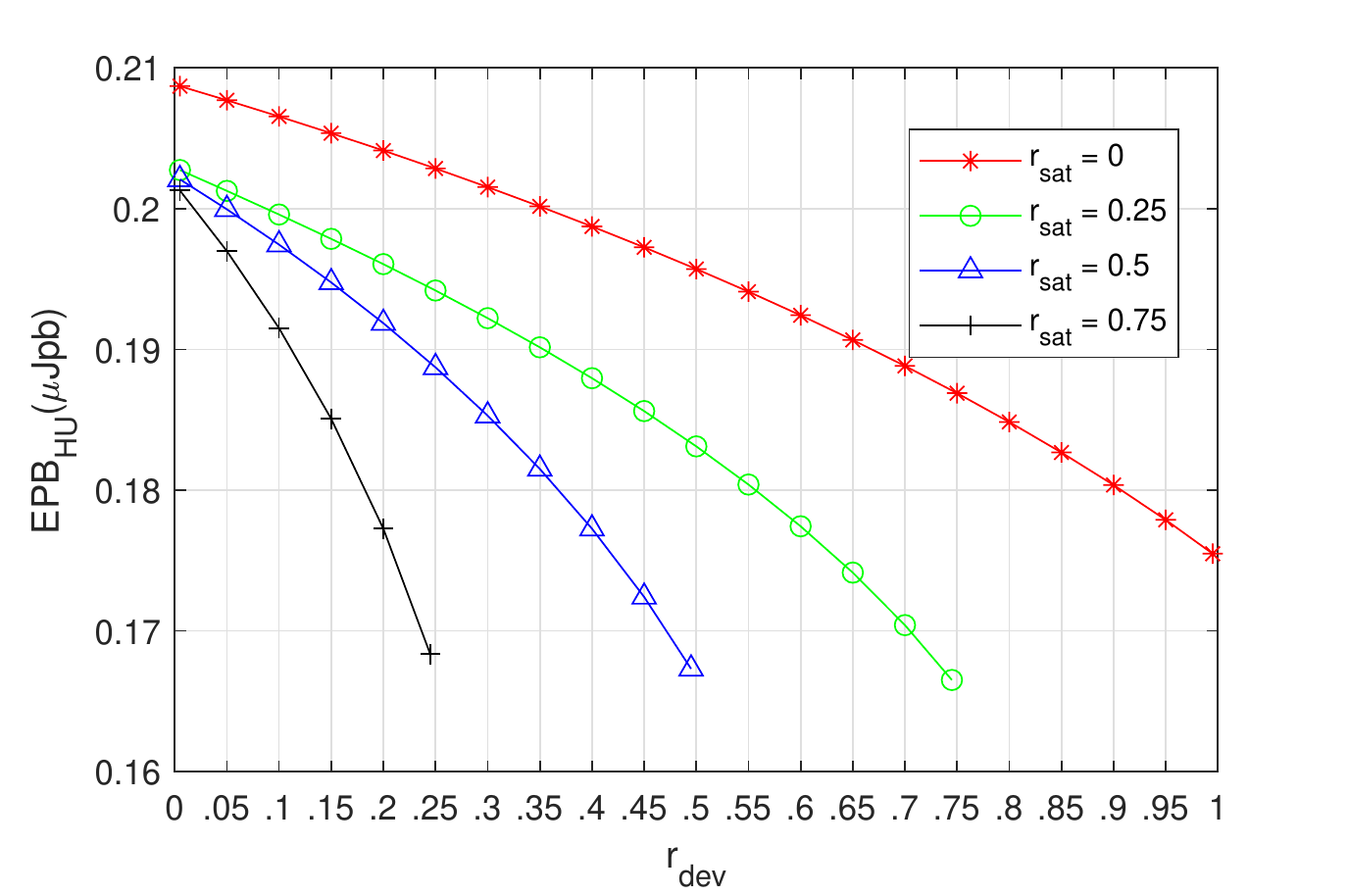}
\label{fig:EE_RSATRBSRDEV_r_sat_fixed_r_dev_x_axis_close}} 
\hfill
\subfloat[Analytical goodput results 
]{\includegraphics[width=0.49\textwidth]{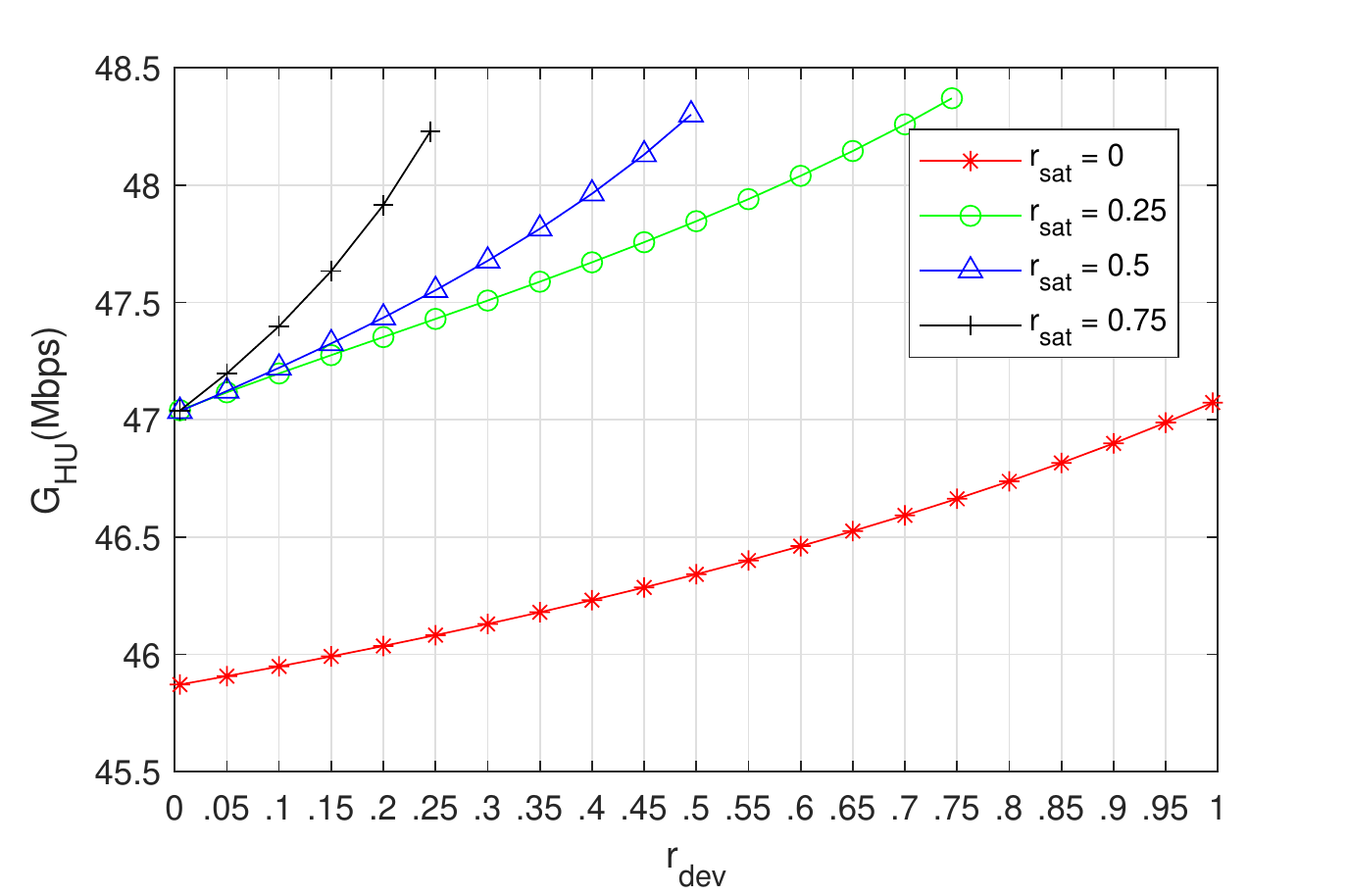}
\label{fig:thr_RSATRBSRDEV_r_sat_fixed_r_dev_x_axis_close}}
\caption{Analytical results for varying $r_{dev}$ values where $r_{sat}$ is fixed ($r_{dev} = 1 - r_{sat} - r_{BS}$).
\label{fig:varyingRDEVResults}}
\end{figure*}

We also inspect the network where both BS and D2D mode are on in terms of analytical $\epbHU$ (Fig.~\ref{fig:EE_RSATRBSRDEV_r_sat_fixed_r_dev_x_axis_close}) and analytical $\goodputHU$ (Fig.~\ref{fig:thr_RSATRBSRDEV_r_sat_fixed_r_dev_x_axis_close}) for $r_{sat}$$\in$$\{0, 0.25, 0.5, 0.75\}$ configurations from a closer perspective. 
In Fig.~\ref{fig:EE_RSATRBSRDEV_r_sat_fixed_r_dev_x_axis_close}, for some fixed D2D mode weight $r_{dev}$, $\epbHU$ increases with decreasing $r_{sat}$ (e.g. for $r_{dev}$= 0.2 when $r_{sat}$ decreases from 0.75 to 0, $\epbHU$ increases from 0.177 $\mu$Jpb to 0.204 $\mu$Jpb.). This is due to the increase in BS usage with decreased $r_{sat}$ where the BS is costly in terms of energy consumption. This leads to more energy being consumed per successful transmitted bit.
In Fig.~\ref{fig:thr_RSATRBSRDEV_r_sat_fixed_r_dev_x_axis_close}, for some fixed $r_{dev}$, the satellite usage decreases with decreasing $r_{sat}$ and the overall goodput decreases (e.g. for $r_{dev}$=0.2 when $r_{sat}$ decreases from 0.75 to 0, $\goodputHU$ decreases from 47.9 Mbps to 46.04 Mbps.). With lower $r_{sat}$, selection of the satellite decreases and the advantage of the large satellite cache and the satellite link is less utilized and this leads to decrease in $\goodputHU$. Especially, when the satellite mode is off ($r_{sat}$=0), the satellite cache and the satellite link are never utilized and this results in evident decrease in $\goodputHU$.

In general, for fixed $r_{sat}$$\in$$\{0,0.25,0.5,0.75\}$, $\epbHU$ decreases with increased $r_{dev}$ (decreased $r_{BS}$) as in Fig.~\ref{fig:EE_RSATRBSRDEV_r_sat_fixed_r_dev_x_axis_close}. Observing a decrease in $\epbHU$ with increased D2D mode and decreased BS mode selection is natural as HU devices consume less energy compared to BS for the transmission of the same content both due to lower power levels ($\DeviceTxPower<\PerChannelBSTxPower$) and shorter service durations. In Fig.~\ref{fig:thr_RSATRBSRDEV_r_sat_fixed_r_dev_x_axis_close}, for fixed $r_{sat}$$\in$$\{0,0.25,0.5,0.75\}$, $\goodputHU$ increases with increasing $r_{dev}$ (decreasing $r_{BS}$). The reason is the short duration of D2D services to HUs and thus new coming HU requests can find the terrestrial frequency that is used for D2D mode in idle state with a higher probability. This way, we observe the increase in the overall system goodput $\goodputHU$. However, this increase is bounded as HU devices have small cache capacities and finding requested content is not always easy due to this cache limitation.  


\section{Conclusions}
\label{sec:conclusion}

In this paper, we model a heterogeneous satellite-terrestrial network with D2D and cognitive communications as a CTMC where the network consists of various unit types (the satellite, the BS and HU devices). We integrate universal source concept and caching into our model. We also enable overlaying in D2D mode operations. Our users operate as PUs over the satellite link and secondary users (SUs) over the terrestrial link. According to our results, the popularity-aware caching algorithm that we integrated into our model outperforms random caching in terms of overall system goodput for high user request rate $\lambdaHU$ regime. Based on our simulations, popularity-aware caching algorithm attains slightly better overall system goodput than LRU and FIFO for larger $\lambdaHU$'s. 
The integration of the universal source does not affect EE especially under light HU content request rate. However, it does not affect overall system goodput performance negatively and besides, allows access to content outside the network boundaries.
Apart from this, the enabling of overlaying in D2D mode improves both energy efficiency (EE) and overall system goodput. 
Our users operate in SU mode in the terrestrial link. Thereof, we inspect how our system is affected by higher PU arrival rate to the terrestrial link $\lambdaPUTer$. In that case, EE is apparently not affected. However, the overall system goodput decreases. 
When we look at our mode selection studies, turning D2D mode weight off ($r_{dev}$ = 0) both deteriorates EE and decreases overall system goodput. In addition, the increase of D2D mode weight improves EE. But during the increase of D2D mode weight when BS mode weight is assigned to zero, we can not utilize BS cache and overall system goodput decreases.

As future work, we plan to include content layering into our CTMC model for inspecting our system from content point of view. Integration of queuing theory is also an option for better modelling of system capacity. We will be focusing on content chunks and look at different chunk prioritization schemes
Advanced prioritization techniques of services in channels and queues provide another research direction. Simple RA scheme and complex content caching algorithm where content chunks will have different priorities for storage is also a promising topic for potential research.



\begin{thebibliography}{10}
\providecommand{\url}[1]{#1}
\csname url@samestyle\endcsname
\providecommand{\newblock}{\relax}
\providecommand{\bibinfo}[2]{#2}
\providecommand{\BIBentrySTDinterwordspacing}{\spaceskip=0pt\relax}
\providecommand{\BIBentryALTinterwordstretchfactor}{4}
\providecommand{\BIBentryALTinterwordspacing}{\spaceskip=\fontdimen2\font plus
\BIBentryALTinterwordstretchfactor\fontdimen3\font minus
  \fontdimen4\font\relax}
\providecommand{\BIBforeignlanguage}[2]{{%
\expandafter\ifx\csname l@#1\endcsname\relax
\typeout{** WARNING: IEEEtran.bst: No hyphenation pattern has been}%
\typeout{** loaded for the language `#1'. Using the pattern for}%
\typeout{** the default language instead.}%
\else
\language=\csname l@#1\endcsname
\fi
#2}}
\providecommand{\BIBdecl}{\relax}
\BIBdecl

\bibitem{6046158}
F.~Alagoz and G.~Gur, ``Energy efficiency and satellite networking: A holistic
  overview,'' \emph{Proceedings of the IEEE}, vol.~99, no.~11, pp. 1954--1979,
  Nov 2011.

\bibitem{CISCOWhitePaper}
C.~W. Paper, ``Cisco visual networking index: Forecast and methodology,
  2016-2021,'' June 2017.

\bibitem{7492943}
S.~Kafiloglu, G.~G\"{u}r, and F.~Alag\"{o}z, ``Modeling and analysis of content
  delivery over satellite integrated cognitive radio networks,'' in \emph{2016
  14th International Symposium on Modeling and Optimization in Mobile, Ad Hoc,
  and Wireless Networks (WiOpt)}, May 2016, pp. 1--8.

\bibitem{7904715}
G.~G\"{u}r and S.~Kaf{\i}lo\u{g}lu, ``Layered content delivery over satellite
  integrated cognitive radio networks,'' \emph{IEEE Wireless Communications
  Letters}, vol.~6, no.~3, pp. 390--393, June 2017.

\bibitem{7752740}
Y.~Long, Y.~Cai, D.~Wu, and L.~Qiao, ``Content-related energy efficiency
  analysis in cache-enabled device-to-device network,'' in \emph{2016 8th
  International Conference on Wireless Communications Signal Processing
  (WCSP)}, Oct 2016, pp. 1--5.

\bibitem{Yao2015}
H.~Yao, C.~Fang, C.~Qiu, C.~Zhao, and Y.~Liu, ``A novel energy efficiency
  algorithm in green mobile networks with cache,'' \emph{EURASIP Journal on
  Wireless Communications and Networking}, vol. 2015, no.~1, p. 139, May 2015.

\bibitem{7919304}
J.~Zhang, X.~Zhang, M.~A. Imran, B.~Evans, Y.~Zhang, and W.~Wang, ``Energy
  efficient hybrid satellite terrestrial {5G} networks with software defined
  features,'' \emph{Journal of Communications and Networks}, vol.~19, no.~2,
  pp. 147--161, April 2017.

\bibitem{7915715}
Y.~Xu and F.~Liu, ``{QoS} provisionings for device-to-device content delivery
  in cellular networks,'' \emph{IEEE Transactions on Multimedia}, vol.~19,
  no.~11, pp. 2597--2608, Nov 2017.

\bibitem{8327582}
L.~Li, G.~Zhao, and R.~S. Blum, ``A survey of caching techniques in cellular
  networks: Research issues and challenges in content placement and delivery
  strategies,'' \emph{IEEE Communications Surveys Tutorials}, pp. 1--1, 2018.

\bibitem{8254837}
M.~C. Lee and A.~F. Molisch, ``Individual preference aware caching policy
  design for energy-efficient wireless {D2D} communications,'' in
  \emph{GLOBECOM 2017 - 2017 IEEE Global Communications Conference}, Dec 2017,
  pp. 1--7.

\bibitem{Yao2014}
H.~Yao, T.~Huang, C.~Zhao, X.~Kang, and Z.~Liu, ``Optimal power allocation in
  cognitive radio based machine-to-machine network,'' \emph{EURASIP Journal on
  Wireless Communications and Networking}, vol. 2014, no.~1, p.~82, May 2014.

\bibitem{7504216}
J.~F. Schmidt, M.~K. Atiq, U.~Schilcher, and C.~Bettstetter, ``Encouraging
  device-to-device communications to improve energy efficiency in cellular
  systems,'' in \emph{2016 IEEE 83rd Vehicular Technology Conference (VTC
  Spring)}, May 2016, pp. 1--5.

\bibitem{7565115}
M.~Br\"{u}ckner, P.~Drie{\ss}, M.~Osdoba, and A.~Mitschele-Thiel, ``A
  dependency-aware {QoS} system for mobile satellite communication,'' in
  \emph{2016 IEEE Wireless Communications and Networking Conference}, April
  2016, pp. 1--6.

\bibitem{4413145}
H.~Su and X.~Zhang, ``Cross-layer based opportunistic {MAC} protocols for {QoS}
  provisionings over cognitive radio wireless networks,'' \emph{IEEE Journal on
  Selected Areas in Communications}, vol.~26, no.~1, pp. 118--129, Jan 2008.

\bibitem{7410054}
A.~Asheralieva and Y.~Miyanaga, ``{QoS}-oriented mode, spectrum, and power
  allocation for {D2D} communication underlaying {LTE-A} network,'' \emph{IEEE
  Transactions on Vehicular Technology}, vol.~65, no.~12, pp. 9787--9800, Dec
  2016.

\bibitem{7483447}
B.~Yu and Q.~Zhu, ``A {QoS}-based resource allocation algorithm for {D2D}
  communication underlaying cellular networks,'' in \emph{2016 Sixth
  International Conference on Information Science and Technology (ICIST)}, May
  2016, pp. 406--410.

\bibitem{6029344}
C.~H. Chen, C.~L. Wang, and C.~T. Chen, ``A resource allocation scheme for
  cooperative multiuser {OFDM}-based cognitive radio systems,'' \emph{IEEE
  Transactions on Communications}, vol.~59, no.~11, pp. 3204--3215, November
  2011.

\bibitem{7542599}
L.~Xu, C.~Jiang, Y.~Shen, T.~Q.~S. Quek, Z.~Han, and Y.~Ren, ``Energy efficient
  {D2D} communications: A perspective of mechanism design,'' \emph{IEEE
  Transactions on Wireless Communications}, vol.~15, no.~11, pp. 7272--7285,
  Nov 2016.

\bibitem{7434298}
B.~Narottama, A.~Fahmi, B.~Syihabuddin, and A.~J. Isa, ``Cluster head rotation:
  a proposed method for energy efficiency in {D2D} communication,'' in
  \emph{2015 IEEE International Conference on Communication, Networks and
  Satellite (COMNESTAT)}, Dec 2015, pp. 89--90.

\bibitem{7556318}
Z.~Wang, H.~Shah-Mansouri, and V.~W.~S. Wong, ``How to download more data from
  neighbors? a metric for {D2D} data offloading opportunity,'' \emph{IEEE
  Transactions on Mobile Computing}, vol.~16, no.~6, pp. 1658--1675, June 2017.

\bibitem{6761239}
K.~Suksomboon, S.~Tarnoi, Y.~Ji, M.~Koibuchi, K.~Fukuda, S.~Abe, N.~Motonori,
  M.~Aoki, S.~Urushidani, and S.~Yamada, ``Popcache: Cache more or less based
  on content popularity for information-centric networking,'' in \emph{38th
  Annual IEEE Conference on Local Computer Networks}, Oct 2013, pp. 236--243.

\bibitem{7841535}
B.~Feng, H.~Zhou, H.~Zhang, J.~Jiang, and S.~Yu, ``A popularity-based cache
  consistency mechanism for information-centric networking,'' in \emph{2016
  IEEE Global Communications Conference (GLOBECOM)}, Dec 2016, pp. 1--6.

\bibitem{7841508}
E.~B. Abdelkrim, M.~A. Salahuddin, H.~Elbiaze, and R.~Glitho, ``A hybrid
  regression model for video popularity-based cache replacement in content
  delivery networks,'' in \emph{2016 IEEE Global Communications Conference
  (GLOBECOM)}, Dec 2016, pp. 1--7.

\bibitem{Nekovee}
M.~Nekovee, ``A survey of cognitive radio access to {TV} white spaces,''
  \emph{International Journal of Digital Multimedia Broadcasting}, vol. 2010,
  2010.

\bibitem{6955966}
H.~J. Kang and C.~G. Kang, ``Mobile device-to-device ({D2D}) content delivery
  networking: A design and optimization framework,'' \emph{Journal of
  Communications and Networks}, vol.~16, no.~5, pp. 568--577, Oct 2014.

\bibitem{7562057}
C.~Liu and B.~Natarajan, ``Average achievable throughput in {D2D} underlay
  networks,'' in \emph{2016 IEEE Conference on Computer Communications
  Workshops (INFOCOM WKSHPS)}, April 2016, pp. 118--123.

\bibitem{8292555}
C.~G\"{u}ven, S.~Bayhan, G.~G\"{u}r, and S.~Eryigit, ``Optimal resource
  allocation for content delivery in {D2D} communications,'' in \emph{2017 IEEE
  28th Annual International Symposium on Personal, Indoor, and Mobile Radio
  Communications (PIMRC)}, Oct 2017, pp. 1--5.

\end{thebibliography}
\end{document}